\documentclass[11pt, oneside]{Thesis} 

\usepackage[square, numbers, comma, sort&compress]{natbib} 
\title{\ttitle} 
\usepackage{amsmath, amsthm, amsfonts, amssymb,mathrsfs}
\usepackage{graphicx,color,epstopdf}
\usepackage{bm}	

\usepackage{cleveref}

\numberwithin{equation}{section}
\numberwithin{figure}{section}

\usepackage{tikz}
\usetikzlibrary{decorations.pathmorphing,shapes}
\usepackage{slashed}
\usepackage[makeroom]{cancel}
\usepackage{caption}
\usepackage{subcaption}
\usepackage{notoccite}
\usepackage{parskip}

\usepackage{lmodern}
\DeclareMathOperator{\Tr}{Tr}

\newcommand{\R}[1]{$R_{#1}$}
\newcommand{\Amp}{\mathscr{M}}

\begin{document}

\frontmatter 

\setstretch{1.3}

\fancyhead{} 
\rhead{\thepage} 
\lhead{}
\pagestyle{fancy} 

\newcommand{\HRule}{\rule{\linewidth}{0.5mm}} 
\hypersetup{pdftitle={\ttitle}}
\hypersetup{pdfsubject=\subjectname}
\hypersetup{pdfauthor=\authornames}
\hypersetup{pdfkeywords=\keywordnames}

%----------------------------------------------------------------------------------------
%	TITLE PAGE
%----------------------------------------------------------------------------------------

\begin{titlepage}
\begin{center}

\textsc{\LARGE \univname}\\[1.5cm] 
\textsc{\Large Masters Thesis}\\[0.5cm] 

\HRule \\[0.4cm] 
{\huge \bfseries \ttitle}\\[0.4cm] 
\HRule \\[1.5cm] 
 
\begin{minipage}{0.4\textwidth}
\begin{flushleft} \large
\emph{Author:}\\
{\authornames} 
\end{flushleft}
\end{minipage}
\begin{minipage}{0.4\textwidth}
\begin{flushright} \large
\emph{Supervisor:} \\
{\supname} 
\end{flushright}
\end{minipage}\\[3cm]
 
\large \textit{A thesis submitted in fulfilment of the requirements\\ for the degree of \degreename}\\[0.3cm] % University requirement text
\textit{in the}\\[0.4cm]
\deptname\\[2cm]
{\large \today}\\[4cm] 
 
\vfill
\end{center}

\end{titlepage}

%----------------------------------------------------------------------------------------
%	DECLARATION PAGE
%----------------------------------------------------------------------------------------

\Declaration{

\addtocontents{toc}{\vspace{1em}} 

I, \authornames, declare that this thesis titled, '\ttitle' and the work presented in it are my own. I confirm that:

\begin{itemize} 
\item[\tiny{$\blacksquare$}] This work was done wholly while in candidature for a research degree at the University of Cape Town.
\item[\tiny{$\blacksquare$}] Where I have consulted the published work of others, this is always clearly attributed.
\item[\tiny{$\blacksquare$}] Where I have quoted from the work of others, the source is always given. With the exception of such quotations, this thesis is entirely my own work.\\
\end{itemize}
 
Signed:\\
\rule[1em]{25em}{0.5pt}
Date:\\
\rule[1em]{25em}{0.5pt}
}

\clearpage

%----------------------------------------------------------------------------------------
%	ABSTRACT PAGE
%----------------------------------------------------------------------------------------

\addtotoc{Abstract}

\abstract{\addtocontents{toc}{\vspace{1em}} 

Recent surprising discoveries of collective behaviour of low-$p_T$ particles in $pA$ collisions at LHC hint at the creation of a hot, fluid-like QGP medium. The seemingly conflicting measurements of non-zero particle correlations and $R_{pA}$ that appears to be consistent with unity demand a more careful analysis of the mechanisms at work in such ostensibly minuscule systems.  We study the way in which energy is dissipated in the QGP created in $pA$ collisions by calculating, in pQCD, the short separation distance corrections to the well-known DGLV energy loss formulae that have produced excellent predictions for $AA$ collisions.  We find that, shockingly, due to the large formation time (compared to the $1/\mu$ Debye screening length) assumption that was used in the original DGLV calculation, a highly non-trivial cancellation of correction terms results in a null short path length correction to the DGLV energy loss formula.  We investigate the effect of relaxing the large formation time assumption in the final stages of the calculation -- doing so throughout the calculation adds immense calculational complexity -- and find, since the separation distance between production and scattering centre is integrated over from $0$ to $\infty$, $\gtrsim 100\%$ corrections, even in the large path length approximation employed by DGLV.

\clearpage 

%----------------------------------------------------------------------------------------
%	ACKNOWLEDGEMENTS
%----------------------------------------------------------------------------------------

\setstretch{1.3} 

\acknowledgements{\addtocontents{toc}{\vspace{1em}}} 

I am  first and foremost indebted to my supervisor, Dr.\ W.\ A.\ Horowitz, for eternally patient guidance and vigorous encouragement, for opening up the global scientific stage to me and for teaching me things I had not realised I wanted to know.  I could never have imagined a more fulfilling year, filled with steep learning curves and reward at every turn.  Thank you for making a year like this a reality.

I am extremely grateful for financial support from the National Institute for Theoretical Physics, the National Research Foundation and the University of Cape Town, that has allowed me to pursue this degree, as well as to SA-CERN for travel support.

Dankie aan my my vriende en familie vir hul ondersteuning en begrip vir my afweesigheid. Vir my huishouer se onvoorwaardelike liefde en daaglikse kospakkies, vir die dinge wat my ouers my geleer het en gegun het en vir my suster se vriendskap, vir hierdie dinge waarsonder ek nie sou kon slaag nie, is ek oneindig dankbaar.
\clearpage

%----------------------------------------------------------------------------------------
%	LIST OF CONTENTS/FIGURES/TABLES PAGES
%----------------------------------------------------------------------------------------

\pagestyle{fancy}

\lhead{\emph{Contents}} 
\tableofcontents 
\lhead{\emph{List of Figures}} 
\listoffigures

%----------------------------------------------------------------------------------------
%	THESIS CONTENT - CHAPTERS
%----------------------------------------------------------------------------------------

\mainmatter 

\pagestyle{fancy}

\chapter{Introduction} 

\label{Introduction}

\lhead{ \emph{Introduction}}

\section{Motivation}
While the SKA will soon be pushing the boundaries of South African physics, particle physics rivals cosmology on the global stage as one of the most fruitful field of discovery physics, and it is truly the heavy ion particle physicists that are pulling down the veil of obscurity surrounding the underlying structure of the universe humanity finds itself in.  The heavy ion community has all but universally accepted that a new state of matter is routinely created in colossal particle colliders such as the Large Hadron Collider (LHC) at CERN in Geneva \cite{Wiedemann2013}, Switzerland and the Relativistic Heavy Ion Collider (RHIC) \cite{Gyulassy2005} at Brookhaven National Labratory (BNL) in the United States.  This new state of matter, called the Quark Gluon Plasma (QGP), is thought to be a deconfined state of the most fundamental constituents of matter, quarks and gluons, that is brought about by the extreme temperatures and energy densities created by colliding heavy nuclei like gold at RHIC and lead at LHC at ultra-relativistic energies. The QGP created in accelerators offers unique insight into the nature of the basic components of matter and it is therefore imperative that mankind investigate its properties.  To date it is known that the QGP is extremely short lived, with a hadronization time on the order of $10$ fm/c \cite{Wiedemann2013}, an attribute that excludes the use of any externally produced probes to discover the properties of this medium.  In order to characterize the QGP then, one must rely on self-generated probes -- probes that are produced along with the medium.  

In a heavy ion collision (hereafter `$AA$ collisions'), thousands of particles are produced ($\sim7000$ in central AuAu at RHIC \cite{Teaney2009}, $\sim10 \ 000$ in central PbPb at the LHC \cite{Grosse-Oetringhaus2014}).  The legion of particles are subjected to an intricate process that starts with a difficult-to-determine initial nuclear parton distribution, followed by `melting' into the QGP which itself evolves obscurely before hadronizing into particles that finally move out to a detector.  Particle physicists are limited to data taken meters away from a medium that is only femto-meters across and must therefore rely on evidence from a conglomerate of phenomena in order to make scientifically sound deductions regarding the nature of the QGP phase of the collision.  In essence, what can be learnt from a detector is only the type and momentum of the particles produced.  Theorists must then attempt to interpret the measured distributions of the particles and their momenta.  We will focus in the Introduction on two major signatures of the QGP that appear in the momentum distributions of the particles.  The first is known as `flow', which indicates collective behaviour, while the second is known as `quenching' or `particle suppression', which indicates that particles have lost energy as they propagated through the medium.  The dual phenomena of elliptic (and higher order) flow of soft (or low transverse momentum, $p_T$) observables and the quenching of hard (or high $p_T$) observables are expected wherever a hot medium is created \cite{Wiedemann2013}.

 An early solution to the problem of controlling the initial state was to simply attempt to measure it;  in principle one should be able to determine the properties of the initial state if one can probe the nucleus without creating a medium.  Since it was inconceivable that hot nuclear matter could be created in a collision between a heavy ion and a proton, RHIC performed a series of experiments involving different combinations of protons, deuterons and a range of heavier nuclei \cite{Back2003} from which it was concluded that $R_{pA}\approx 1$ (or $R_{dAu}$), suggesting that there were no hot nuclear matter effects present in these mixed experiments and that the results served as a decent control measure with which to carefully analyse $AA$ data.    Following suit, the LHC then performed a p-Pb (proton--lead) run in order to reproduce these results, but, shockingly, discovered signatures of hot QCD matter present in their $pA$ experiments.  Most remarkably, eight particle correlations seen by CMS \cite{GranierdeCassagnac2014} (shown in Figure \ref{picCMSCorr}), but also measurements of $v_2$, $v_3$ and even $v_4$ (the Fourier modes of particle correlations) at ATLAS \cite{Steinberg2014}.

\begin{figure}
\centering
\includegraphics[scale=0.4]{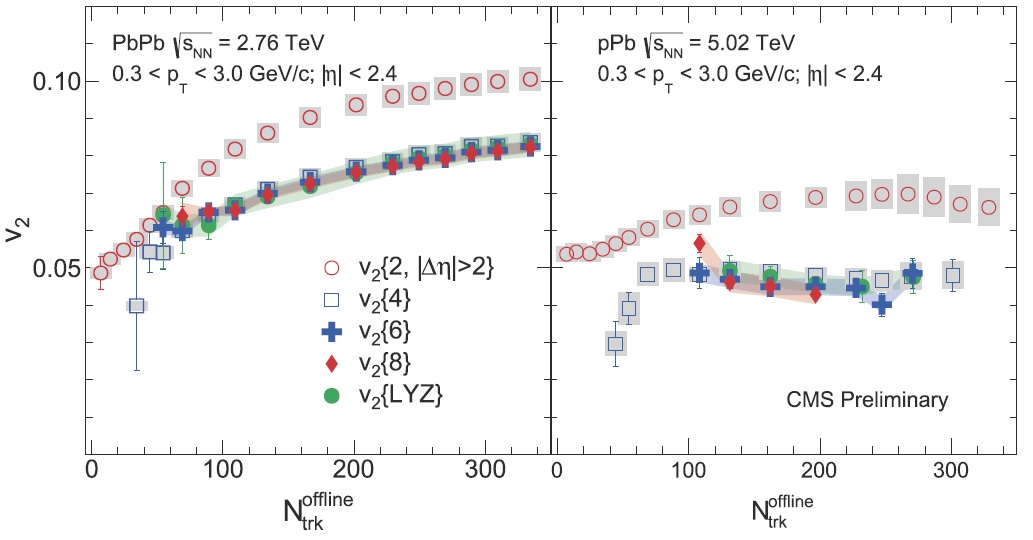}
\caption{Multi-particle correlations in pPb experiments at the CMS detector \cite{GranierdeCassagnac2014}.\label{picCMSCorr}}
\end{figure}

The presence of what appears to be collective behaviour in seemingly impossibly small systems (as any system of interacting particles was expected to be in a $pPb$ or $dAu$ experiment) has raised a number of uncomfortable questions:  Firstly, collective behaviour is well understood within the framework of viscous hydrodynamics \cite{Teaney2009}, but hydrodynamics is a statistical approach and  effectively assumes an infinitely large system of particles in order to describe their collective behaviour \cite{Huovinen2006}; Is a system of no more than $\sim 2$ fm large enough for the assumptions in hydrodynamics to hold?  Secondly, could it be possible that  the number of interacting nucleons does not always scale in the same way?  That is, is it clear that the scaling is \textit{always} by the number of participants (the sum of the particles entering into the collision) or the number of binary collisions (each participant might collide more than once)?  So do we have legitimate confidence in constraints placed on the extent of the effects of geometry and nucleon distributions from $R_{pA}$?  In light of the fact that centrality and multiplicity in $pPb$ show a much broader correlation (left hand panel of Figure \ref{picpAmultiplicity}), can we even say that we have determined the centrality, and with it the number of participants, in a $pPb$ collision clearly?  In short this second question boils down to asking whether or not measuring $R_{pA}$ is as simple as originally thought, whether the measurement can in fact be made reliably, and if not, the extent to which the $R_{AA}$ results are under experimental control.  A third problem, though not new and certainly not unique to the $pA$ case:  Virtually all QGP phenomena can be described by either low $p_T$, strong coupling physics like hydrodynamics \cite{Huovinen2006}, or high $p_T$, weak coupling physics like perturbative QCD (pQCD)\cite{Gyulassy1999a}, but if we believe that both formalisms describe nature accurately, then the coupling must run, and yet it is still by no means clear \textit{how} the coupling is to run. We must attempt to discover whether or not high momentum observables can be understood in pA experiments within the framework of perturbative QCD (pQCD), as they have been in AA collisions \cite{Djordjevic2004}.  If a QGP does exist, the experiments must show energy loss.

While not monopolizing QGP observables, $v_2$ and $R_{AA}$ offer important insight into collective behaviour and jet quenching, the careful study of which is the key to gleaning information about the QGP.  However, the experimental data are mute without a rigorous theoretical understanding of the underlying mechanisms that produce the observed phenomena.   Jet quenching has been exceptionally well described by a number of energy loss formalisms, but in this work the GLV (Gyulassy, Levai, Vitev) setup \cite{Gyulassy1999, Gyulassy1999a, Gyulassy2001} was followed as it is pedagogically insightful.  The GLV description of radiative energy loss has been immensely successful in describing the energy loss of partons moving through a QGP that is created in very central $AA$ collisions, as can be seen in Figure \ref{picTShirt}. Figure \ref{picTShirt} shows that initial state effects are well under control since we see no suppression of the (colorless) photons.  Furthermore, particles with very different masses, the $\pi^0$ and $\eta$ ($\sim$ 135 MeV and $\sim$ 548 MeV, respectively), show similar suppression, indicating that energy is lost partonically rather than hadronically - that is, the constituent quarks have already experienced energy loss by the time they are able to hadronize into observable hadrons.

\begin{figure}
\centering
\includegraphics[scale=0.6]{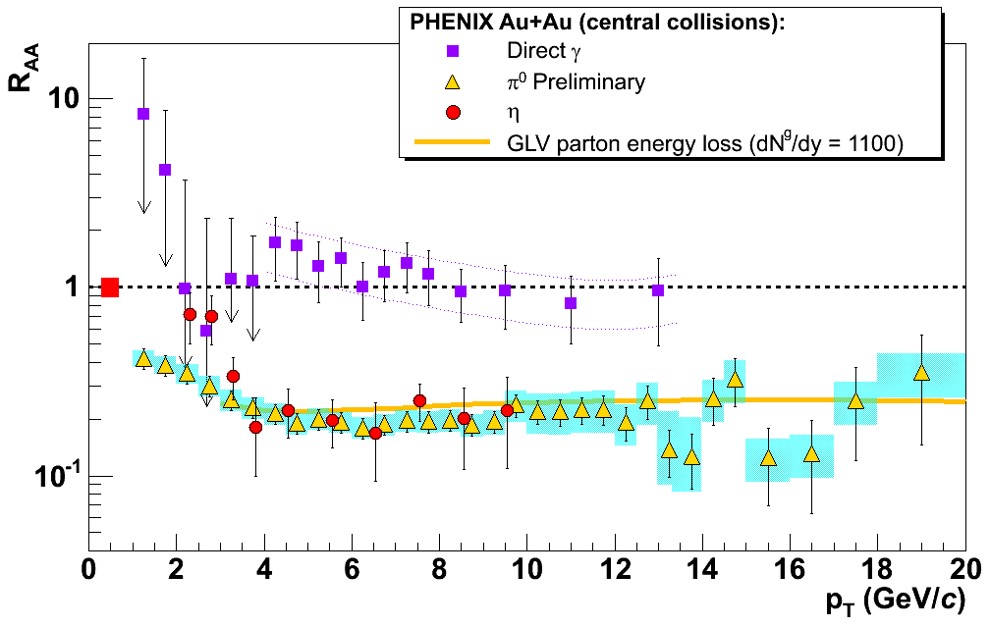}
\caption{$R_{AA}$ for AuAu at RHIC compared to GLV energy loss for $\gamma$, $\pi^0$ and $\eta$ \cite{Isobe2006,Lajoie2006}.  \label{picTShirt}}
\end{figure}

The GLV predictions are the result of a number of calculations and corrections to previous works \cite{Gyulassy1999,Gyulassy1999a,Gyulassy2001}, resulting in a sophisticated calculation that takes into account finite time effects (looking at the energy loss of a particle created at $t=0$ inside the medium rather than at $t=-\infty$) and ultimately also generalizing to massive quarks, the DGLV (Djordjevic GLV) result \cite{Djordjevic2004}.  The GLV result makes use of the Gyulassy-Wang model \cite{Gyulassy1993}, which describes in-medium interactions by considering a number of static scattering centres and then looking at a very high momentum parton (a quark or a gluon) that scatters off the centres and is thereby engendered to radiate a gluon.  However, the calculations rely heavily on a simplification that arises when one considers only large separation distances between the static scattering centres.  Considering only large separation distances means that the calculation is only valid for large system sizes (as the separation distance must necessarily be smaller than the system size in order for any scattering to occur).  Therefore, even though the result obtained by DGLV \cite{Djordjevic2004} is fully general for massive quarks and indeed depends linearly on the size of the system, it is in fact not valid for small system sizes since the derivation of the result makes repeated use of a large separation distance assumption. The monumental success of energy loss studies in $AA$ collisions, that have been extended even beyond the DGLV result to include dynamical scattering centres \cite{Djordjevic2009, Djordjevic2008}, provide an excellent starting point for carrying out a short path length extension.

A need has therefore arisen for a quantitative analysis of energy loss in $pA$ collisions –- not only to determine whether or not a droplet of QGP exists in $pA$ experiments, but also to ensure that all relevant effects are properly under control.  In order then to allow the field of high energy particle physics –- in particular heavy ion physics, to advance, it is vital that the effects seen in $pA$ be understood wholly.  A precise statement regarding whether the coupling of the medium is strong or weak, the applicability of competing theories and the extent to which current interpretations of $R_{pA}\sim1$ hold, cannot be made without a falsifiable energy loss calculation.  It is exactly the absence of such a prediction that we aim to address.

\pagebreak
\section{Relevant Concepts and Current Problems}
In this section I review the various concepts and current difficulties that inspired the execution of the calculation in this dissertation.  Although some of the topics below do not enter explicitly into the calculation, they serve as important aspects of the context in which the main calculation is performed.

\subsection*{Centrality and Multiplicity}

\begin{figure}
\centering
\begin{subfigure}[b]{0.45\textwidth}
\centering
\includegraphics[scale=0.3]{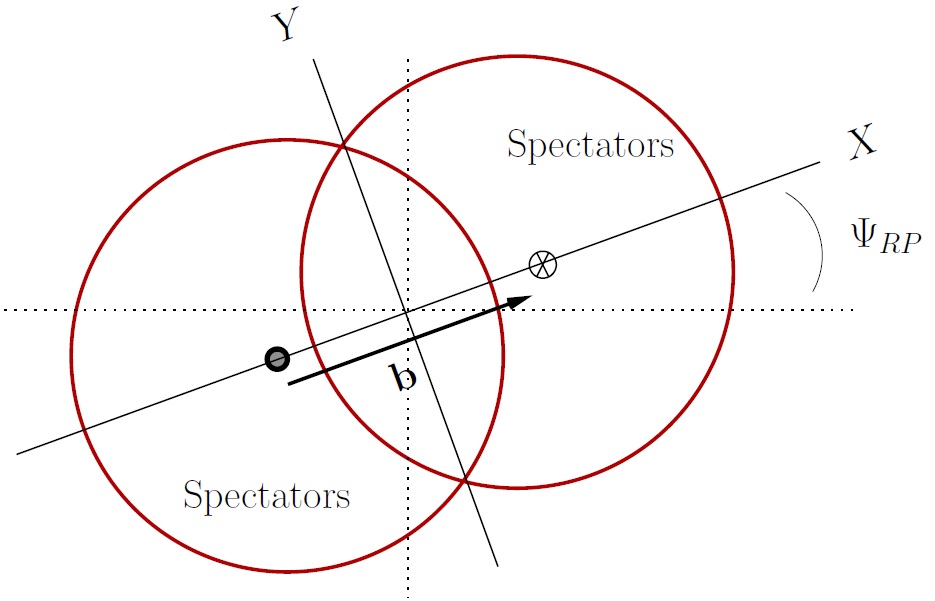}
\caption{\label{picImpactParam}}
\end{subfigure} \quad \begin{subfigure}[b]{0.45\textwidth}
\centering
\includegraphics[scale=0.4]{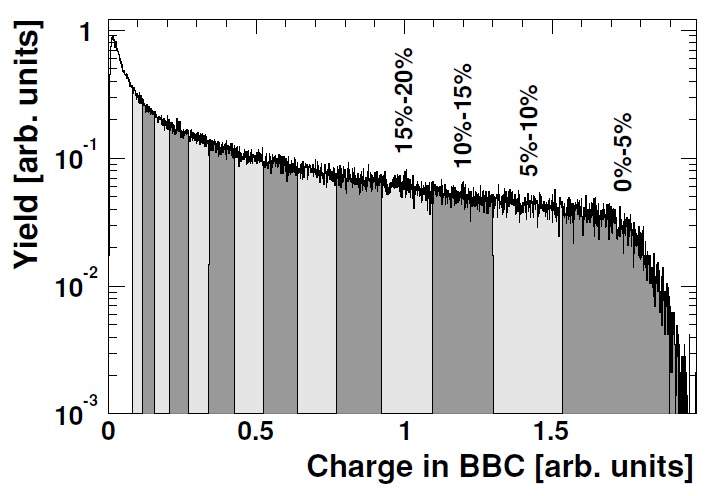}
\caption{\label{picCentralityProb}}
\end{subfigure}
\caption{Impact parameter and centrality. (a) The reaction plane defined by $\Psi_{RP}$ and impact parameter \textbf{b}.\cite{Teaney2009}. (b)Yield as a function of total charge collected in the beam-beam counters \cite{Adler2005}. }
\end{figure}

Figure \ref{picImpactParam} shows the definition of the impact parameter \textbf{b} which is the spatial vector between the centres of the two colliding nuclei.  The magnitude of the impact parameter defines the `centrality' of an event while its direction defines the event raction plane.  For historical reasons, centrality is given as a percentage in such a way that the most central events (smallest impact parameter) have the lowest percentage while more peripheral events (large impact parameter) have a larger percentage.  Figure \ref{picCentralityProb} is a plot of the probability of an event occurring in a given centrality bin, with most of the events that occur in a collider being peripheral ($95\%$ centrality) and only very few events in the $0-5\%$ centrality bin.    It is impossible to know \textit{a priori} what the centrality of any given event is and centrality must therefore be determined on an `event-by-event' basis.  To determine centrality, experimentalists often use an observable called `multiplicity'\footnote{Other determinations of centrality come from, e.g.\ Zero Degree Calorimeters.} which is simply the number of charged hadrons (charged mesons or baryons) measured in the detectors.  In $AA$ collisions, the centrality is directly related to the multiplicity, as shown in the right hand panel of Figure \ref{picpAmultiplicity}, which is a plot of the relationship between multiplicity and the number of participants as calculated in a Glauber Monte-Carlo simulation \cite{Grosse-Oetringhaus2014}.  
 
The number of participant nucleons is closely related to the centrality (and the impact parameter), something that is easily seen geometrically. In the right hand panel of Figure \ref{picpAmultiplicity}, we clearly see that in $AA$, the centrality is very closely related to the multiplicity -- an extremely useful fact since centrality is an important quantity theoretically but must be measured indirectly experimentally through measurements of multiplicity.  However, in the left hand panel, we see a broadening of the correlation, making it much more difficult to accurately relate an experimental measurement of multiplicity in $pA$ to impact parameter.  

\begin{figure}
\centering
\includegraphics[scale=1]{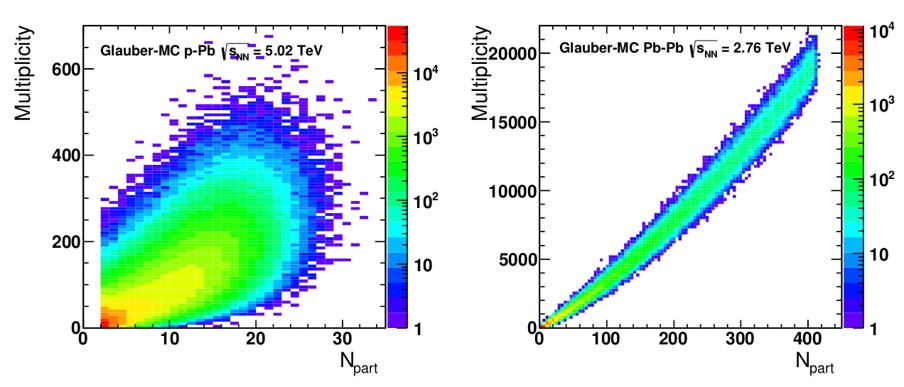}
\caption{The relationship between the number of participating nucleons and the multiplicity in pA collisions (on the left) and AA collisions (on the right).\cite{Grosse-Oetringhaus2014}\label{picpAmultiplicity}}\
\end{figure}

The poor correspondence between centrality and multiplicity in $pA$ is often bypassed by considering `minimum bias' results in which experimental observables are not measured for a given centrality.  Instead, a kind of average over centrality bins is taken of all events, giving a result that is not centrality dependent.  This smearing of centrality dependence is problematic in searches for the creation of `droplets' of QGP in $pA$ experiments because it is only expected that the QGP be created in very central $pA$ collisions - since it is only in these events that one can expect a high enough energy density and enough particles to form an interacting system \cite{Huovinen2006}.  However, the converse solution is equally ambiguous; if one were to consider only high multiplicity $pA$ events, the small number of particles involved introduces a selection bias that enters into the uncertainty related to a high $p_T$ jet - one cannot say whether a large number of low momentum particles in a given region are the decay products of one very high momentum particle (a jet) or are simply a collection of soft particles collected in one place due to some geometrical factor. 

We must therefore be clear when discussing multiplicity and centrality:  In $AA$ experiments, the terms are interchangeable, but they must be treated with more care when discussing $pA$ experiments.

\subsection*{$R_{AA}$ and $R_{pA}$}

\begin{figure}
\centering
\includegraphics[scale=0.5]{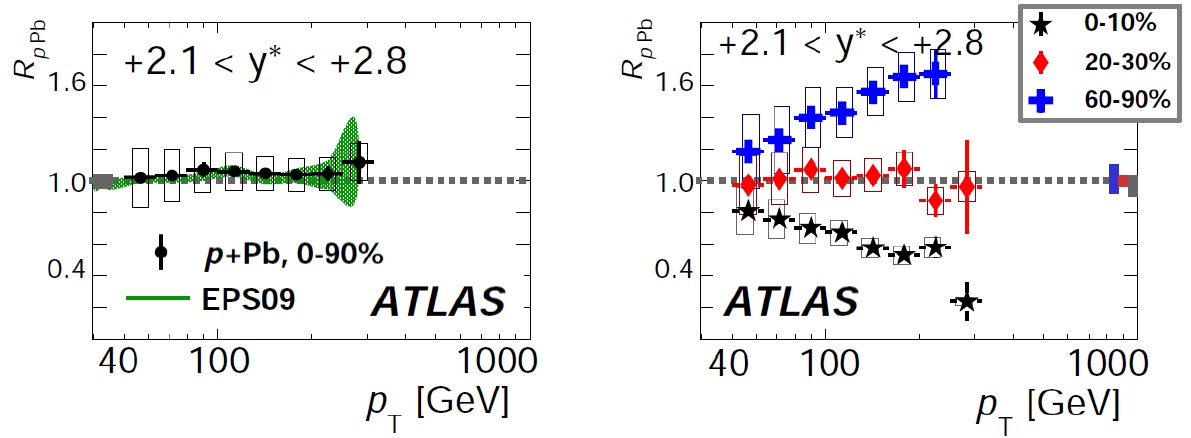}
\caption{A minimum bias $R_{pPb}$ result (left hand panel) vs.\ the same evaluation for different centralities (right hand panel) at forward rapidity presented by the ATLAS collaboration at Hard Probes 2015 \cite{Cole2015}\label{PicATLASRpPb}}.
\end{figure}

$R_{AA}$ is an experimental observable that compares the particle yields in an $AA$ collision to those in $pp$ collisions and can be plotted as a function of, for example, momentum or rapidity.  Equation \eqref{DefRAA} shows how $R_{AA}$, as a function of transverse momentum $p_T$, is expressed in terms of the differential particle number seen in $AA$ collisions ($dN_{AA}/dp_T$), the differential cross section $d\sigma_{pp}/dp_T$ in pp collisions and the average thickness function\footnote{Integrating over the average thickness function gives the average number of binary collisions.} $\langle T_{AA}\rangle$\cite{Miller2007}.
\begin{equation}\label{DefRAA}
R_{AA}(p_T)=\frac{1}{\langle T_{AA}\rangle}\frac{dN_{AA}/dp_T}{d\sigma_{pp}/dp_T}
\end{equation}

An $R_{AA}= 1$ should indicate then that not only is the initial state well understood, but also that there are no final state (medium induced) effects, while an $R_{AA}\neq 1$ should in principle speak to the presence of a medium, provided that all unrelated effects are well under control. One such effect that is not under control is the non-trivial distribution of nucleons within a nucleus which directly affects the counting of $\langle T_{AA}\rangle$ while another is the afore-mentioned determination of multiplicity.   The ambiguities associated with $R_{pA}$ are illustrated in Figure \ref{PicATLASRpPb} which was presented by the ATLAS collaboration \cite{Grosse-Oetringhaus2014} and clearly shows the difficulties in interpretation connected to a claim that $R_{pA} \approx 1$.

The interpretational complexities associated with any $R_{pA}$ results require assurances that energy loss in a small systems is well under control - if one wants to claim that $R_{pA}\sim 1$ and that the scaling and initial state effects are therefore well under control, one must be absolutely sure that the effects governing energy loss are well understood.  Furthermore, the presence of multi-particle correlations (Figure \ref{picCMSCorr})\cite{GranierdeCassagnac2014}, although by no means sufficient support for the existence of hot QCD matter, does prompt the search for a QGP in these small systems.

All successful energy loss calculations result in a (positive power) dependence on the size of the system and the expectation must then be that \R{pA} measurements will yield small results even in the presence of a hot QCD medium. In light of the additional experimental difficulties, it is \textit{crucial} that energy loss for short path lengths be understood from a theoretical standpoint in order to properly interpret \R{pA} measurements.

\subsection*{Other Energy-loss formalisms}

The current work is performed within the GLV framework, but I will discuss briefly here the features of some of the other formalisms that have been developed.  

There are four major formalisms that have been developed for computing partonic energy loss in a hot QCD medium using perturbative QCD (pQCD), each attempting to solve the problem using a different set of assumptions. The first radiative energy loss calculations were performed by BDMPS (Baier, Dokshitzer, Mueller, Peign\'{e} and Schiff) \cite{Baier1995,Baier1996,Baier1997,Baier1998} in which multiple soft gluon scattering was considered.  The major break from multiple soft gluon scattering was made by GLV in a series of papers \cite{Gyulassy1993,Gyulassy1999,Gyulassy1999a} in which they developed a formalism to handle hard scattering in a thin plasma (where only a few scatterings are expected to occur).  The `Reaction Operator' formalism that GLV developed led to two more methods; ASW (Armesto, Salgado and Wiedemann)\cite{ASW} and the 'Higher Twist' \cite{Majumder2011} formalism.  An alternative method was developed by AMY (Arnold, Moore, Yaffe) which is an effective Hard Thermal Loop theory \cite{Arnold2002}.  Djordjevic and Heinz \cite{Djordjevic2008} made the next major leap in energy loss calculations by considering dynamical scattering centres \cite{Djordjevic2009} and it has recently been shown \cite{Blagojevic2015} that indeed, a dynamical approach is necessary to describe data more accurately. However, the static case provided theorists with excellent intuition in $AA$ studies and therefore constitutes a useful first step that may provide insight with which to proceed.

The analysis of energy loss in $AA$ collisions is exhaustive and extensive and has in recent years been developed to an extremely sophisticated level \cite{Blagojevic2015,Buzzatti2013} taking into account a range of different effects including finite time considerations, both collisional and radiative energy loss, energy loss in both static and dynamic media, finite magnetic mass effects and the running of the coupling.  In the near future it will be the task of theorists to extend this range of studies to short path lengths.  We endeavour to lead this new analysis by starting, as was done in $AA$, with the static scattering case and generalizing the DGLV result for heavy quarks to include short separation distances.

\chapter{Formalism and Setup} 

\label{Formalism} 

\lhead{\emph{Formalism}} 

\section{Notation and Conventions}\label{notationsAndConventions}

In this dissertation, for consistency with both \cite{Djordjevic2004,Gyulassy2001}, I have used the following notation for vectors:

\begin{itemize}
\item	\textbf{p}:	Transverse 2D vectors.
\item	$\vec{\textbf{p}}=(p_z,\textbf{p})$:	3D vectors
\item	$p=(p^0,\vec{\textbf{p}})=[p^0+p^z,p^0-p^z,\textbf{p}]$: four vectors in Minkowski and Light Cone coordinates respectively.
\end{itemize}

The calculations are performed in Minkowski space, but we will use many values derived in light cone coordinates.  The dot products are, in `mostly-minus' Minkowski and light cone respectively,
\begin{align}
p\cdot k	&=p^0k^0-p^zk^z-\textbf{p}\cdot\textbf{k}\nonumber\\
			&=\frac{1}{2}\bigg(p^+k^-+p^-k^+\bigg)-\textbf{p}\cdot\textbf{k},
\end{align}
and the transformation rule
\[\begin{cases}
p^0&=\frac{1}{2}\big(p^++p^-\big)\\
p^z&=\frac{1}{2}\big(p^+-p^-\big)\\
\textbf{p}&=\textbf{p}
\end{cases} 
\Rightarrow \left.
  \begin{cases}
p^+&=p^0+p^z\\
p^-&=p^0-p^z\\
\textbf{p}&=\textbf{p}.
\end{cases}
\right.
\]

Since the calculation is done in QCD, there will necessarily be color factors.  Notations differ and so, in an attempt to be absolutely clear, the color exchanges are handled using the applicable $SU(N_c)$ generator $T_a(n)$ in the $d_n$ dimensional representation of the target.  That is $d_n$ refers either to the dimension of the fundamental representation ($d(N)$ in \cite{Peskin1995}) in which the quarks live, or the dimension of the adjoint representation ($d(G)$ in \cite{Peskin1995}) in which the gluons live.  The generators $T_a(n)$ are traceless, but obey (\cite{Djordjevic2004, Peskin1995})
\begin{equation}
\Tr \big(T_a(i)T_b(j)\big)=\delta_{ij}\delta_{ab}C_2(i)\frac{d_i}{d_A}\Leftrightarrow
	\Tr(t^at^b)=\delta_{ab}\frac{d(r)}{d(G)}C_2(r),
\end{equation}
from the definition of the Casimir operator (in DGLV notation on the left and Peskin notation on the right).  Therefore, in the notation used in this calculation (we will follow DGLV), $d_A$ is the dimension of the adjoint representation and $C_2(i)$ the Casimir operator of whichever representation (adjoint or fundamental) has dimension $d_i$.  In fact, when the color algebra is done explicitly, we will use the simplifying notation $a\equiv t_a$ for the generator of the representation of the parton, with $aa=C_R\textbf{1}_R$.  Although these numbers are known (and can be found in \cite{Peskin1995}), the constants will be kept in algebraic form for clarity.

\begin{figure}
\centering
\includegraphics[scale=0.25]{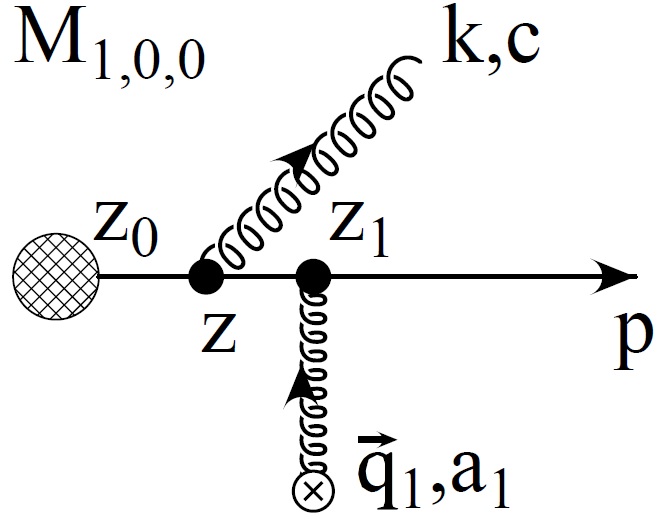}
\caption{$\Amp_{1,0,0}$.\cite{Gyulassy1993}\label{picDiagramM100}}
\end{figure}

A short hand for energy ratios will prove useful notationally as it has in the DGLV calculation.  Consider therefore the diagram \ref{picDiagramM100} defining the relevant momenta. The energy of the radiated gluon (with transverse momentum $\textbf{k}$) is given by $\omega\approx\frac{1}{2}xE^+=xE=xp^z$, with $x$ the fraction of energy (or momentum) carried away by the radiated gluon and E the energy of the hard parton. We will then use

\begin{align}\label{omegas}
\omega				&\approx\frac{xE^+}{2}\approx\frac{xP^+}{2}\nonumber\\
\omega_0			&=\frac{\textbf{k}^2}{2\omega}\nonumber\\
\omega_i			&=\frac{(\textbf{k}-\textbf{q}_i)^2}{2\omega}\\
\omega_{(ij)}		&=\frac{(\textbf{k}-\textbf{q}_i-\textbf{q}_j)^2}{2\omega}\nonumber\\
\tilde{\omega}_m	&=\frac{m_g^2+M^2x^2}{2\omega}\nonumber
\end{align}

Lastly, we will use a number of relations in the color manipulations.  These are from \cite{Peskin1995}
\begin{align}
[a,b]	&=ab-ba=if^{abc}c\nonumber\\
tr[a]	&=0\nonumber\\
f^{acd}f^{bcd}	&=C_2(G)\delta^{ab}\\
aa		&=C_2(r)\hat{\mathbf{1}}\nonumber\\
tr[ab]	&=C(r)\delta^{ab}\nonumber\\
f^{abc}bc	&=\frac{1}{2}iC_2(G)a\nonumber\\
\Tr(T_{a_1}T_{a_2})	&=\frac{C_2(T)d_T}{d_A}\delta_{a_1,a_2}\nonumber
\end{align}

\pagebreak
\section{Assumptions}
The present calculation is performed within the GLV formalism, the calculational details of which are discussed in Section \ref{Formalism}.  Within this formalism, a number of assumptions are made that affect both the ease of the calculation and the physical interpretation of the result. In this section I discuss these assumptions briefly.

\subsection*{The Eikonal Approximation (High energy)}
In the eikonal approximation, the largest scale of the problem is the original energy of the hard parton.  The momentum vectors that are relevant here are derived in section \ref{SecCompkpq} to be
\begin{align}
k	&=\bigg[xP^+, \frac{m_g^2+\textbf{k}^2}{xP^+},\textbf{k}\bigg]\nonumber\\
p	&=\bigg[(1-x)P^+,\frac{M^2+\textbf{k}^2}{(1-x)P^+},-\textbf{k}\bigg]\nonumber\\
q	&=[q^+,q^-,\textbf{q}].
\end{align}
Now, $E^+$ the largest scale of the problem means that $P^+\gg P^-$. The eikonal approximation therefore leads to the relation \cite{Gyulassy2001}
\begin{equation}\label{scaleSetup}
E^+\gg k^+\gg k^- \equiv \omega_0 \sim\omega_{(i\ldots j)}
\end{equation}

\subsection*{Soft gluon and soft rescattering}
The radiated gluon carries away a momentum fraction $x$ of the momentum of the parent parton.  The soft gluon approximation, $x\ll 1$, allows for a number of simplifications. (Note that, although $x$ is a small number, $E^+$ is so large that we may assume $xE^+\gg\vert\textbf{k} \vert$.)

First, the eikonal approximation allows us to assume that the source current that describes the hard production of the parent parton varies slowly with momentum $p$.  The scale on which this variation occurs is the Debye screened length $\frac{1}{\mu}$, the range of which is determined by the momentum transfers from the scattering potential. Therefore we have that, in light of both the soft radiation and eikonal approximations,
\begin{equation}
J(p-q+k)\approx J(p+k)\approx J(p).
\end{equation}

We can also say the following because of $x\ll1$:
\begin{align}
\epsilon_\mu(k)(p-q+k)^\mu&\approx\frac{\boldsymbol\epsilon\cdot\textbf{k}}{x}\nonumber\\
	k\cdot(p-q)&\approx k\cdot(p)\approx\frac{\textbf{k}^2}{2x}
\end{align}

We may also now extend equation \eqref{scaleSetup} to read
\begin{equation}
E^+\gg k^+\gg k^- \equiv \omega_0 \sim\omega_{(i\ldots j)}\gg\frac{(\textbf{p}+\textbf{k})^2}{E^+}.
\end{equation}
\subsection*{Impact parameter }
Here we assume that the impact parameter (describing the overlap of the nuclei in the collision) varies over a large transverse area $A_\perp$ relative to the interaction area $1/\mu^2$.  This assumption allows for a simplification to occur when performing the ensemble average over initial states.  The impact parameter average reduces to
\begin{align}
\langle\cdots\rangle=\int\frac{d^2\textbf{b}}{A_\perp}\ldots,
\end{align}
which allows the ensemble average over the phase factor to become
\begin{align}
\langle e^{-i(\textbf{q}-\textbf{q}')\cdot \textbf{b}}\rangle=\frac{(2\pi)^2}{A_\perp}\delta^2(\textbf{q}-\textbf{q}').
\end{align}

It is worth noting that the ensemble average over the phase factor containing the impact parameter can still be reduced in this manner in $pA$ since the transverse area over which the impact parameter varies is still large in comparison to the reaction area:  the reaction area is $1/\mu^2\sim 1/(0.5 GeV)^2\sim0.1$ fm$^2$ while the overlap is, in central $pPb$ which we consider, $\pi (1fm)^2 \approx 3$fm$^2$
\subsection*{Large formation time $\omega_i \ll \mu_1$}\label{tauFormSection}

A finite formation time \cite{Gyulassy1993},
\begin{align}
\tau(k)\sim\frac{\hbar}{\Delta E(k)}\sim \frac{2 \omega}{\textbf{k}^2}\sim\frac{2}{\omega \theta},
\end{align}
for $\theta =\textbf{k}/\omega$, the angle at which the gluon is radiated, leads to a destructive interference phenomenon for particles emitted at small angles (with large momenta).  The effect is known as the Landau-Pomeranchuck-Migdal (LPM) effect \cite{Baier1996a, Zakharov1996}, or coherent radiation limit, and causes a strong suppression of induced gluon bremsstrahlung \cite{Gyulassy1999} (not including the `self-quenching' phenomenon discussed in section \ref{Formalism}) because a particle will scatter multiple times before radiating. LPM suppression is reduced in the case of a massive quark \cite{Djordjevic2004} because the formation time of a gluon emitted from a massive quark is reduced due to the non-zero mass effects.

As in the calculation performed by \cite{Djordjevic2004}, we will also assume that the formation time is much larger than the Debye screened length.  
\begin{align}
\mu \gg \omega_i \equiv \frac{\textbf{k}^2}{2 \omega}=\frac{1}{\tau}
\end{align}
This assumption only featured lightly in the computation of the DGLV result, but it will play a crucial simplifying role in the calculation presented in this dissertation and I will therefore elaborate on it briefly.  

By assuming that the formation time is large compared to the inverse Debye screened length, one is simply restating the GW model \cite{Gyulassy1993} - partons scatter off Debye screened scattering centres and therefore cannot resolve any structure of the scattering centre that might affect the interaction in a complicated manner.  One expects to find that the energy loss due to radiation goes to zero for separation distances (separation between scattering centres) going to zero. However, if the formation length is larger than the mean free path one enters the LPM regime where a particle must scatter multiple times before radiating.

\pagebreak
\section{Formalism}\label{Formalism}
The calculation presented in the subsequent sections is a short path-length generalization of the result obtained by \cite{Djordjevic2004}.  Therefore, the entire calculation is performed within the GLV formalism, based on the Gyulassy and Wang model \cite{Gyulassy1993}, then developed in a number of papers in which various assumptions were refined and corrected \cite{Gyulassy1999,Gyulassy1999a,Gyulassy2001}, and finally generalized by Magdalena Djordjevic in a paper \cite{Djordjevic2004} in which the GLV energy loss formula was generalized for heavy quarks and whose results are reproduced and added to in this dissertation.  In this section I briefly review the GW (Gyulassy-Wang) model, following \cite{Gyulassy1993,Gyulassy1999,Gyulassy1999a,Gyulassy2001}.

The GLV model considers the induced gluon radiation of a hard parton as it interacts with a random color field produced by a color neutral ensemble of static partons \cite{Gyulassy1993}.  The setup therefore requires a means to handle (1) the production of a highly off-shell hard parton, (2) the vacuum radiation associated with such a hard parton and (3) the interactions with the medium that stimulate the emission of a soft gluon.  Since we work in the high energy limit with a low momentum transfer, the spin characteristics of the parent parton can be ignored and we therefore deal with scalar QCD, a theory that is similar to scaler Quantum Electrodynamics (sQED), with the difference residing only in the addition of a treatment of color exchange.  The setup below is therefore analogous to the sQED setup treated in \cite{Itzykson1980}.

For (1), consider the production (at finite time, in contrast to a jet that has been prepared in the infinite past as in \cite{Gunion1982})  of a hard, highly off-shell parton with momentum $p$.  The parton is created along with the medium and is therefore (initially) localized at $x_0^\mu=(0,\vec{\textbf{x}})^\mu$.  The amplitude is assumed to vary slowly with $p$ and can be written as \cite{Gyulassy2001}
\begin{equation}
\label{MJ}
\Amp_J= iJ(p)e^{ipx_0}\mathbf{1}_R
\end{equation}
\begin{figure}
\centering
\includegraphics[scale=0.25]{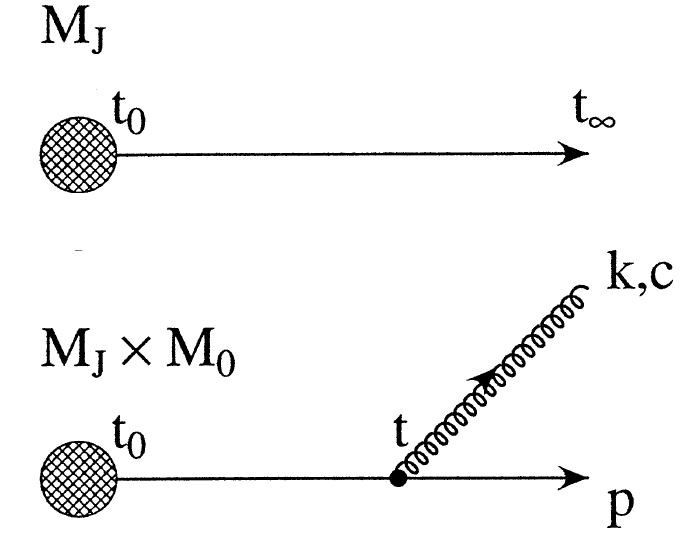}
\caption{$\Amp_{J}$ and $\Amp_{J}\otimes\Amp_{0}$.\cite{Gyulassy1993}\label{picDiagramMJ0}}
\end{figure}

Even in the absence of a medium, a highly off-shell parton will radiate gluons, thereby losing energy (`self-quenching') and softening the spectrum of hard jets.  To first order in $g_s$, the softening of the spectrum of hard jets implied by equation \eqref{MJ} is effected as illustrated in Figure \ref{picDiagramMJ0}, and given (in the eikonal limit), by \cite{Gyulassy1993, Djordjevic2004}

\begin{align}\label{Mvac}
\Amp_{\text{vac}}=(-2ig_s)\frac{\boldsymbol\epsilon\cdot\textbf{k}}{k^2+m_g^2+M^2x^2}e^{i\omega_oz_0}c
\end{align} 

We can therefore express the amplitude for a jet and soft gluon in the final state as \cite{Djordjevic2004}
\begin{align}\label{DefM0}
\Amp_0\equiv\Amp_J\otimes\Amp_{\text{vac}}&=iJ(p+k)e^{i(p+k)x_0}(ig_s)(2p+k)_\mu\epsilon^\mu(k)i\Delta_M(p+k)c\nonumber\\
		&\approx iJ(p)e^{ipx_0}(-2ig_s)\frac{\boldsymbol\epsilon\cdot\textbf{k}}{k^2+m_g^2+M^2x^2}e^{i\omega_0z_0}c
\end{align}

The medium through which this highly off-shell parton is moving will now induce the additional radiation of a gluon that will carry away a fraction $x$ of the parent parton's momentum.  The medium is modelled by GLV \cite{Gyulassy1999} as an ensemble of static scattering centres (that one can perhaps think of as a static, heavy parton), localized at say $\vec{\textbf{x}}_i=(z_i,\textbf{b}_i)$, that are all distributed with the same density
\begin{align}
\rho(\vec{\textbf{q}})=\frac{N}{A_\perp}\bar{\rho}(z).
\end{align} 

The exact form of $\bar{\rho}(z)$ will have to be considered carefully, but for the time being it is only necessary that it be normalized such that $\int dz \rho(z)=1$. Each scattering centre is modelled as a Debye screened (or color screened Yukawa) potential with Fourier and color structure given by
\begin{align}
V_n &= V(\vec{\textbf{q}}_n)e^{-i\vec{q}_n\cdot \vec{x}_n}\nonumber\\
	&= 2\pi\delta(q^0)v(\textbf{q}_n,q^z_n)e^{-i\vec{q}_n\cdot \vec{x}_n}T_{a_n}(R)\otimes T_{a_n}(n),
\end{align}

where
\begin{align}\label{potential}
v(\vec{\textbf{q}}_n)&=v(\textbf{q}_n,q^z_n)\equiv\frac{4\pi\alpha_s}{\vec{\textbf{q}}^2+\mu^2}\nonumber\\
	&=\frac{4\pi\alpha_s}{(q_1^z)^2+\mu_n^2}=\frac{4\pi\alpha_s}{(q_n^z-i\mu_n)(q_n^z+i\mu_n)},
\end{align}

and $\mu_n$ is defined as
\begin{align}
\mu_n^2=\mu_{n\perp}^2\equiv\mu^2+\textbf{q}_n^2.
\end{align}

The effective small transverse momentum, differential elastic cross section is then given by
\begin{align}
\frac{d^2\sigma_{el}}{d^2\textbf{q}}&=\frac{C_RC_2(n)}{d_A}\frac{4\alpha_s^2}{(q_n^2+\mu^2)^2}\nonumber\\
	&=\frac{C_RC_2(n)}{d_A}\frac{\vert v(\textbf{q})\vert^2}{(2\pi)^2}
\end{align}

We are doing an opacity expansion (an expansion in the number of scatterings that occur) and therefore, to first order in opacity (scattering only once), 
\begin{align}
d^3N&=\frac{1}{d_t}\Tr \vert \Amp_0+\Amp_1+\Amp_2 +  \ldots \vert^2
		\frac{d^3\vec{\textbf{p}}}{ (2\pi)^3 2 \vert \vec{\textbf{p}}\vert}\nonumber\\
	&=d^3N_0+\frac{1}{d_T}\Tr\big[\{\Amp_1\Amp_1^*\}+2\Re\Tr\{\Amp_2\Amp_0^*\}
		\frac{d^3\vec{\textbf{p}}}{ (2\pi)^3 2 \vert \vec{\textbf{p}}\vert} + \ldots,
\end{align}

where $\Amp_1$ is the sum of all diagrams with one interaction with a scattering centre and $\Amp_2$ the sum of all diagrams with two interactions with a (single) scattering centre.
First, to obtain the unperturbed inclusive distribution of jets in the wave packet, take the color trace of $\vert \Amp_0\vert^2$ multiplied by the invariant one particle phase space:
\begin{align}\label{d3NJ}
d^3N_J&=\Tr \vert \Amp_0\vert^2\frac{d^3\vec{\textbf{p}}}{ (2\pi)^3 2 \vert \vec{\textbf{p}}\vert}\nonumber\\
	&=\vert J(p)\vert^2 d_R\frac{d^3\vec{\textbf{p}}}{ (2\pi)^3 2 \vert \vec{\textbf{p}}\vert},
\end{align}
the spectrum of which can now be extracted as in \cite{Djordjevic2004}
\begin{align}\label{spectrum}
\vert \Amp_0\vert^2  \frac{d^3\vec{\textbf{q}}}{2E(2\pi)^3}\frac{d^3\vec{\textbf{k}}}{2\omega(2\pi)^3}
		\approx d^3N_Jd^3N_g^{(0)}.
\end{align}
Using equations \eqref{Mvac}, \eqref{d3NJ} and \eqref{spectrum}, one can obtain the radiation spectrum for the vacuum radiation
\begin{align}
\omega\frac{dN_g^{(0)}}{d^3\vec{\textbf{k}}}\approx x\frac{dN_g^{(0)}}{dxd^2\textbf{k}}
		\approx\frac{C_R\alpha_s}{\pi^2}\frac{\textbf{k}^2}{(\textbf{k}^2+m_g^2+x^2M^2)^2}.
\end{align}
In a similar fashion one can obtain the radiation spectrum for the first order in opacity energy loss due to final state medium interactions:
\begin{align}\label{Magda10}
d^3N_g^{(1)}d^3N_J=\Bigg( \frac{1}{d_R}\Tr\big\langle\vert\Amp_1\vert^2\big\rangle
		+\frac{2}{d_R}\Re\Tr\big\langle\Amp_0^*\Amp_2\big\rangle\Bigg)
		\frac{d^3\vec{\textbf{q}}}{2E(2\pi)^3}\frac{d^3\vec{\textbf{k}}}{2\omega(2\pi)^3},
\end{align}
for which the energy loss is given by
\begin{align}\label{ELossExp}
dE_{ind}^{(1)}&=\omega d^3N_g =\cancel{\omega} \bigg(\frac{1}{d_R}\Tr\langle\vert\mathcal{M}_{1}\vert^2
			\rangle+\frac{2}{d_R}\Re\Tr\langle\mathcal{M}^*_0\mathcal{M}_2\rangle\bigg)
			\frac{\cancel{\frac{d^3\vec{\textbf{p}}}{(2\pi)^32p^0}}\frac{d^3\vec{\textbf{k}}}{(2\pi)^32\cancel{\omega}}}
			{d_R\vert J(p)\vert ^2\cancel{\frac{d^3\vec{\textbf{p}}}{(2\pi)^32p^0}}}	\nonumber\\
		&=\frac{1}{d_R\vert J(p)\vert ^2}	\Bigg( \frac{1}{d_R}\Tr\big\langle\vert\Amp_1\vert^2\big\rangle
		+\frac{2}{d_R}\Re\Tr\big\langle\Amp_0^*\Amp_2\big\rangle\Bigg)	\frac{d^3\vec{\textbf{k}}}{(2\pi)^32}.
\end{align}

\chapter{Derivation of Energy Loss Formula}
\lhead{ \emph{Derivation of Energy Loss Formula}}
\label{EnergyLoss}

In this chapter we will derive an expression for the first order in opacity energy loss of a high momentum particle as it propagates through a plasma of thickness $L$, mean free path $\lambda_{mfp}=1/(\sigma\rho)$ and with color electric fields screened on a scale $\mu$.  In order to do so, one must calculate  the invariant matrix elements of each relevant diagram and perform a number of manipulations in order to compute equation \eqref{ELossExp}. Here we present the computed invariant matrix elements of the relevant diagrams and then sketch an outline of the process whereby equation \ref{ELossExp} is computed.  The details of the calculation are presented in the Appendices, sections \ref{AppendixAmplitudes} and \ref{AppendixFormula}.

We will first compute the invariant matrix elements of the Feynman diagrams associated with the leading order terms of the Dyson series expansion.  The amplitudes are calculated by computing the $\int dq_1^z$ integrals, closing the contour below the real axis (because $z_1>z_0$, meaning that $z_1-z_0>0$ and so the exponents that appear will only converge if the contour is closed below the real line).  To perform these integrals, one must find the poles, their residues and therefore the integral.  I present here only the results of the computation of the amplitudes, please see Appendix \ref{AppendixAmplitudes} for details of the results below.
\begin{figure}
\centering
\includegraphics[scale=0.5]{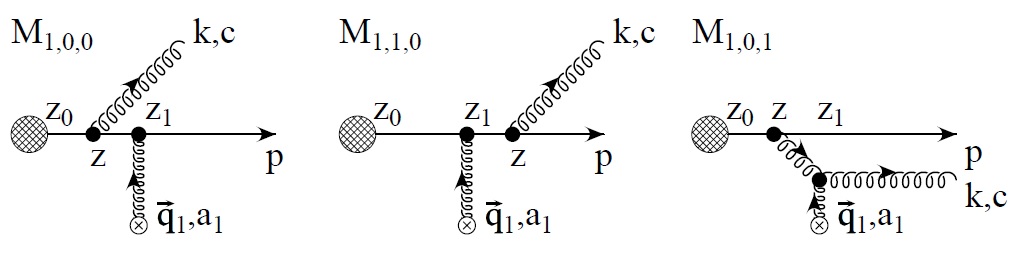}
\caption{Three ``direct'' terms $\Amp_{1,0,0}$, $\Amp_{1,1,0}$, $\Amp_{1,0,1}$ contribute to the soft gluon radiation amplitude to first order in opacity $L/\lambda\propto\sigma_{el}/A_{\perp}$ \label{M1s}}

\end{figure}

The matrix elements for the diagrams in Figure \ref{PicM1Diagram1} are calculated in section \ref{CalcM100}, \ref{CalcM110} and \ref{CalcM110} to be 

\begin{alignat}{2}\label{PicM1Diagram1}
\Amp_{1,0,0}		&\approx &&J(p)e^{i(p+k)x_0}(2ig_s)a_1cT_{a_1}(-i)\int
			\frac{d^2\textbf{q}_1}{(2\pi)^2}e^{-\textbf{q}_1\cdot\textbf{b}_1}v(0,\textbf{q}_1)\times\nonumber\\
	&	&&\times\frac{\boldsymbol\epsilon\cdot\textbf{k}}{\textbf{k}^2+m_g^2+M^2x^2}
			\big[e^{i(\omega_0+\tilde{\omega}_m)}-1\big] \\
\Amp_{1,1,0}	&=&&J(p)e^{ipx_0}(-i)\int\frac{d^2\textbf{q}_1}{(2\pi)^2}v(0,\textbf{q}_1)
			e^{-i\textbf{q}_1\cdot\textbf{b}_1}(-2ig_s)\times\nonumber\\
	&	&&\times\frac{\textbf{k}\cdot\boldsymbol\epsilon}{m_g^2+\textbf{k}^2+x^2M^2}
			\bigg[e^{i(\omega_0+\tilde{\omega}_m)(z_1-z_0)}
			-\frac{1}{2}e^{-\mu_1(z_1-z_0)}\bigg]T_{a_1}ca_1\\
\Amp_{1,0,1}	&\approx	&&J(p)e^{i(p+k)x_0}(-i)\int\frac{d2\textbf{q}_1}{(2\pi)^2}v(0,\textbf{q}_1)
			e^{-i\textbf{q}_1\cdot\textbf{b}_1}2ig_s\times\nonumber\\
	&	&&\times\frac{\epsilon\cdot(\textbf{k}-\textbf{q}_1)}{(\textbf{k}-\textbf{q}_1)^2+M^2x^2+m_g^2} \times\nonumber\\
	&	&&\times \Big(e^{i(\omega_0+\tilde{\omega}_m)(z_1-z_0)}-e^{i(\omega_0-\omega_1)(z_1-z_0)}\Big)[c,a_1]T_{a_1}
\end{alignat}

\begin{figure}
\centering
\begin{subfigure}{1\textwidth}
\includegraphics[scale=0.45]{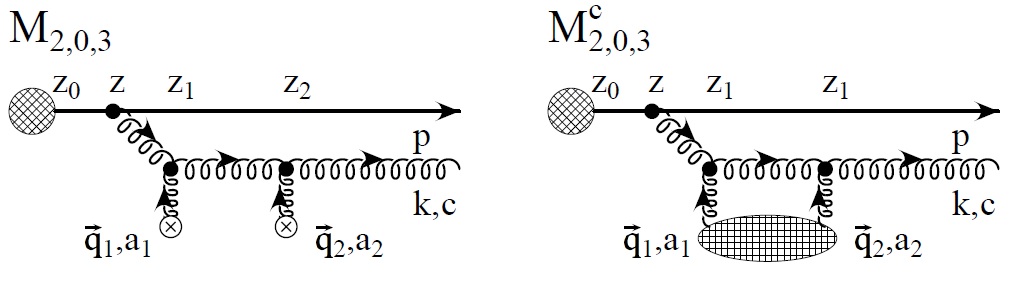}
\caption{$\Amp_{2,0,3}$ ``direct'' contributes to second order in opacity while $\Amp_{2,0,3}^c$ contributes to first order in opacity. }
\end{subfigure}

\begin{subfigure}{1\textwidth}
\includegraphics[scale=0.5]{M200}
\caption{$\Amp_{2,0,0}$ graphs for the well separated case as well as the contact limit.}
\end{subfigure}

\begin{subfigure}{1\textwidth}
\includegraphics[scale=0.5]{M220}
\caption{$\Amp_{2,2,0}$ diagrams in both the well separated and contact limit case.}
\end{subfigure}

\begin{subfigure}{1\textwidth}
\includegraphics[scale=0.4]{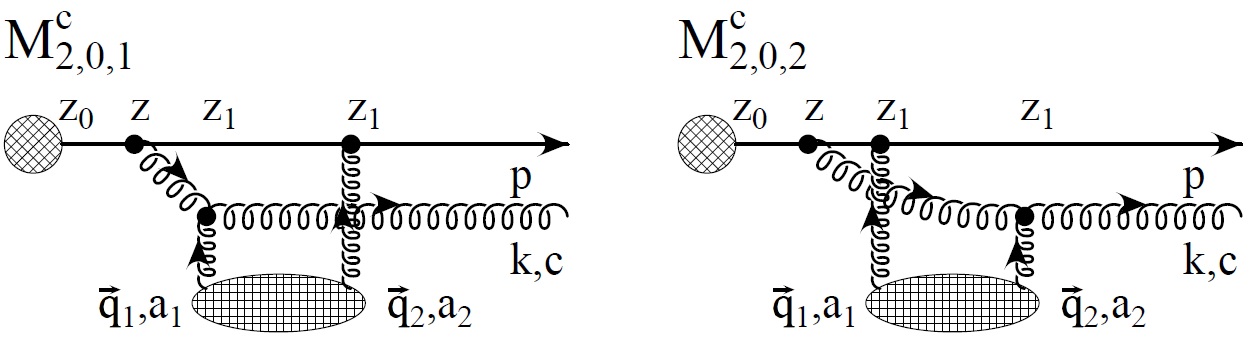}
\caption{Diagrams $\Amp_{2,0,1}$ and $\Amp_{2,0,2}$ showing only the special contact case.}
\end{subfigure}

\caption{Two scattering centre diagrams, taken from \cite{Djordjevic2004}\label{PicM2Diagram1}}.
\end{figure}

Notice that it is only the $\Amp_{1,1,0}$ diagram that contributes a short separation distance correction.

In order to properly perform the Dyson series expansion, we must consider diagrams with two interactions with one scattering centre, described in DGLV as two scattering centres in the `contact limit' for the first order in opacity calculation. There are $2^n-1$ diagrams for $n$ scatterings, but we will find that not all $7$ contribute.  The diagrams that do contribute are shown in Figure \ref{PicM2Diagram1}, with the contact limit taken in cases where both the contact limit and well separated cases are shown.

The amplitudes for the contact limits of the relevant diagrams are calculated in Sections \ref{CalcM203}, \ref{CalcM200}, \ref{CalcM220} and \ref{CalcM201M202} and given by equations \eqref{ResM203}, \eqref{ResM200}, \eqref{ResM220}, \eqref{ResM201} and \eqref{ResM202}. Surprisingly, only the $\Amp_{2,2,0}$ diagram contributes to the short separation distance generalization.

\begin{alignat}{2}
\Amp_{2,0,3}^c	&\approx &&J(p)e^{i(p+k)x_0}(-i)^2\int\frac{d^2\textbf{q}_1}{(2\pi)^2}v(0,\textbf{q}_1)e^{-i\textbf{q}_1\cdot\textbf{b}_1}
					\int\frac{d^2\textbf{q}_2}{(2\pi)^2}\times\nonumber\\
		&	&&\times v(0,\textbf{q}_2)e^{-i\textbf{q}_2\cdot\textbf{b}_2}
				\frac{1}{2}(2ig_s)\frac{\epsilon\cdot(\textbf{k}-\textbf{q}_1-\textbf{q}_2)}
				{\big[(\textbf{k}-\textbf{q}_1-\textbf{q}_2)^2+M^2x^2+m_g^2\big]}\times\nonumber\\
		&	&&\times\big[[c,a_2],a_1\big]\big(T_{a_2}T_{a_1}\big)\big\lbrace 
				e^{i(\omega_0+\tilde{\omega}_m)(z_1-z_0)}-e^{i(\omega_0-\omega_{(12)})(z_1-z_0)}\big\rbrace\\
\Amp_{2,0,0}^c		&\approx &&J(p)e^{i(p+k)x_0}\int\frac{d^2\textbf{q}_1}{(2\pi)^2}
				\int\frac{d^2\textbf{q}_2}{(2\pi)^2}v(0,\textbf{q}_1)v(0,\textbf{q}_2)e^{-i(\textbf{q}_1+\textbf{q}_1)\cdot\textbf{b}_1}\times\nonumber\\
		&	&&\times\frac{1}{2}\frac{-2ig_s(\epsilon\cdot\textbf{k})}{\textbf{k}^2+m_g^2+M^2x^2}
				\Big(e^{i(\omega_0+\tilde{\omega}_m)(z_1-z_0)}-1\Big)a_2a_1cT_{a_2}T_{a_1}	\\
\Amp_{2,2,0}^c		&\approx	&& J(p)e^{i(p+k)x_0}\int\frac{d^2\textbf{q}_1}{(2\pi)^2}\int\frac{d^2\textbf{q}_2}{(2\pi)^2}
					e^{-i(\textbf{q}_1\textbf{q}_2)\cdot\textbf{b}_1}
					v(0,\textbf{q}_1)v(0,\textbf{q}_1)\times\nonumber\\
		&	&&\times ca_2a_1(T_{a_2}T_{a_1})\frac{1}{2}\frac{-(-2ig_s)(\epsilon\cdot\textbf{k})}{m_g^2+\textbf{k}^2+x^2M^2}\times\nonumber\\ 
		&	&&\times\bigg[e^{i(\omega_0+\tilde{\omega}_m)(z_1-z_0)}+e^{-\mu_1(z_1-z_0)}
				\bigg(1-\frac{\mu_1e^{-\mu_2(z_1-z_0)}}{2(\mu_1+\mu_2)}\bigg)\bigg]\\
\Amp_{2,0,1}^c	&=&&J(p)e^{i(p+k)x_0}(-i)^2\int	\frac{d^2\textbf{q}_1}{(2\pi)^2}e^{-i\textbf{q}_1\cdot\textbf{b}_1}v(0,\textbf{q}_1)v(0,\textbf{q}_2)
				\int\frac{d^2\textbf{q}_2}{(2\pi)^2}\times\nonumber\\
		&	&&\times e^{-i\textbf{q}_2\cdot\textbf{b}_2}2ig_s\frac{\epsilon\cdot(\textbf{k}-\textbf{q}_1)}
				{(\textbf{k}-\textbf{q}_1)^2+M^2x^2+m_g^2}\times\nonumber\\
		&	&&\times\bigg[e^{i(\omega_0+\tilde{\omega}_m)(z_1-z_0)}-e^{i(\omega_0-\omega_1)(z_1-z_0)}\bigg]a_2[c,a_1](T_{a_1}T_{a_2})\\
\Amp_{2,0,2}^c	&=&&J(p)e^{i(p+k)x_0}(-i)^2\int	\frac{d^2\textbf{q}_1}{(2\pi)^2}e^{-i\textbf{q}_1\cdot\textbf{b}_1}v(0,\textbf{q}_1)v(0,\textbf{q}_2)
				\int	\frac{d^2\textbf{q}_2}{(2\pi)^2}\times\nonumber\\
		&	&&\times e^{-i\textbf{q}_2\cdot\textbf{b}_2}2ig_s
				\frac{\epsilon\cdot(\textbf{k}-\textbf{q}_1)}{(\textbf{k}-\textbf{q}_1)^2+M^2x^2+m_g^2}\nonumber\\
		&	&&\bigg[e^{i(\omega_0+\tilde{\omega}_m)(z_1-z_0)}-e^{i(\omega_0-\omega_1)(z_1-z_0)}\bigg]a_1[c,a_2](T_{a_2}T_{a_1})	.	
\end{alignat}

Here we will calculate the first order radiative energy loss.  According to equation (10) \cite{Djordjevic2004}, the starting point is the following quantity
\begin{equation}\label{Magda10}
d^3N_g^{(1)}d^3N_J=\bigg(\frac{1}{d_T}\Tr\langle\vert\Amp_{1}\vert^2\rangle+\frac{2}{d_T}\Re\Tr\langle\Amp^*_0\Amp_2\rangle\bigg)\frac{d^3\vec{\textbf{p}}}{(2\pi)^32p^0}\frac{d^3\vec{\textbf{k}}}{(2\pi)^32\omega}.
\end{equation}
$\Amp_{1}$ is the sum of all diagrams with one scattering centre and $\Amp_{2}$ the sum of all diagrams with two scattering centres.  Therefore, in order to calculate this quantity, we must sum the relevant diagrams first.   Once the summation is done, we square the amplitude (and take the real part of the trace), average over initial and sum over final states.  This means that there is an averaging over impact parameter (introducing a factor of $1/A_\perp$) and a sum over scattering centres.

In this section I outline the process whereby one obtains an expression for the first order (in opacity) energy loss by using equation \eqref{ELossExp}.

To simplify the process below, consider the following short hand:
\begin{align}
f_k			&\equiv	\frac{\boldsymbol\epsilon\cdot\textbf{k}}{m_g^2+\textbf{k}^2+x^2M^2}\nonumber\\
f_q			&\equiv	\frac{\boldsymbol\epsilon\cdot(\textbf{k}-\textbf{q}_1)}{(\textbf{k}-\textbf{q}_1)^2+M^2x^2+m_g^2}\nonumber\\
\omega_{0m}	&\equiv (\omega_0-\tilde{\omega}_m)(z_1-z_0)\nonumber\\
\omega_{01}	&\equiv (\omega_0-\omega_1)(z_1-z_0)\\
-\omega_{1m}	&\equiv \omega_{01}- \omega_{0m}=(\omega_0-\omega_1)(z_1-z_0)-(\omega_0-\tilde{\omega}_m)(z_1-z_0)\nonumber\\
			&=-(\omega_{1}--\tilde{\omega}_m)(z_1-z_0)\nonumber\\
\alpha_3	&\equiv \frac{1}{2} e^{-\mu\Delta z}\nonumber\\
\alpha		&\equiv \Tr(c^2a^2-caca).\nonumber
\end{align}
Using the results from sections \ref{CalcM100}, \ref{CalcM101} and \ref{CalcM110}, given in equations (\ref{ResM100}), (\ref{ResM101}) and (\ref{ResM110}) and performing a rearrangement of terms so as to group like phases together, we obtain the sum of the one scattering centre diagrams, 

\begin{alignat}{3}
\frac{1}{d_T}\Tr&\langle\vert& \Amp_{1}&\vert^2\rangle =N\vert J(p)\vert^2(4g_s^2)\frac{1}{A_T}\int
				\frac{d^2\textbf{q}_1}{(2\pi)^2}\frac{C_2(T)}{d_A}\times\nonumber\\	
		&	&\times	&\Bigg\{4\alpha f_q^2(1-\cos\omega_{0m})+2\alpha f_k^2(1-\cos\omega_{0m})\nonumber\\
		&	&	&-4\alpha f_kf_q(1-\cos\omega_{1m})+\alpha f_qf_k2\cos\omega_{01}+\Tr c^2a^2f_k^2+\nonumber\\
		&	&+	&e^{-\mu\Delta z}\bigg[f_k^2\Tr c^2a^2(\cos\omega_{0m}-1)-\Tr c^2a^2f_k^2\cos\omega_{0m}\nonumber\\
		&	&	&+f_kf_q\alpha(\cos\omega_{0m}-\cos\omega_{01})\bigg]\nonumber\\
		&	&+	&\frac{1}{4}f_k^2\Tr c^2a^2 e^{-2\mu\Delta z}\Bigg\}. \label{M1Full,M1New}
	\end{alignat}

Consider now the results in equations (\ref{ResM203}), (\ref{ResM200}),  (\ref{ResM220}), \eqref{ResM201} and (\ref{ResM202}). They can be summed, squared and averaged in a similar fashion to obtain

\begin{alignat}{3}
\frac{2}{d_T}\langle\Amp_{0}^*\Amp_{2}&\rangle &=N&\vert J(p)\vert^2(4g_s^2)\frac{1}{A_T}\int	
			\frac{d^2\textbf{q}_1}{(2\pi)^2}\frac{C_2(T)}{d_A}\nonumber\\
		&	&\times &\bigg[f_k^2\big(2\alpha\cos\omega_{0m}-2\alpha-\Tr c^2a^2\big)\nonumber\\
		&	&	&+2\alpha f_kf_q(\cos\omega_{0m}-\cos\omega_{01})\nonumber\\
		&	&	&+e^{-\mu \Delta z}f_k^2\Tr c^2a^2\cos\omega_{0m}-\frac{1}{4} e^{-2\mu \Delta z}f_k^2\bigg].\label{M2Full}
\end{alignat}

It now remains simply to add the contributions from equations \eqref{M1Full,M1New} and \eqref{M2Full}, which, after a number of cancellations and substitution into equation \eqref{Magda10}, finally gives

\begin{alignat}{2}\label{ELossShort}
\Delta &E&&_{ind}^{(1)} =\frac{C_R\alpha_sLE}{\pi\lambda_g}	\int  \frac{d^2\textbf{q}_1}{\pi}\frac{\mu^2}{(\mu^2+\textbf{q}_1^2)^2}
				\frac{d^2\textbf{k}}{4\pi} \int dz_1\bar{\rho}(z_1)\times\nonumber\\
	&\times	&&\Bigg[-2f_q(f_k-f_q)(1-\cos\omega_{1m})\nonumber\\
	&		&&+\frac{e^{-\mu(z_1-z_0)}}{2}\bigg\{f_k^2\bigg(1-\frac{2C_R}{C_A}\bigg)\bigg(1-\cos\omega_{0m}\bigg)\nonumber\\
	&		&&+f_kf_q\big(\cos\omega_{0m}-\cos\omega_{01}\big)\bigg\}\Bigg],
\end{alignat}
where  $\bar{\rho}(z_1)$ is the distribution of the scattering centres which we will choose at a later stage (see equation \eqref{rhobar}) and $\lambda_g$ is the gluon mean free path.  The second and third lines of equation \eqref{ELossShort} are then the short separation distance correction terms to the DGLV result. A cursory glance will reveal that our result behaves as can be reasonably expected:  the dimensions of the correction terms are correct and the correction terms are suppressed exponentially in the limit that either $\Delta z \rightarrow \infty$ or $\mu\rightarrow \infty$.  That is, equation \eqref{ELossShort} reduces to the DGLV result in the large separation distance limit.  Furthermore, as expected, equation \eqref{ELossShort} also goes to zero as $\Delta z \rightarrow 0$ due to destructive LPM interference.

However, a surprising and wholly unpredictable cancellation occurs here; coincidentally, for the very same reason that many short separation distance correction terms were suppressed at the amplitude level, we see again that the correction terms are suppressed under the assumption that the formation length be longer than the Debye screening length of the scattering centres.  It is noteworthy that, if one continues to assume that $\mu\gg\omega_i$, a term arising from the sum over single interaction diagrams cancels exactly with a term from the double interaction diagrams, leading to a correction term equal to $0$ under the assumption that $\mu\gg\omega_i$.  To relax the assumption that $\mu\gg\omega_i$ and rederive all expressions, will constitute a monumental task as the assumption led to simplifications already in the DGLV calculations and caused a massive reduction in the number of terms that need to be considered at the amplitude level in the present calculation.

Lastly, note the breakdown of color triviality due to the $C_R/C_A$ term in the second line of equation \eqref{ELossShort}, suggesting that, should $\mu\gg\omega_i$ not hold (causing the retention of the correction terms), one will see a disintegration of the purely gluon final state interpretation in \cite{Gyulassy2001}.

For completeness, the full energy loss formula (that remains to be integrated), is

\begin{alignat}{2}
\Delta E_{ind}^{(1)}& =&&\frac{C_R\alpha_sLE}{\pi\lambda_g}	\int \frac{d^2\textbf{q}_1}{\pi}\frac{\mu^2}{(\mu^2+\textbf{q}_1^2)^2}
				\frac{d^2\textbf{k}}{4\pi} \int d\Delta z\bar{\rho}(\Delta z)\times\nonumber\\
		&\times && \Bigg[-\frac{2\big(1-\cos\big\{(\omega_1+\tilde{\omega}_m)\Delta z\big\}\big)}
				{(\textbf{k}-\textbf{q}_1)^2+M^2x^2+m_g^2}\times\nonumber\\
		&	&&\times\bigg(\frac{(\textbf{k}-\textbf{q}_1)\cdot\textbf{k}}{m_g^2+\textbf{k}^2+x^2M^2}
				-\frac{(\textbf{k}-\textbf{q}_1)^2}{(\textbf{k}-\textbf{q}_1)^2+M^2x^2+m_g^2}\bigg)\nonumber\\
		&	&&+\frac{1}{2}e^{-\mu\Delta z)}\Bigg\{\bigg(\frac{\textbf{k}}{m_g^2+\textbf{k}^2+x^2M^2}\bigg)^2\times\nonumber\\
		&	&&\times\bigg(1-\frac{2 C_R}{C_A}\bigg)
				\bigg(1-\cos\{(\omega_0-\tilde{\omega}_m)\Delta z\}\bigg)\nonumber\\
		&	&&+	\frac{\textbf{k}\cdot (\textbf{k}-\textbf{q}_1)}
				{\big(\textbf{k}^2+m_g^2+x^2M^2\big)\big((\textbf{k}-\textbf{q}_1)^2+M^2x^2+m_g^2\big)}\times\nonumber\\
		&	&&\times\big(\cos\{(\omega_0-\tilde{\omega}_m)\Delta z\}-\cos\{(\omega_0-\omega_1)\Delta z\}\big)\Bigg\}\Bigg]\label{ELossFull}
\end{alignat}
% Chapter Template

\chapter{Results and Conclusions} % Main chapter title

\label{Conclusions} % Change X to a consecutive number; for referencing this chapter elsewhere, use \ref{ChapterX}

\lhead{ \emph{Results and Conclusions}} % Change X to a consecutive number; this is for the header on each page - perhaps a shortened title

In this section I present the results from a numerical analysis of equation \eqref{ELossShort}, an improved energy loss formula, building on the DGLV result to include the effects of short separation distances between production and scattering.  The computation in the previous sections of the energy loss was motivated by a desire to better understand energy loss in small systems, prompting us to consider possible modifications to existing energy loss formulae that could take small system sizes into account.  Our na\"ive approach was to equate the concept of a small system (that is, small $L$) to the idea of small separation distances (small $\Delta z$), justified by the intuition that \textit{if} energy loss occurs in a short, thin medium, the distance between production and scattering of the hard parton must necessarily be small as well. That is to say that, in order to generalize DGLV, we changed the length scale of the problem, taking:
\begin{align}
\frac{1}{\mu}\ll\lambda_{mfp}\ll L \quad\Rightarrow\quad \frac{1}{\mu}\ll L
\end{align}

where $\lambda_{mfp}$ is the mean free path in the QGP.  Therefore, while the calculation was performed with a focus on separation distances, the separation distance is eventually integrated over and one must look at the effect that the correction terms have on the length dependence of the energy loss formula.

Presented here are plots of the numerical evaluation of equation \eqref{ELossShort} after analytically performing the integral over scattering centres.  In computing the $dz_1$ integral, we assumed a distribution of scattering centres $\bar{\rho}(z_1)$ such that
\begin{align}\label{rhobar}
\int_{0}^{\infty}d\Delta z\bar{\rho}(\Delta z)=\int_{0}^{\infty}d\Delta z\exp\bigg\{-2\frac{\Delta z}{L}\bigg\}\frac{2}{L}.
\end{align}

Analyses of light quark quenching at $130$ GeV $AuAu$ \cite{Levai2002} suggest that the effective static opacity of the plasma can reasonably be fixed at $L/\lambda_{mfp}\approx 4$.  Note also that the finite masses of the quarks considered in \cite{Djordjevic2004} shield the singularity at $\textbf{k}\rightarrow 0$ which allows for numerical integration with momentum cut-off at $0$. The numerics employ the following values:  $\mu = 0.5$ GeV, $L=4$ fm where a constant for the length of the system was required, $E=10$ GeV or $E=100$ GeV where a constant parton energy was required, $\lambda_{mfp}=1$ fm, $C_R=4/3$, $C_A=3$, $\alpha_s=0.3$, mass of the charm quark was assumed to be $1.3$ GeV while that of the bottom qark $4.75$ GeV and lastly the QCD analogue of the Ter-Mikayelian plasmon  effect was taken into account by setting the mass of the gluon as $\mu/\sqrt{2}$ \cite{Djordjevic2003,Dokshitzer2001}.  All values were chosen to be consistent with the DGLV plots \cite{Djordjevic2004}.  As in \cite{Djordjevic2004}, kinematic upper limits were used for the momentum integrals such that $0 \leq k\leq 2x(1-x)E$, ensuring collinearity, and $0\leq q \leq \sqrt{3 E \mu}$, due to finite kinematics.  The fraction of momentum carried away by the radiated gluon, $x$ is integrated over from $0$ to $1$.

\section{Numerical Results}

\begin{figure}

\centering
\includegraphics{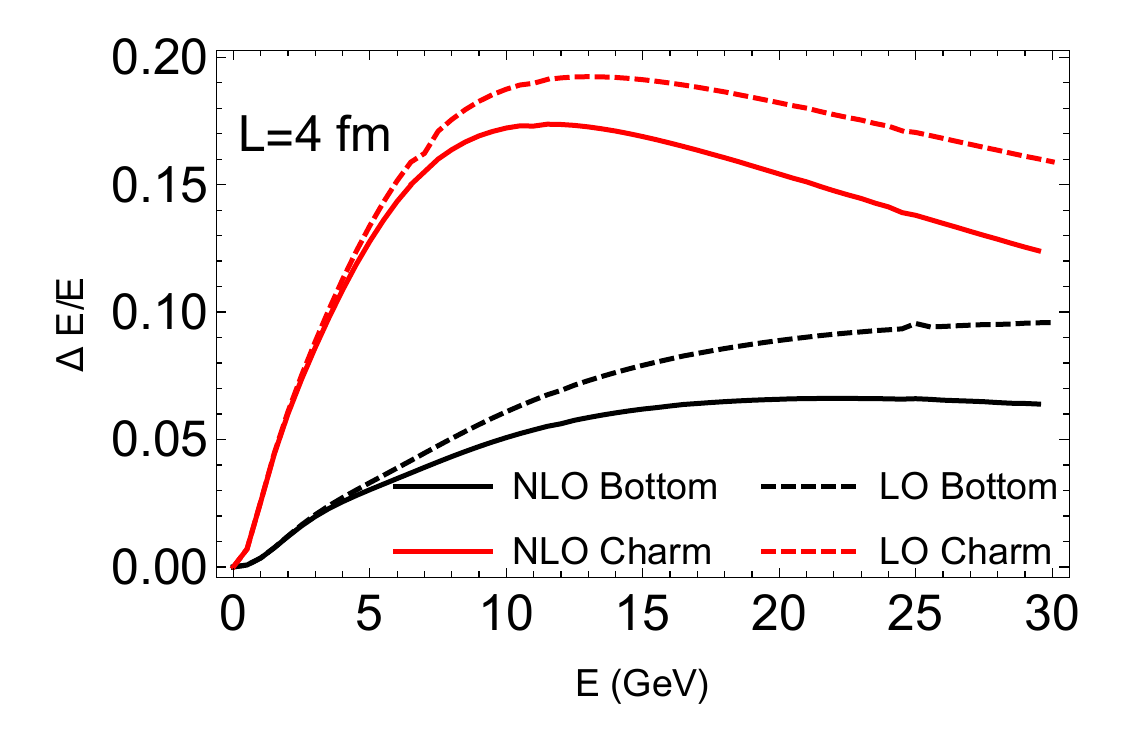}
\caption{Energy loss per unit energy as a function of the parton's initial energy for Leading Order (dashed) and Leading Order with NLO corrections (solid) for bottom and charm quarks.\label{dEbyEtoE}}
\end{figure}

Figure \ref{dEbyEtoE} shows results from numerically integrating equation \eqref{ELossFull} as a function of the initial energy of the hard parton.  We compare the leading order energy loss of charm (because it is an experimentally clear probe \cite{Djordjevic2014}) and bottom quarks to their energy loss when small separation distances are considered.  In the figure captions, NLO refers to Next to Leading Order in separation distance (as compared to the Debye screening length).  Note that the results show a \textit{decrease} in energy loss due to a negative correction to the LO result.  

The plot in Figure \ref{dEbyEtoE} differs slightly from its counterpart in \cite{Djordjevic2004} due to somewhat different kinematic limits in the numerical integration.
\begin{figure}
\centering
\includegraphics{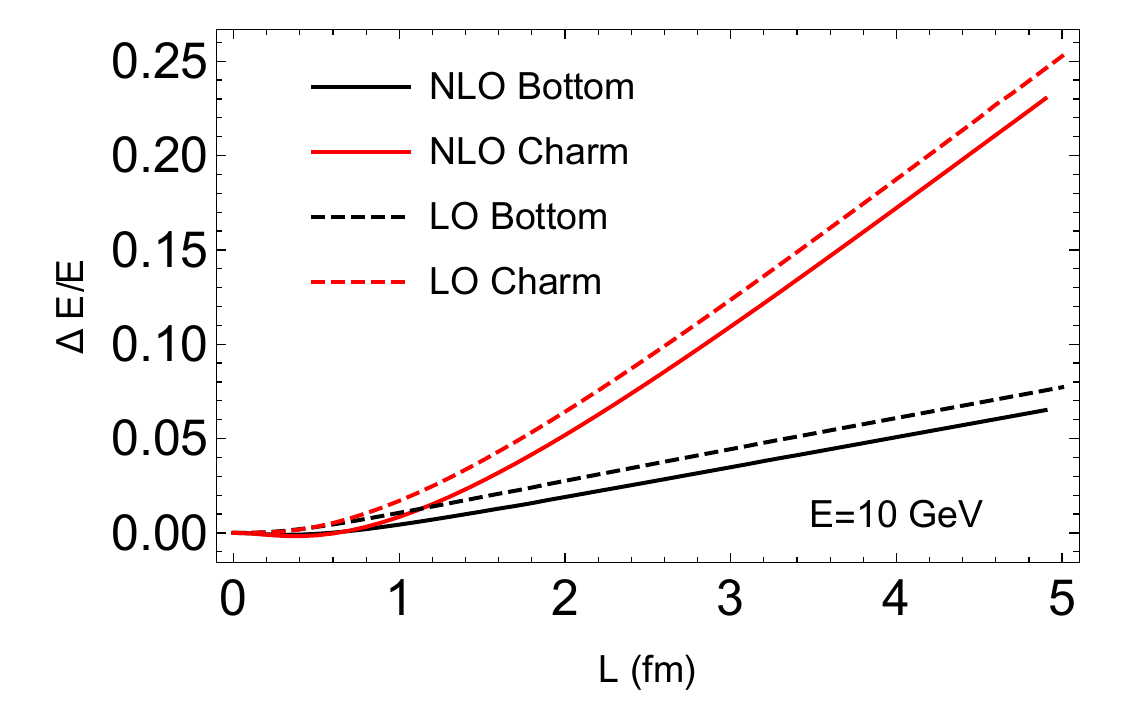}
\caption{Energy loss per unit energy as a function of the static thickness of the medium for Leading Order (dashed) and Leading Order with NLO corrections (solid) for bottom and charm quarks at 10 GeV.\label{dEbyEtoL}}
\end{figure}

Figure \ref{dEbyEtoL} shows the differential energy loss of charm and bottom quarks as a function of the static thickness of the medium for both the LO and the NLO (LO with small separation distance correction).  Again we see here a \textit{reduction} in energy loss to the point that we expect to see a small amount of particle enhancement for systems less than a femtometer across.  A striking feature of the thickness ($L$) dependence of the correction term is that the effect of including short separation distances does not seem to decrease with increasing system size, suggesting that the effect should be taken into account even when performing calculations in the large system size approximation.  The fact that the NLO correction applies even to large systems is not a complete surprise as the $dz_1$ integral is performed over \textit{all} $z_1$, from $0$ to $\infty$.

However, when one considers the same plot for $100$ GeV partons, shown in Figure \ref{dEbyEtoL100}, the negativity of the energy loss becomes so large that one must revisit the premises of the calculation.  The numerics seem to suggest no suppression whatsoever, even for sizeable systems - a postdiction that simply does not describe the plethora of available data for particle suppression in $AA$ \cite{Xu2014,Grosse-Oetringhaus2014,GranierdeCassagnac2014}.

\begin{figure}
\centering
\includegraphics{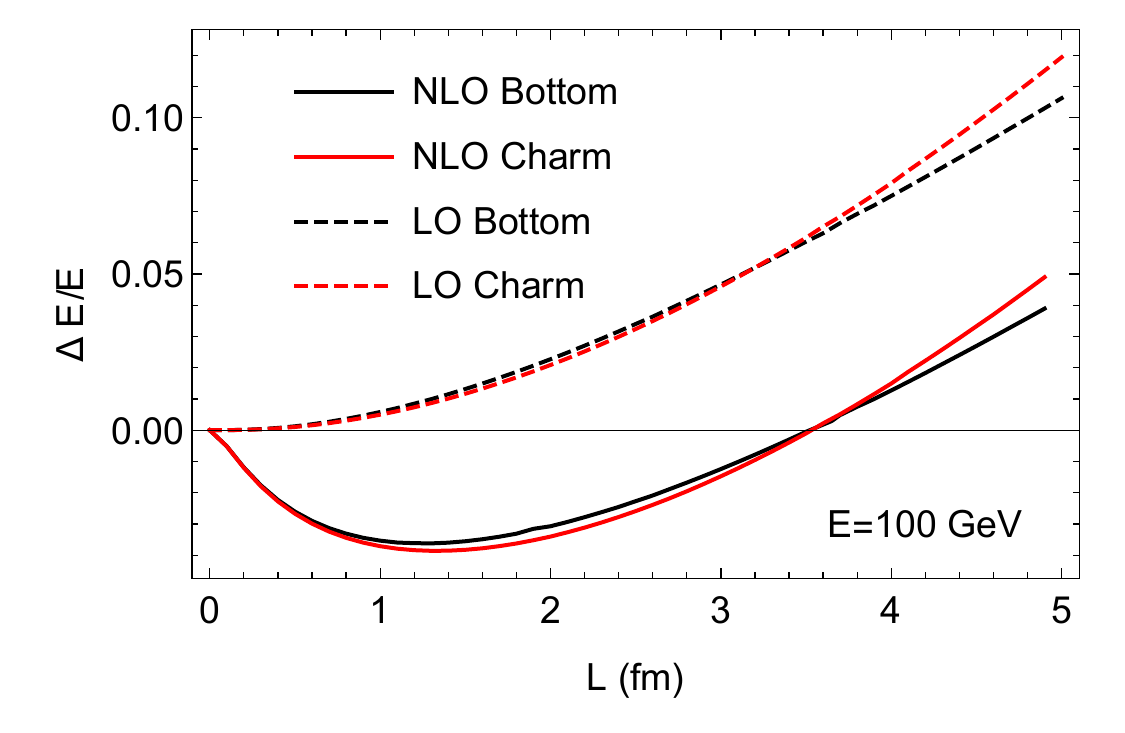}
\caption{Energy Loss per unit energy as a function of the static thickness of the medium for Leading Order (dashed) and Leading Order with NLO corrections (solid) for bottom and charm quarks at 100 GeV.  \label{dEbyEtoL100}}
\end{figure}

\section{Discussion and Conclusion}

In order to interpret these results, it is important to highlight a key feature of the calculation of the correction terms, which involves the assumption that the formation time of the radiated gluon is much larger than the Debye screened length of the scattering centres.  

In the large separation distance limit that was employed by DGLV \cite{Djordjevic2004}, the effects of the large formation time assumption were fairly subdued due to the much larger and more immediately evident $\exp\{-\Delta z\mu\}$ exponential suppression.  However, when the calculation is performed under the assumption that the separation distances are \textit{on the order} of the Debye screened length of the scattering centres, one must deal explicitly with a regime of $\Delta z$ values that are small enough that the formation length becomes the largest length scale\footnote{To be precise, when considering small separation distances, the mean free path is the largest length scale of the problem, but it is not relevant to the calculation because the mean free path is a fixed, calculable quantity that is considered to be much larger than the Debye length of the scattering centres, and therefore larger than the system size, which is the scale we work on.} of the problem.  The small separation distance regime necessarily enhances the effect of the large formation length assumption.  

The large formation length assumption introduced a considerable calculational simplification, leading to a surprising and non-trivial cancellation effect.  The suppression of a large number of NLO terms appeared already at the amplitude level of the calculation, leaving only two amplitudes, one single scattering and one double scattering, with non-zero NLO terms -- even under the assumption that  $\mu_1 \gg \omega_1$.  However, when performing the ensemble average within the Dyson series, the  $\mu_1 \gg \omega_1$ assumption leads to a non-trivial cancellation,  resulting finally in a null correction term to the end result.  Numerically, if one does not assume that the large formation time approximation holds (as we have done here), the correction term contributes more than $100\%$ as seen in Figure \ref{dEbyEtoL100}.  Relaxing the  $\mu_1 \gg \omega_1$ assumption  would entail an immense complication of the calculation (even for large path lengths, the relaxation of the large formation time limit would include extra terms), but it is clear that the effect must be taken into account since, as will be elaborated on in the next paragraph, the small separation distance limit also affects the large system size calculation.

Figure \ref{dEbyEtoL} clearly shows that the correction to the energy loss due to the inclusion of small separation distances remains sizable even for large systems.  Upon closer inspection of the details of the calculation, the reader will notice that the integral over separation distances is in fact performed starting at $\Delta z =0$, meaning that small separation distances are included even if they are neglected in the calculation at the amplitude level.  The fact that the separation distances are integrated over the entire range of distances suggests that the corrections are, in principle, important for large system sizes as well.  The numerics presented here explicitly show the importance of the small separation distance correction terms, particularly at very high energies.

The heavy ion physics community has stumbled across an excellent opportunity to scrutinize existing interpretations of data due to surprising recent discoveries in what should have been calibrative exercises.  The use of $pA$ results as a probe of nature of the QCD coupling and the validity of competing theories relies heavily on a robust understanding of known processes.  Such an understanding is contingent upon extensive analyses of the two most common signatures of the QGP, collective behaviour and particle suppression. 

We have endeavoured to catalyze the analysis of energy loss in $pA$ by calculating the small separation distance corrections to the large path length static energy loss formulae and have found that:
\begin{enumerate}
\item Under the assumption that the Debye screened length of the scattering centres is much larger than the collinearity of the radiated gluon, a remarkable cancellation occurs and there is no correction.  In future calculations this assumption will have to be relaxed since it is not clear that  $\mu_1 \gg \omega_1$ holds throughout the momentum and collinearity ranges that are considered.
\item If one relaxes the large formation length assumption after the amplitude stage of the calculation, one finds that, since the separation distance between production and first scattering is integrated over the entire separation range, the effect of including short separation distances is $\gtrsim 100\%$ even for large system sizes and must be taken into account.
\end{enumerate}

%----------------------------------------------------------------------------------------
%	THESIS CONTENT - APPENDICES
%----------------------------------------------------------------------------------------

\addtocontents{toc}{\vspace{2em}} 

\appendix 

\chapter{Useful Results}

\label{AppendixImportantResults}

\lhead{\emph{Useful Results}}

\section{$\Amp_{0}$}

From \cite{Djordjevic2004}, the amplitude for hard jet radiation to emit a transverse plasmon with momentum, polarization and color $(k,\epsilon,c)$ is

\begin{align}\label{M0}
\Amp_{0}	&=iJ(p+k)e^{i(p+k)x_0}(ig_s)(2p+k)_{\mu}\epsilon^{\mu}(k)i \vartriangle_M(p+k)c\nonumber\\
				&\approx J(p)e^{ipx_0}(-2ig_s)\frac{\epsilon\cdot\textbf{k}}{k^2+m_g^2+M^2x^2}.
\end{align}
In (\ref{M0}) we have used that $x_0(0,z_0,\textbf{0})$ is the jet production point inside the plasma.  We also have that $J$ varies slowly with $p$ which allows for the above approximation since $k\ll p$.

\section{Writing down the Matrix element}\label{M100Section}
In this section I look briefly at the `Feynman Rules' for the invariant matrix elements we consider.  Take for example the matrix element for $\Amp_{1,0,0}$, which is given by

\begin{align}\label{M100}
\Amp_{1,0,0}	&=	 \int \frac{d^4 q_1}{(2\pi)^4}i J(p+k-q_1)e^{i(p+k-q_1)x_0}(ig_s)\epsilon_\alpha(2p-2q+k)^\alpha\times\nonumber\\
					&\times i\vartriangle_M(p-q_1+k)i\vartriangle_M(p-q_1)(2p-q_1)^0V(q1)e^{iq_1x_1}T_{a_1} a_1 c\nonumber\\
					&\approx J(p+k)e^{i(p+k)x_0}(-ig_sa_1cT_{a_1}) I_1,
\end{align}

where
\begin{align}
I_1(p,k,\textbf{q}_1,z_1-z_0)	&=\int \frac{dq^z_1}{2\pi}\frac{\epsilon_\alpha(2p-2q+k)^\alpha}{(p-q_1+k)^2-M^2+i\epsilon}\times\nonumber\\
	&\times\frac{1}{(p-q_1)^2-M^2+i\epsilon}v(\vec{\textbf{q}}_1)e^{-iq^z_1(z_1-z_0)}.
\end{align}

To see this, consider the first diagram in Figure (\ref{M1s}).  Equation (\ref{M100}) can be `read' off by going from left to right and considering the following set of `Feynman rules':

\begin{itemize}
\item	$iJ(p+k-q_1)e^{i(p+k-q_1)x_0}$, which is the highly localized wave packet produced in free space, propagating with momentum $(p+k-q_1)$.
\item	$(ig_s)\epsilon_\alpha(2p-2q+k)^\alpha a_1cT_{a_1}$, the radiation vertex with color factor .  The form comes from `Peskin's trick' described in section \ref{PeskinsTrick}.
\item 	$\vartriangle _M(p-q_1+k)$ and $\vartriangle _M(p-q_1)$.  These are the propagators of particles with mass $M$ and with momentum $(p-q_1+k)$ and $(p-q_1)$ respectively.  They have the form 
\begin{equation}\label{propagatorExpression}
\vartriangle _M(p-q_1+k)=\frac{1}{(p-q_1+k)^2-M^2+i\epsilon}
\end{equation}
\item $V(q_1)e^{iq_1x_1}$.  The potential, the random color screened potential that is given in \cite{Djordjevic2004} as
\begin{equation}\label{potential}
V_n=V(q_n)e^{iq_nx_n}=2\pi \delta(q^0)v(\vec{\textbf{q}}_n)e^{-i\vec{\textbf{q}}_n\cdot\vec{\textbf{x}}_n}T_{a_n}(R)\otimes T_{a_n}(n), 
\end{equation}
where $\vec{\textbf{x}}_n$ is the location of the $n^{th}$ scattering centre and $v(\vec{\textbf{q}}_n)\equiv4\pi\alpha_s/(\vec{\textbf{q}}_n^2+\mu^2)$

\end{itemize}

\section{Components of k, p and q}\label{SecCompkpq}
Here I present the derivation of the components of the momentum (four) vectors $k$, $p$ and $q$. We are considering the diagram shown in Figure \ref{RadMom}.  Recall that the invariant masses are such that $p\cdot p=M^2$ and $k\cdot k=m_g^2$.  We also know from conservation of energy that $p+k=P$.  Suppose now that the gluon carries away a fraction of the initial quark momentum $x$.  We then have, in light cone coordinates:
\begin{align}
k	&= [xP^+, k^-, \textbf{k}];\\
p	&= [(1-x)P^+, p^-, -\textbf{k}];
\end{align}
It remains only to determine $k^-$ and $p^-$.  Therefore
\begin{align}
p\cdot p	&= p^+p^--\textbf{p}\cdot\textbf{p}\nonumber\\
			&= (1-x)P^+p^--(\textbf{k})-\textbf{k}\nonumber\\
			&= (1-x)P^+p^--(\textbf{k})^2= M^2\nonumber\\
			&\Rightarrow p^-=\frac{M^2+\textbf{k}^2}{(1-x)P^+}
\end{align}
\begin{align}
k\cdot k	&= k^+k^--\textbf{k}\cdot\textbf{k}\nonumber\\
			&= xP^+k^--\textbf{k}^2 = m_g^2\nonumber\\
			&\Rightarrow k^-=\frac{m_g^2+\textbf{k}^2}{xP^+}
\end{align}

Since $\epsilon\cdot k=0$ by the Ward identity, we can find the polarization vector for the emitted gluon by choosing $\epsilon^+=0$;
\begin{align}
\epsilon\cdot k	&=\frac{1}{2}\big(\epsilon^+k^--\epsilon^-k^+\big) - \boldsymbol\epsilon\cdot\textbf{k}=0\nonumber\\
				&\Rightarrow\epsilon^-	=\frac{2\boldsymbol\epsilon\cdot\textbf{k}}{k^+}\nonumber\\
				&=\frac{2\boldsymbol\epsilon\cdot\textbf{k}}{xP^+}
				\end{align}

We therefore have that 
\begin{align}
k	&=\bigg[xP^+, \frac{m_g^2+\textbf{k}^2}{xP^+},\textbf{k}\bigg],\label{k}\\
p	&=\bigg[(1-x)P^+,\frac{M^2+\textbf{k}^2}{(1-x)P^+},-\textbf{k} \label{p}\bigg]\\
\epsilon	&=\bigg[0,\frac{2\boldsymbol\epsilon\cdot\textbf{k}}{xP^+},\boldsymbol\epsilon\bigg]
\end{align}

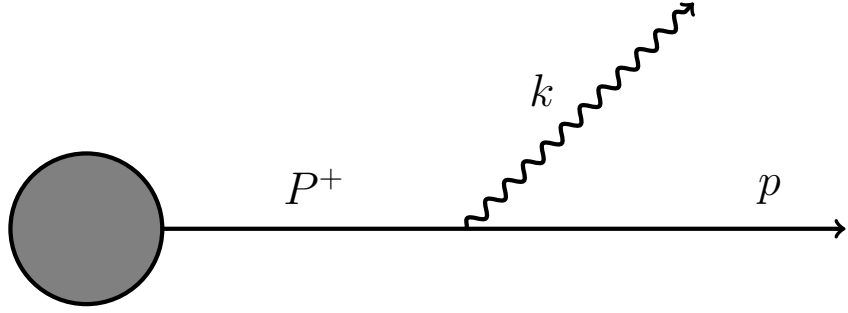
\begin{figure}
\centering\
\begin{tikzpicture}[decoration=snake]
\draw [fill=gray,ultra thick] (0,0) circle [radius=1];
\draw [->,ultra thick] (1,0) --(10,0);
\draw [->, decorate, ultra thick] (5,0)--(8,3);

\node [above] at (3,0.2) {\LARGE $P^+$};
\node [above] at (9,0.2) {\LARGE $p$};
\node [above] at (6,1.5) {\LARGE $k$};
\end{tikzpicture}
\caption{A gluon with momentum $k$ radiated from a quark with initial momentum $P^+$.\label{RadMom}}

\end{figure}

\pagebreak
\section{Recurring dot Products}
In this section I present results for three dot products that appear regularly.  These are the dot products of \textit{four vectors}.
\subsection{$2p \cdot k$}\label{2pk}
Using the definition of the dot product in light cone coordinates given in Section \ref{notationsAndConventions}
\begin{align}
2p\cdot k	&= p^+k^-+p^-k^+-2\textbf{p}\cdot\textbf{k}\nonumber\\
			&=\frac{1}{xP^+}\bigg[(1-x)P^+(m_g^2+\textbf{k}^2)\bigg]+\frac{1}{(1-x)P^+}\bigg[(M^2+\textbf{k}^2)(xP^+)\bigg]-2(-\textbf{k})(\textbf{k})\nonumber\\
			&\approx\frac{m_g^2+\textbf{k}^2}{x}+(M^2+\textbf{k}^2)x-2\textbf{k}^2, \qquad (x\ll1 \Rightarrow (1-x)A\approx A)\nonumber\\
			&=\frac{1}{x}\big[m_g^2+\textbf{k}^2+x^2M^2+\textbf{k}^2x^2-2\textbf{k}^2x^2\big]\nonumber\\
			&\approx\frac{1}{x}\big[m_g^2+\textbf{k}^2+x^2M^2\big], \qquad \qquad(\textbf{k}^2(1+x+x^2)\approx\textbf{k}^2)
\end{align}

\subsection{$2p \cdot q$} 

Dot products with $q$ ned to be started off slightly differently due to a slight simplification.  We use the definitions set out  in Section \ref{notationsAndConventions} to obtain.

\begin{align}\label{2pq}
2p\cdot q	&= 2\big(\cancel{p^0q^0}-p^zq^z-\textbf{p}\cdot\textbf{q}\big), \qquad (q^0=0)\nonumber\\
			&= 2\bigg[-\frac{1}{2}(p^+-p^-)\bigg]q^z-2\textbf{p}\cdot\textbf{q}\nonumber\\
			&=-\bigg[(1-x)P^+-\frac{M^2+\textbf{k}^2}{(1-x)P^+}\bigg]q^z-2\textbf{p}\cdot\textbf{q}\nonumber\\
			&\approx -P^+q^z+2\textbf{k}\cdot\textbf{q}
\end{align}

In the last line we have used that 

\begin{align}
p^z	&=\frac{1}{2}\big(p^+-p^-)\nonumber\\
	&=\frac{1}{2}\bigg[(1-x)P^+-\frac{M^2+\textbf{k}^2}{(1-x)P^+}\bigg]\nonumber\\
	&\approx\frac{1}{2}\bigg[P^+-\frac{M^2+\textbf{k}^2}{P^+}\bigg]\approx\frac{1}{2}P^+
\end{align}

\subsection{$2k \cdot q$} \label{2kq}

In a similar fashion to the process that yields equation (\ref{2pq}), we obtain

\begin{align}
2k\cdot q	&= 2\big(\cancel{k^0q^0}-k^zq^z-\textbf{k}\cdot\textbf{q}\big), \qquad (q^0=0)\nonumber\\
			&=-2\bigg[\frac{1}{2}\bigg(xP^+-\frac{m_g^2+\textbf{k}^2}{xP^+}\bigg)q^z\bigg]-2\textbf{k}\cdot\textbf{q}\nonumber\\
			&\approx -xP^+q^z-2\textbf{k}\cdot\textbf{q}
\end{align}

\subsection{$2p \cdot\epsilon$}

Recalling that $\epsilon^+=0$, we have the following, again relying on the definitions in Section \ref{notationsAndConventions}.

\begin{align}\label{2pepsilon}
2p\epsilon	&=2\bigg[\frac{1}{2}\big(p^+\epsilon^-+\cancel{p^-\epsilon^+}\big)-\textbf{p}\cdot\boldsymbol\epsilon\bigg]\nonumber\\
			&=\bigg[\bigg((1-x)P^+\frac{2\boldsymbol\epsilon\cdot\textbf{k}}{xP^+}\bigg)+2\boldsymbol\epsilon\cdot\textbf{k}\bigg], \qquad (\textbf{p}=-\textbf{k})\nonumber\\
			&=\frac{(1-x)}{x}2\boldsymbol\epsilon\cdot\textbf{k}+2\boldsymbol\epsilon\cdot\textbf{k}\nonumber\\
			&=\frac{2\boldsymbol\epsilon\cdot\textbf{k}}{x}-2\boldsymbol\epsilon\cdot\textbf{k}+2\boldsymbol\epsilon\cdot\textbf{k}\nonumber\\
			&=\frac{2\boldsymbol\epsilon\cdot\textbf{k}}{x}
\end{align}

\section{Poles from propagators}
\subsection{$\vartriangle_M(p-q_1)$}\label{polepq}

Since it is the denominator of the propagator which has the form given in (\ref{propagatorExpression}), we need to solve the following equation for $q_1^z$ (I drop the subscript $1$ for simplicity since it will not cause any confusion):
\begin{equation}
(p-q)^2-M^2+i\epsilon=0
\end{equation}
This implies that
\begin{align}
0	&=p^2-2p\cdot q+q^2-M^2+i\epsilon \nonumber\\
	&=\cancel{M^2} -2(-p^zq^z- \textbf{p}\cdot\textbf{q})+(-(q^z)(q^z)-\textbf{q}^2)-\cancel{M^2}\nonumber\\
	&=2p^zq^z+2\textbf{p}\cdot\textbf{q}-(q^z)^2-\textbf{q}^2+i\epsilon\nonumber\\
	&\approx P^+q^z+2\textbf{p}\cdot\textbf{q}-(q^z)^2-\textbf{q}^2+i\epsilon
\end{align}
One can now do one of two things:  One can solve the above for $q^z$ directly, or one can use the quadratic formula.  The latter option is not only more rigorous, but also allows for a simpler handling of cancellations that occur when the residues are calculated.  Therefore, I will follow this method:
\begin{align}
q^z	&=\frac{-P^+\pm\sqrt{(P^+)^2-4(-1)[\bullet]}}{2(-1)}, \qquad [\bullet]=2\textbf{p}\cdot\textbf{q}-\textbf{q}^2+i\epsilon\nonumber\\
	&=\frac{P^+}{2}\mp\frac{\sqrt{(P^+)^2+4[\bullet]}}{2}\nonumber\\
	&=\frac{P^+}{2}\mp\frac{P^+}{2}\sqrt{1+\frac{4[\bullet]}{(P^+)^2}}\nonumber\\
	&=\frac{P^+}{2}\Bigg[1\mp \sqrt{1+\frac{4[\bullet]}{(P^+)^2}}\Bigg]\nonumber\\
	&\approx\frac{P^+}{2}\Bigg[1\mp \bigg(1+\frac{2}{(P^+)^2}[\bullet]\bigg)\Bigg], \qquad \text{(By the Binomial Theorem)}\nonumber\\
	&= \frac{P^+}{2}\Bigg[1\mp \bigg(1+\frac{2}{(P^+)^2}\big[2\textbf{p}\cdot\textbf{q}-\textbf{q}^2+i\epsilon\big]\bigg)\Bigg]\nonumber\\
\end{align}
Which gives us (as expected) two solutions.  We ignore the solution from the plus sign because it gives $+i\epsilon$ which lies in the upper half of the complex plane and is therefore not included as we close the contour \textit{below} the real axis.  The solution from the minus sign is then
\begin{align}
q^z	&= \frac{P^+}{2}\bigg[1- 1-\frac{2}{(P^+)^2}\big(2\textbf{p}\cdot\textbf{q}-\textbf{q}^2+i\epsilon\big)\bigg]\nonumber\\
	&= \frac{P^+}{2}\bigg[-\frac{2}{(P^+)^2}\big(2\textbf{p}\cdot\textbf{q}-\textbf{q}^2+i\epsilon\big)\bigg]\nonumber\\
	&= \frac{-1}{P^+}\bigg(2\textbf{p}\cdot\textbf{q}-\textbf{q}^2+i\epsilon\bigg)\nonumber\\
	&= \frac{2\textbf{k}\cdot\textbf{q}}{P^+}+\frac{\textbf{q}^2}{P^+}-i\epsilon, \qquad \textbf{p}=-\textbf{k} \text{ , see equation \ref{p}}\\
	&=\frac{(\textbf{k}+\textbf{q})^2-\textbf{k}62}{P^+}-i\epsilon
\end{align}

Although this result differs slightly in form from the corresponding result in \cite{Djordjevic2004}, the difference does not affect later results.
\pagebreak
\subsection{$\vartriangle_M(p-q_1+k)$}\label{polepqk}

As in the previous section, we need to solve the following equation for $q_1^z$:
\begin{align}
0	&=(p-q_1+k)^2-M^2+i\epsilon\nonumber\\
	&= p^2+k^2+q^2+2p\cdot k-2p\cdot q-2k\cdot q-M^2+i\epsilon\nonumber\\
	&=\cancel{M^2}+m_g^2+q^2+\frac{m_g^2+\textbf{k}^2+M^2x^2}{x}+P^+q^z\nonumber\\
	&-\cancel{2\textbf{k}\cdot \textbf{q}}+\cancel{xP^+q^z}+\cancel{2\textbf{k}\cdot \textbf{q}}-\cancel{M^2}+i\epsilon\nonumber\\
	&=\cancel{m_g^2}-(q^z)^2-\textbf{q}^2+\frac{m_g^2+\textbf{k}^2+M^2x^2}{x}+P^+q^z+i\epsilon\nonumber\\
\end{align}

In the third and fourth lines we have used the fact that $x\ll 1\Rightarrow xA \ll A$ to neglect certain terms.  We can now solve for $q^z$ in much the same way as in the previous section:

\begin{align}
q^z	&=\frac{-P^+ \pm \sqrt{(P^+)^2-(4)(-1)[\ast]}}{2(-1)}, \qquad [\ast]=\frac{m_g^2+\textbf{k}^2+M^2x^2}{x}-\textbf{q}^2+i\epsilon\nonumber\\
	&=\frac{P^+}{2}\mp\frac{P^+}{2}\sqrt{1+\frac{4}{(P^+)^2}[\ast]}\nonumber\\
	&=\frac{P^+}{2}\bigg[1 \mp \bigg(1+\frac{2}{(P^+)^2}[\ast]\bigg)\bigg]
\end{align}

Which again has both a positive and a negative root and we will again ignore the root in the upper half of the complex plane.  The root in the lower complex half-plane is then

\begin{align}
q^z	&\approx\frac{P^+}{2}\bigg[\cancel{1}-\cancel{1}-\frac{2}{(P^+)^2}\bigg(\frac{m_g^2+\textbf{k}^2+M^2x^2}{x}-\textbf{q}^2+i\epsilon\bigg)\bigg]\nonumber\\
	&=-\frac{1}{(P^+)}\bigg(\frac{m_g^2+\textbf{k}^2+M^2x^2}{x}-\textbf{q}^2+i\epsilon\bigg)\nonumber\\
	&\approx-\frac{m_g^2+\textbf{k}^2+M^2x^2}{xP^+}-i\epsilon, \qquad (\textbf{q}^2\propto\textbf{k}^2, \qquad \textbf{k}^2(1+x)\approx\textbf{k}^2)\nonumber\\
	&=-\omega_0-\tilde{\omega}_m - i\epsilon
\end{align}

\pagebreak
\subsection{$\vartriangle_{m_g}(k-q_1)$}\label{polekq}
As in previous sections, we must solve the following equation for $q_1^z$ and we will drop the subscript $1$.
\begin{align}
0	&=(k-q)^2-m_g^2+i\epsilon\nonumber\\
	&=k^2-2k\cdot q+q^2-m_g^2+i\epsilon\nonumber\\
	&=\cancel{m_g^2}-2\big[\cancel{k^0q^0}-k^zq^z-\textbf{k}\cdot\textbf{q}\big]+q_1^2-\cancel{m_g^2}+i\epsilon, \qquad (q^0=0)\nonumber\\
	&=-2\bigg[-\frac{1}{2}\big(k^+-k^-\big)q^z-\textbf{k}\cdot\textbf{q}\bigg]+(q^z)(-q^z)-\textbf{q}\cdot\textbf{q}+i\epsilon\nonumber\\
	&=q^zxP^++2\textbf{k}\cdot\textbf{q}-(q^z)^2-\textbf{q}^2+i\epsilon
\end{align}

Again we can use the quadratic formula to obtain two solutions:

\begin{align}
q^z	&=\frac{P^+x\pm\sqrt{(P^+x)^2-4(1)[\ast]}}{2}, \qquad [\ast]=2\textbf{k}\cdot\textbf{q}-\textbf{q}^2+i\epsilon\nonumber\\
	&=\frac{P^+x}{2}\Bigg(1\pm \sqrt{1+\frac{4}{(P^+x)2}[\ast]}\Bigg)
\end{align}

Again we must choose the root that lies below the real axis:
\begin{align}
q^z	&=\frac{P^+x}{2}\bigg(1-1-\frac{2}{(xP^+)^2}[\ast] \bigg)\nonumber\\
	&=\frac{1}{P^+x}\big(-2\textbf{k}\cdot\textbf{q}+\textbf{q}\big)-i\epsilon\nonumber\\
	&=\frac{1}{P^+x}\big[\big(\textbf{k}-\textbf{q}\big)^2-\textbf{k}^2\big]-i\epsilon\nonumber\\
	&=\omega_1-\omega_0-i\epsilon
\end{align}

\pagebreak
\section{Poles from the Potential}\label{polepotential}
The potential is given by equation (\ref{potential}) as
\begin{equation}
V_n=V(q_n)e^{iq_nx_n}=2\pi \delta(q^0)v(\vec{\textbf{q}}_n)e^{-i\vec{\textbf{q}}_n\cdot\vec{\textbf{x}}_n}T_{a_n}(R)\otimes T_{a_n}(n),
\end{equation}
The parts of this that we are concerned with are $v(\vec{\textbf{q}}_n)$, given by
\begin{equation}
v(\vec{\textbf{q}}_n)=\frac{4\pi\alpha_s}{(\vec{\textbf{q}}_n)^2+\mu^2} =\frac{4\pi\alpha_s}{-(q_n^z)^2-\textbf{q}_n^2+\mu^2}:=v(q_n^z,\textbf{q}_n) ,
\end{equation}
for which we have to find the $q_1^z$'s that result in infinities.  Using the definition $\mu _i^2\equiv\mu _{i\perp}^2=\textbf{q}_i^2+\mu^2$, we have that
\begin{align}
\vec{\textbf{q}}_n^2+\mu^2	&= \vec{\textbf{q}}_1^2+\mu^2	\nonumber\\
							&=(q_1^z)^2-\textbf{q}_1^2+\mu_1^2 = 0\nonumber\\
\Rightarrow(q_1^z)^2		&=\textbf{q}_1^2-\mu_1^2\nonumber\\	
							&=\textbf{q}_1^2-\textbf{q}_1^2-\mu_1^2=-\mu_1^2	\nonumber\\
\Rightarrow		q_1^z		&=\pm 			i\mu_1.											
\end{align}

However, both of these poles were neglected in the large separation distance limit.  The positive pole is neglected because it lies in the upper half of the complex plane and the contour is closed below the real axis.  The negative pole is neglected in some cases because its contribution will be accompanied by an exponential term  of the form $\exp{-i\mu(z_1-z_0)}$ which is then exponentially suppressed because we deal with the well separated case in which $\mu(z_1-z_0)=\mu\lambda\gg 1$.  In the contact case related to the diagrams with more than one scattering centre this is not the case.

In this treatment, where the size of the system is of the order of the Debye screening length, we will consider the pole in the lower half of the complex plane since its contribution will no longer be exponentially suppressed.

\pagebreak
\section{Factorizations within residues}
\subsection{$(p-q)^2-M^2$}\label{factorpq}

\begin{align*}
(p-q)^2-M^2	&=\cancel{p^2}-2p\cdot q+q^2-\cancel{M^2}\\
			&=-2(\cancel{p^0q^0}-p^zq^z-\textbf{p}\cdot\textbf{q})+ (\cancel{q^0q^0}-q^zq^z-\textbf{q}\cdot\textbf{q}), \qquad (q^0=0)\\
			&=2p^zq^z+2\textbf{p}\cdot\textbf{q}-(q^z)^2-\textbf{q}^2\\
			&=2\bigg[\frac{1}{2}\big(p^+-p^-\big)\bigg]q^z-2\textbf{k}\cdot\textbf{q}-(q^z)^2-\textbf{q}^2\\
			&=(1-x)P^+q^z-2\textbf{k}\cdot\textbf{q}-(q^z)^2-\textbf{q}^2\\
			&\approx P^+q^z+\textbf{k}^2-(\textbf{k}+\textbf{q})^2\\
			&=P^+x\bigg(\frac{q^z}{x}+\omega_0+\tilde{\omega}_m\bigg)
\end{align*}

\subsection{$(p-q+k)^2-M^2$}\label{factorpqk}
\begin{align*}
(p-q+k)^2-M^2	&=\cancel{p^2}+q^2+k^2-2p\cdot q+2p\cdot k -2q\cdot k -\cancel{M^2}\\
				&=-(q^z)^2-\textbf{q}^2+m_g^2+P^+q^z-\cancel{2\textbf{k}\cdot\textbf{q}}\nonumber\\
				&+1/x\big(m_g^2+\textbf{k}^2+x^2M^2\big)+xP^+q^z+\cancel{2\textbf{k}\cdot\textbf{q}}\\
				&\approx P^+q^z+1/x(m_g^2+\textbf{k}^2+x^2M^2), \qquad (x\ll 1, \qquad P^+\gg 1)\\
				&=P^+q^z+P^+(\omega_0+\tilde{\omega}_m)\\
				&=P^+(q^z+\omega_0+\tilde{\omega}_m)
\end{align*}

In the third line we have made a number of cancellations based on approximations of large $P^+$ and small $x$.  It is important to note that they can only be made as comparisons.  That is, terms can only be neglected if they are much smaller than a similar term that is different only by a factor of $x$ or $P^+$.

\subsection{$(k-q)^2-m_g^2$}\label{factorkq}
\begin{align}
(k-q)^2-m_g^2	&=\cancel{k^2}-2k\cdot q +q^2 -\cancel{m_g^2}\nonumber\\
		&=-\big(xP^+q^z-2\textbf{k}\cdot\textbf{q}\big)-(q^z)^2-\textbf{q}^2, \qquad \text{Section \ref{2kq}}\nonumber\\
		&=-xP^++2\textbf{k}\cdot\textbf{q}-(q^z)^2-\textbf{q}^2\nonumber\\
		&\approx -xP^+q^z-(\textbf{k}-\textbf{q})^2+\textbf{k}^2\nonumber\\
		&=-xP^+q^z-xP^+(\omega_0-\omega_1)=-xP(q^z -\omega_1+\omega_0)
\end{align}

\subsection{$v(\vec{\textbf{q}}_n)$}\label{factorv}
\begin{align*}
v(\vec{\textbf{q}}_n)	&=\frac{4\pi\alpha_s}{\vec{\textbf{q}}^2_n+\mu^2} := v(q_n^z,\textbf{q}_n)\\
						&=\frac{4\pi\alpha_s}{(q^z)^2+\textbf{q}^2+\mu^2}\approx v(0,\textbf{q}_n)
\end{align*}

We will encounter two kinds of residues: Ones that contain $\omega$'s and ones that contain $-i\mu$'s.  The above result will hold in both cases.  For the former, it holds due to the small $x$ approximation.  In the latter case it is identically true.  We will also see, when dealing with diagrams that have two scattering centres, situations where the poles containing $-i\mu$ are not suppressed.  We will then see two cases:  one where the $z$-component of one $q$ vector is used along with the perpendicular component of another and a second case where the same $q$ vector is involved. In the first case we have

\begin{align}\label{factorvMixed}
v(q_2^z,\textbf{q}_1) &=\frac{4\pi\alpha_s}{-(q_2^z)^2-\textbf{q}_1^2+\mu^2}\nonumber\\
		&=\frac{4\pi\alpha_s}{(q_2^z)^2+\textbf{q}_1^2+\mu_1^2-\textbf{q}_1^2}, \qquad (\mu_i^2\equiv \mu_{i\perp}=\textbf{q}_i^2+\mu^2)\nonumber\\
		&=\frac{4\pi\alpha_s}{(q_2^z+i\mu_1)(q_2^z-i\mu_1)},
\end{align}

while in the second case we have that

\begin{align}\label{factorvSame}
v(q_2^z,\textbf{q}_2) &=\frac{4\pi\alpha_s}{-(q_2^z)^2-\textbf{q}_2^2+\mu^2}\nonumber\\
		&=\frac{4\pi\alpha_s}{(q_2^z)^2+\textbf{q}_2^2+\mu_1^2-\textbf{q}_2^2}, \qquad (\mu_i^2\equiv \mu_{i\perp}=\textbf{q}_i^2+\mu^2)\nonumber\\
		&=\frac{4\pi\alpha_s}{(q_2^z+i\mu_2)(q_2^z-i\mu_2)},
\end{align}

\section{Color Matrix results}\label{colorRes}

Using the relations set out in section \ref{notationsAndConventions}, I prove here that $f^{abc}f^{abd}da$=$-2if^{acb}cba$

\begin{align*}
\text{LHS}	&=f^{abc}f^{abd}da=C_2(G)\delta^{cd}cd\nonumber\\
		&=C_2(G)cc\nonumber\\
		&=-C_2(G)C_2(r)\hat{\mathbf{1}}\\
\text{RHS}	&=2if^{acb}cba=2i\bigg(\frac{1}{2} iC_2(G)a\bigg)a\\
		&=C_2(G)C_2(r)\hat{\mathbf{1}}=-\text{LHS}\nonumber\\
		&\therefore f^{abc}f^{abd}da+2if^{acb}cba=0
\end{align*}

It is also worth noting that

\begin{align*}
[A,B]^*	&=(AB-BA)^*\\
		&=(AB)^*-(BA)^*\\
		&=B^*A^*-A^*B^*\\
		&=BA-AB	\qquad \text{Hermitian}\\
		&=[B,A]=-[A,B]
\end{align*}

We will also need to know that $c^2a^2=a^2c^2$.  It is trivially true by change of variables, but can also be shown from the definitions:

\begin{align*}
ccaa	&=c(ac+if^{cab}b)=caca +if^{cab}cba\\
		&=(ac+if^{cab}b)ca+if^{cab}cba\\
		&=acca+if^{cab}bca+if^{cab}cba\\
		&=ac(ac+if^{cab}b)if^{cab}bca+if^{cab}(bc+if^{cbd}d)a\\
		&=acac+if^{cab}acb+if^{cab}bca+if^{cab}bca-f^{cab}f^{cbd}da\\
		&=a(ac+if^{cab}b)c+if^{cab}a(bc+if^{cbd}d)+2if^{cab}bca-f^{cab}f^{cbd}da\\
		&=a^2c^2+if^{cab}abc+if^{cab}abc-f^{cab}f^{cbd}ad+2if^{cab}bca-f^{cab}f^{cbd}da\\
		&=a^2c^2+2if^{cab}abc+2if^{cab}bca-f^{cab}f^{cbd}ad-f^{cab}f^{cbd}da\\
		&=a^2c^2+4if^{abc}-2f^{cab}f^{cbd}da=a^2c^2
\end{align*}
Also note that, by change of variables, $acac=caca$.
\section{Peskin's Trick}\label{PeskinsTrick}
We will be making use of `Peskin's trick' as it is called in \cite{Horowitz2008}.  It pertains to the Feynman rules of diagrams with bremstrahlung and appears in Peskin on page 183.:
\begin{align}
\bar{u}(p')\gamma^\mu\epsilon_\mu^*(\slashed{p}'+m)	&=\bar{u}(p')\gamma^\mu\epsilon_\mu^*(\gamma^\nu p_{\nu}'+m)\nonumber\\
		&=\bar{u}(p')\big[\gamma^\mu\gamma^\nu \epsilon_\mu^*p_\nu'+\gamma^\mu\epsilon_\mu^*m\big]\nonumber\\
		&=\bar{u}(p')\big[\big(-\gamma^\nu\gamma^\mu+2g^{\mu\nu}\big)\epsilon_\mu^*p_\nu'+\gamma^\mu\epsilon_\mu^*m\big]
			,\qquad \text{(Antiommutation relations)}\nonumber\\
		&=\bar{u}(p')2\epsilon^{*\nu}p_\nu'-\gamma^\nu\gamma^\mu \bar{u}(p')\epsilon_\mu^*p_\nu'+\bar{u}(p')\gamma^\mu\epsilon_\nu^*m\nonumber\\
		&=\bar{u}(p')\big[2\epsilon^*p'+(-\slashed{p}+m)\gamma\cdot\epsilon^*\big]\nonumber\\
		&=\bar{u}(p)2\epsilon\cdot p', \qquad \text{(Dirac Equation $\bar{u}(p')(\slashed{p}-m)=0$ )}
\end{align}
The conjugate relation follows exactly the same reasoning.

\section{Major Simplification Effect}\label{MajorCancel}
The vast majority of contributions from previously exponentially suppressed terms are still suppressed but by a different factor.  I discuss here that effect:

This entire calculation (even the original) worked under the assumption that $\omega_i \ll \mu_1$.  That is to say that 
\begin{align*}
\frac{\textbf{k}^2}{2\omega} \ll \mu_1
\end{align*}
This in turn implies that 
\begin{align}
\frac{1}{\omega_1}\gg \frac{1}{\mu_1}
\end{align}
This fact is used very often in the calculation and is true for all $\omega_i$ and $\mu_i$.

\chapter{Detailed calculation of Amplitudes} 

\lhead{\emph{Detailed calculation of Amplitudes}} 
\label{AppendixAmplitudes} 
In this chapter I present the details of the calculation of the amplitudes of the relevant Feynman Diagrams.  These calculations closely follow the Appendices of \cite{Djordjevic2004}, with the exception of additional terms that are considered in the small separation distance limit.
\section{$\Amp_{1,0,0}$, $\Amp_{1,1,0}$, $\Amp_{1,0,1}$ }
The basis results appear in \cite{Djordjevic2004} in Appendix B. The diagrams are shown in Figure (\ref{M1s}).  In what follows, I will often drop the subscript $1$ when using $q$ ($q_1\equiv q$). 

The general idea behind what is done here is to calculate the $\int dq_1^z$ integrals by closing the contour below the real axis (because $z_1>z_0$, meaning that $z_1-z_0>0$ and so the exponents that appear will only converge if the contour is closed below the real line).  To be able to do this, one must find the poles, their residues and hence the integral.  Many of the results are repeated and are therefore presented in Appendix \ref{AppendixImportantResults}.  
\begin{figure}
\centering
\includegraphics[scale=0.5]{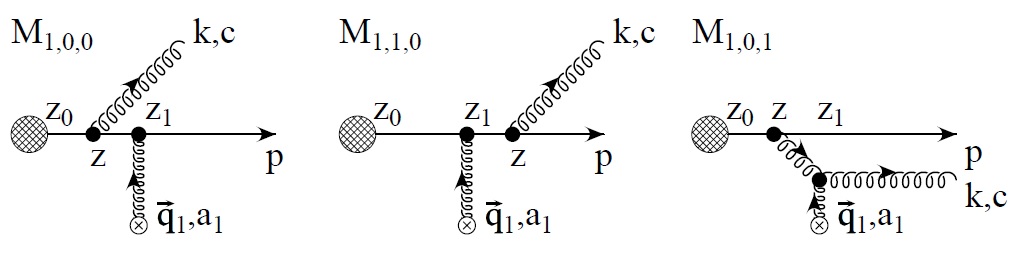}
\caption{Three ``direct'' terms $\Amp_{1,0,0}$, $\Amp_{1,1,0}$, $\Amp_{1,0,1}$ contribute to the soft gluon radiation amplitude to first order in opacity $L/\lambda\propto\sigma_{el}/A_{\perp}$\cite{Djordjevic2004}. \label{M1s}}

\end{figure}

\subsection{Computation of $\Amp_{1,0,0}$}\label{CalcM100}

The matrix element for $\Amp_{1,0,0}$ is given by section \ref{M100Section} as

\begin{alignat}{2}
\Amp_{1,0,0}	&=	 &&\int \frac{d^4 q_1}{(2\pi)^4}i J(p+k-q_1)e^{i(p+k-q_1)x_0}(ig_s)\epsilon_\alpha(2p-2q+k)^\alpha\times\nonumber\\
					& &&\times i\Delta_M(p-q_1+k)i\Delta_M(p-q_1)(2p-q_1)^0V(q1)e^{iq_1x_1}T_{a_1} a_1 c\nonumber\\
					&\approx &&J(p+k)e^{i(p+k)x_0}(-ig_sa_1cT_{a_1}) 2E\int \frac{d^2\textbf{q}_1}{(2\pi)^2} e^{-\textbf{q}_1\cdot\textbf{b}_1}  I_1,
\end{alignat}

where
\begin{align}
I_1(p,k,\textbf{q}_1,z_1-z_0)	&=\int \frac{dq^z_1}{2\pi}\frac{\epsilon_\alpha(2p-2q+k)^\alpha}{(p-q_1+k)^2-M^2+i\epsilon}\times\nonumber\\
	&\times\frac{1}{(p-q_1)^2-M^2+i\epsilon}v(\vec{\textbf{q}}_1)e^{-iq^z_1(z_1-z_0)}
\end{align}
We can simplify $I_1$ slightly through noting that
\begin{align}
\epsilon_\alpha(2p-2q+k)^\alpha	&\approx 2p\epsilon\nonumber \approx \frac{2\boldsymbol\epsilon\cdot\textbf{k}}{x}, \qquad (\text{see Section \ref{2pepsilon}})
\end{align}

To solve $I_1$, note that there are three poles; one from each propagator and one from the potential. These poles are calculated in Sections \ref{polepq} and \ref{polepqk}.  (The third pole, calculated in Section \ref{polepotential} is no longer neglected for the reasons explained there).  The poles are then, from the propagators
\begin{align*}
q_1^{z(1)}	&=-\omega_0-\tilde{\omega}_m-i\epsilon\\
q_1^{z(2)}	&=-\frac{(\textbf{k}+\textbf{q})^2-\textbf{k}^2}{P^+}-i\epsilon\\
q_1^{z(3)}	&=-i\mu_1
\end{align*}

We must now calculate the residues due to these poles, taking the limit as $\epsilon\rightarrow 0$.  The residue for the pole at $q_1^{z(1)}=-\omega_0-\tilde{\omega}_m-i\epsilon$ is given by:
\begin{align}
\text{Res}[q_1^{z(1)} &=-\omega_0-\tilde{\omega}_m-i\epsilon]\nonumber\\
	&=\lim_{q_1^z\to q_1^{z(1)}}\Bigg[ \frac{\frac{2\boldsymbol\epsilon\cdot\textbf{k}}{x}(q_1^z-q_1^{z(1)})}{(p-q+k)^2-M^2+i\epsilon}
		\frac{v(\vec{\textbf{q}}_1)e^{-q_1^z(z_1-z_0)}}{(p-q_1)^2-M^2+i\epsilon}\Bigg].
\end{align}

Each part of this can be evaluated individually, by sections \ref{factorpqk}, \ref{factorpq} and \ref{factorv}

\begin{align}\label{M100I2Resq1}
\text{Res}[I_1,q_1^{z(1)}]	&\approx\frac{2\boldsymbol\epsilon\cdot\textbf{k}}{x}
		\frac{\cancel{(q^z+\omega_0+\tilde{\omega}_m)}}{(-P^+\cancel{\big(q^z+\omega_0+\tilde{\omega}_m)}\big)}
		\frac{e^{-i(-\omega_0-\tilde{\omega}_m)(z_1-z_0)} v(0,\textbf{q}_1)}{P^+q^z+\textbf{k}^2-(\textbf{k}+\textbf{q})^2}\nonumber\\
	&\approx \frac{2\boldsymbol\epsilon\cdot\textbf{k} e^{-i(-\omega_0-\tilde{\omega}_m)(z_1-z_0)} v(0,\textbf{q}_1)}{x(P^+)^2(\omega_0+\tilde{\omega}_m)}.
\end{align}

In a similar way we can calculate the residue for the pole at $q_1^{z(2)}=\frac{(\textbf{k}+\textbf{q})^2-\textbf{k}}{P^+}$, which is given by
\begin{align}\label{M100I2Resq2}
\text{Res}[I_1,&q_1^{z(2)}]=\lim_{q_1^z\to q_1^{z(1)}}\Bigg[ \frac{\frac{2\boldsymbol\epsilon\cdot\textbf{k}}{x}(q_1^z-q_1^{z(1)})}{(p-q+k)^2-M^2+i\epsilon}\frac{v(\vec{\textbf{q}}_1)e^{-q_1^z(z_1-z_0)}}{(p-q_1)^2-M^2+i\epsilon}\Bigg]\nonumber\\
		&\approx
		\frac{2\boldsymbol\epsilon\cdot\textbf{k}}{x}
		\frac{\cancel{q^z-\frac{(\textbf{k}+\textbf{q})^2-\textbf{k}}{P^+}}}{P^+\cancel{q^z+\textbf{k}^2-(\textbf{k}+\textbf{q})^2}}
		\frac{\exp\big \{-i\frac{(\textbf{k}+\textbf{q})^2-\textbf{k}}{P^+}(z_1-z_0)\big\} v(0,\textbf{q})}{P^+\big(-\frac{(\textbf{k}+\textbf{q})^2-\textbf{k}}{P^+}+\omega_0+\tilde{\omega}_m\big)}\nonumber\\	
		&\approx 
		\frac{2\boldsymbol\epsilon\cdot\textbf{k}}{x(P^+)^2}
		\frac{\exp\big \{i\frac{(\textbf{k}+\textbf{q})^2-\textbf{k}}{P^+}(z_1-z_0)\big\} v(0,\textbf{q})}{(\omega_0+\tilde{\omega}_m)}\nonumber\\
		&\approx\frac{2\boldsymbol\epsilon\cdot\textbf{k}}{x(P^+)^2}\frac{v(0,\textbf{q})}{(\omega_0+\tilde{\omega}_m)}
\end{align}

We now consider a residue that was not calculated in \cite{Djordjevic2004} because it is exponentially suppressed in the large separation distance approximation.  The residue due to the pole at $q_1^{z(3)}	=-i\mu_1$, is given by
\begin{align}
\text{Res}[&q_1^{z(3)}=-i\mu_1,I_1]	=\lim_{q_1^z\rightarrow q_1^{z(3)}} \frac{\cancel{q_1^z+i\mu_1} (4\pi\alpha_s)}{\cancel{(q_1^z+i\mu_1)}(q_1^z-i\mu_1)}
			\frac{2\boldsymbol\epsilon\cdot\textbf{k}}{x} \times\nonumber\\
		&\times\frac{e^{-iq_1^z(z_1-z_0)}}{(p-q+k)^2-M^2+i\epsilon}\frac{1}{(p-q_1)^2-M^2+i\epsilon}\nonumber\\
		&\approx \frac{(4\pi\alpha_s)}{-2i\mu_1} \frac{2\boldsymbol\epsilon\cdot\textbf{k}}{x}
			\frac{1}{P^+(-i\mu_1+\omega_0+\tilde{\omega}_m)}\frac{e^{-\mu_1(z_1-z_0)}}{P^+(-i\mu_1)+\textbf{k}^2-(\textbf{k}+\textbf{q})^2}\nonumber\\
		&\approx \frac{(4\pi\alpha_s)}{-2i\mu_1} \frac{2\boldsymbol\epsilon\cdot\textbf{k}}{x}
			\frac{1}{(P^+)^2}\frac{1}{(-i\mu_1)^2}e^{-\mu_1(z_1-z_0)}\nonumber\\
		&= \frac{4\pi\alpha_s}{2x(P^+)^2}\frac{2\boldsymbol\epsilon\cdot\textbf{k}}{i(\mu_1^3)}e^{-\mu_1(z_1-z_0)}\nonumber\\
		&= \frac{4\pi\alpha_s}{x(P^+)^2}\frac{2\boldsymbol\epsilon\cdot\textbf{k}}{\mu_1^3}(-i)e^{-\mu_1(z_1-z_0)}\nonumber\\
\end{align}

These results can be multiplied together to find $I_1$.  Also note that the small $x$ assumption reduces the exponent in equation (\ref{M100I2Resq2}) to $0$ in comparison to the exponent in equation (\ref{M100I2Resq1}).  Since the integral is from $-\infty$ to $\infty$ and we close in the lower half of the complex plane, the contour is negatively orientated. We therefore have 
\begin{alignat}{2}
I_1	&= &&\frac{-2\pi i}{-2\pi} \sum \text{Res}\nonumber\\
	&\approx &&-i \Bigg[\frac{2\boldsymbol\epsilon\cdot\textbf{k}}{x(P^+)^2}\frac{v(0,\textbf{q})}{(\omega_0+\tilde{\omega}_m)}
		\big(1-e^{i(\omega_0+\tilde{\omega}_m)(z_1-z_0)}\big)\nonumber\\
	&	&&+\frac{4\pi\alpha_s}{x(P^+)^2}\frac{2\boldsymbol\epsilon\cdot\textbf{k}}{\mu_1^3}(-i)e^{-\mu_1(z_1-z_0)}\Bigg]\nonumber\\
	&= 	&&\frac{2\boldsymbol\epsilon\cdot\textbf{k}}{x(P^+)^2}\frac{4\pi\alpha_s}{\mu_1^2}
		\Bigg[\frac{e^{i(\omega_0+\tilde{\omega}_m)(z_1-z_0)}-1}{(\omega_0+\tilde{\omega}_m)}
		-i\frac{e^{-\mu_1(z_1-z_0)}}{\mu_1}\Bigg],
\end{alignat}
where we have used the definitions of $\omega$, $E$, and $E^+$ in section \ref{notationsAndConventions}. Here we encounter for the first time an effect that will cancel many of the contributions in the small separation distance limit.  The effect is discussed more fully in Section \ref{MajorCancel}.  Since $\omega_0\ll\mu_1$, the second term in the large brackets -- which is also the `new' term -- is suppressed in relation to the first.  This means that $I_1$ remains unchanged under the small separation distance assumption and therefore so does the amplitude:  We now have the matrix element $\Amp_{1,0,0}$:
\begin{alignat}{2}\label{ResM100}
\Amp_{1,0,0}&\approx &&J(p)e^{i(p+k)x_0}(-ig_s)a_1cT_{a_1}2\cancel{E}\int\frac{d^2\textbf{q}_1}{(2\pi)^2}
			e^{-\textbf{q}_1\cdot\textbf{b}_1}\times\nonumber\\
	&	&&i\frac{\boldsymbol\epsilon\cdot\textbf{k}}{\cancel{E}}\frac{v(0,\textbf{q}_1)}{(\textbf{k}^2+m_g^2+M^2x^2)}
			\big[1-e^{i(\omega_0+\tilde{\omega}_m)}\big]\nonumber\\
	&=	&&J(p)e^{i(p+k)x_0}(2ig_s)a_1cT_{a_1}(-i)\int\frac{d^2\textbf{q}_1}{(2\pi)^2}e^{-\textbf{q}_1\cdot\textbf{b}_1}v(0,\textbf{q}_1)\times\nonumber\\
	&	&&\times\frac{\boldsymbol\epsilon\cdot\textbf{k}}{\textbf{k}^2+m_g^2+M^2x^2}\big[e^{i(\omega_0+\tilde{\omega}_m)}-1\big] \nonumber
\end{alignat}

\subsection{Computation of $\Amp_{1,1,0}$}\label{CalcM110}
The matrix element for $\Amp_{1,1,0}$ is
\begin{alignat}{2}
\Amp_{1,1,0}	&= &&\int \frac{d^4q_1}{(2\pi)^4}iJ(p+k-q_1)e^{i(p+k-q_1)x_0}(2p+2k-q_1)^0\times\nonumber\\
			& &&\times i\Delta_M(p+k-q_1)i\Delta_M(p+k)(ig_s)\epsilon_\alpha(2p+k)^\alpha V(q_1)e^{iq_1x_1}T_{a_1}ca_1\nonumber\\
			&\approx &&J(p+k)e^{i(p+k)x_0}(-ig_sT_{a_1}ca_1)(2E+2\omega)\int\frac{d^3\textbf{q}_1}{(2\pi)^3}e^{-i\vec{\textbf{q}}_1(\vec{\textbf{x}}_1-\vec{\textbf{x}}_0)}v(\vec{\textbf{q}}_1)\times\nonumber\\
			& &&\times\frac{1}{(p+k-q_1)^2-M^2+i\epsilon}\frac{1}{(p+k)^2-M^2+i\epsilon}\epsilon_\alpha(2p+k)^\alpha
\end{alignat}
The same simplifications apply as before.  We have that 
\begin{align}
\epsilon_\alpha(2p+k)^\alpha	&\approx 2p\epsilon\nonumber \approx \frac{2\boldsymbol\epsilon\cdot\textbf{k}}{x}, \qquad (\text{see Section \ref{2pepsilon}}),
\end{align}
while the second propagator does not depend on $q_1$ and simplifies as per equations (\ref{2pk})and (\ref{2pepsilon}) to give 
\begin{align}
\frac{\epsilon_\alpha(2p+k)^\alpha}{(p+k)^2-M^2+i\epsilon}	&\approx \frac{2p\cdot\epsilon}{\cancel{p^2}+k^2+2p\cdot k-\cancel{M^2}}\nonumber\\
		&=\frac{2p\cdot\epsilon}{k^2+2p\cdot k}\nonumber\\
		&\approx \frac{2\textbf{k}\cdot\boldsymbol\epsilon}{x}\frac{1}{\frac{1}{x}(m_g^2+\textbf{k}^2+x^2M^2)+m_g^2}\nonumber\\
		&\approx \frac{2\textbf{k}\cdot\boldsymbol\epsilon}{m_g^2+\textbf{k}^2+x^2M^2}.
\end{align}
We also know that, since $\omega\ll E$, the simplification $(2E+2\omega)\approx 2E$ applies.
We therefore have that
\begin{align}
\Amp_{1,1,0}	&=J(p)e^{ipx_0}(-ig_s)T_{a_1}ca_14E\frac{\textbf{k}\cdot\boldsymbol\epsilon}{m_g^2+\textbf{k}^2+x^2M^2}\int\frac{d^2\textbf{q}}{(2\pi)^2}e^{-i\textbf{q}\cdot\textbf{b}_1}I_2(p,k,\textbf{q},z_1-z_0),
\end{align}
where
\begin{align}
I_2(p,k,\textbf{q},z_1-z_0)&=\int\frac{dq^z}{2\pi}\frac{1}{(p+k-q_1)^2-M^2+i\epsilon}e^{-iq^z(z_1-z_0)}v(\vec{\textbf{q}}_1).
\end{align}
There are only two poles:  One from the remaining propagator and one from the potential.  Again we now retain here the pole from the potential that had previously been suppressed exponentially. The pole from the propagator is calculated in section \ref{polepqk} to be $q^{z(1)}\approx -\omega_0-\tilde{\omega}_m - i\epsilon$.  Taking the limit as $\epsilon\rightarrow 0$ and using results from sections \ref{factorpqk} and \ref{factorv}, the Residue due to the pole at $q^{z(1)}$ is
\begin{align}
\text{Res }[I_2,&q^{z(1)}]=\lim_{q^z\to q^{z(1)}}\Bigg[ \frac{(q^z-q^{z(1)})}{(p-q+k)^2-M^2+i\epsilon}v(\vec{\textbf{q}})e^{-iq^{z}(z_1-z_0)}\Bigg]\nonumber\\
		&\approx \frac{q^z+\omega_0+\tilde{\omega}_m}{P^+(q^z+\omega_0+\tilde{\omega}_m)}e^{i(\omega_0+\tilde{\omega}_m)(z_1-z_0)}v(0,\textbf{q}), \nonumber\\
		&= \frac{1}{P^+}e^{i(\omega_0+\tilde{\omega}_m)(z_1-z_0)}v(0,\textbf{q}).
\end{align}

For the same reasons as for the previous residue, we have the residue due to the pole at $q_1^{z(2)}=-i\mu_1$
\begin{align}
\text{Res}[q_1^{z(2)}]	&=\lim_{q^z\to q^{z(2)}}
			\Bigg[\frac{\cancel{(q_1^z+i\mu)}(4\pi\alpha_s)}{(\cancel{q_1^z+i\mu})(q_1^z-i\mu)}
			\frac{e^{-q_1^z(z_1-z_0)}}{P^+(-q_1^z+\omega_0+\tilde{\omega}_m)}\Bigg]\nonumber\\
		&\approx \frac{(4\pi\alpha_s)e^{-\mu_1(z_1-z_0)}}{P^+(-i\mu_1)(-2i\mu_1)}
			=-\frac{(4\pi\alpha_s)e^{-\mu_1(z_1-z_0)}}{2P^+(\mu_1)^2}
\end{align}
We therefore have that $I_2$ is given by
\begin{align}
I_2(p,k,\textbf{q},z_1-z_0) &=-\frac{2\pi i}{2\pi}
			\Bigg[\frac{1}{P^+}e^{i(\omega_0+\tilde{\omega}_m)(z_1-z_0)}v(0,\textbf{q})
			-\frac{(4\pi\alpha_s)e^{-\mu_1(z_1-z_0)}}{2P^+(\mu_1)^2}\Bigg]\nonumber\\
		&=(-i)\frac{(4\pi\alpha_s)}{\mu_1^2(2p^z)+}\bigg[e^{i(\omega_0+\tilde{\omega}_m)(z_1-z_0)}
			-\frac{1}{2}e^{-\mu_1(z_1-z_0)}\bigg]
\end{align}
Where the negative sign is again due to the negative orientation of the contour (that closes in the lower half of the complex plane). The matrix element is therefore
\begin{alignat}{2}\label{ResM110}
\Amp_{1,1,0}	&\approx J(p)&&e^{ipx_0}(-ig_s)T_{a_1}ca_14E\frac{\textbf{k}\cdot\boldsymbol\epsilon}{m_g^2+\textbf{k}^2+x^2M^2}
			\int\frac{d^2\textbf{q}_1}{(2\pi)^2}e^{-i\textbf{q}_1\cdot\textbf{b}_1}\frac{(-i)}{2p^z}\times\nonumber\\
	&	&&\times\bigg[e^{i(\omega_0+\tilde{\omega}_m)(z_1-z_0)}-\frac{1}{2}e^{-\mu_1(z_1-z_0)}\bigg]v(0,\textbf{q}_1)\nonumber\\
	&\approx J(p)&&e^{ipx_0}(-i)\int\frac{d^2\textbf{q}_1}{(2\pi)^2}v(0,\textbf{q}_1)e^{-i\textbf{q}_1\cdot\textbf{b}_1}(-2ig_s)\times\nonumber\\
	&	&&\times\frac{\textbf{k}\cdot\boldsymbol\epsilon}{m_g^2+\textbf{k}^2+x^2M^2}\bigg[e^{i(\omega_0+\tilde{\omega}_m)(z_1-z_0)}
			-\frac{1}{2}e^{-\mu_1(z_1-z_0)}\bigg]T_{a_1}ca_1
\end{alignat}
Here we see then for the first time the retention of a term in the small separation distance treatment that is exponentially suppressed in the large separation distance approximation:  Notice that the second term in square brackets is exponentially suppressed in the large separation distance limit, meaning that \ref{ResM110} reduces to the large path length expression in the limit that $1/\mu_1\ll(z_1-z_0)$.

\subsection{Computation of $\Amp_{1,0,1}$}\label{CalcM101}

In a similar fashion, the matrix element for $\Amp_{1,0,1}$ is given by
\begin{alignat}{2}
\Amp_{1,0,1}	&=&&\int \frac{d^4q_1}{(2\pi)^2}iJ(p+k-q_1)e^{i(p+k-q_1)x_0}\Lambda_1(p,k,q_1)V(q_1)e^{iq_1x_1}\times\nonumber\\
	&	&& i\Delta_M(p+k-q_1)(-i)\Delta_{m_g}(k-q_1)\nonumber\\
	&\approx &&J(p+k)e^{i(p+k)x_0}[c,a_1]T_{a_1}(-i)\int \frac{d^2\textbf{q}_1}{(2\pi)^2}e^{-i\textbf{q}_1\cdot(\textbf{x}_1-\textbf{x}_0)}\times\nonumber\\
	&	&&\times 2g_s\epsilon\cdot(\textbf{k}-\textbf{q}_1)2E I_3,		
\end{alignat}
where
\begin{align}
I_3(p,k,\textbf{q}_1,z_1-z_0)	&=\int\frac{dq_1^z}{2\pi}v(q_1^z,\textbf{q}_1)\Delta_M(p+k-q_1)\Delta_{m_g}(k-q_1)e^{-q_1^z(z_1-z_0)}.
\end{align}
To solve $I_3$, we again consider the poles from the potential and the two propagators.  We now consider all three poles, including the one from the potential.  These are, as per sections \ref{polepqk} and \ref{polekq}, at (dropping the subscript)
\begin{align}
q^{z(1)}	&=-\omega_0-\tilde{\omega}_m-i\epsilon\nonumber\\
q^{z(2)}	&=-\omega_0+\omega_1-i\epsilon\nonumber\\
q^{z(3)}	&=-i\mu_1
\end{align}
We need to calculate the residues due to each pole.  Taking $\epsilon\rightarrow 0$, the residue due to the pole at $q^{z(1)}$
\begin{align}
\text{Res }[q^{1(z)}]	&=\lim_{q^z\rightarrow q^{z(1)}} \Bigg[\frac{(q^z-q^{z(1)})v(-\omega_0-\tilde{\omega}_m)}{(p+k-q)^2-M^2+i\epsilon}\frac{e^{-iq^{z(1)}(z_1-z_0)}}{(k-q)^2-m_g^2+i\epsilon}\bigg]
\end{align}
Again we can simplify using sections \ref{factorkq}, \ref{factorv} and \ref{factorpqk} so we have that
\begin{align}
\text{Res }[q^{z(1)}=-\omega_0-\tilde{\omega}_m]	&\approx\frac{\cancel{(q^z+\omega_0+\tilde{\omega}_m)}}{xP^+\cancel{(q^z+\omega_0+\tilde{\omega}_m)}}\frac{v(0,\textbf{q})e^{-i(-\omega_0-\tilde{\omega}_m)(z_1-z_0)}}{xP^+(-\omega_0-\tilde{\omega}_m +\omega_0-\omega_1)}\nonumber\\
	&=\frac{-1}{x(P^+)^2}\frac{v(0,\textbf{q})}{\tilde{\omega}_m +\omega_1}e^{-i(-\omega_0-\tilde{\omega}_m)(z_1-z_0)}
\end{align}

We can use exactly the same methodology to find the residue due to the pole at $q^{z(2)}=-\omega_0+\omega_1-i\epsilon$.
\begin{align}
\text{Res }[q^{z(2)}]	&=\lim_{q^z\rightarrow q^{z(2)}} \Bigg[\frac{(q^z-q^{z(2)})v(-\omega_0+\omega_1)}{(p+k-q)^2-M^2+i\epsilon}\frac{e^{-iq^{z(2)}(z_1-z_0)}}{(k-q)^2-m_g^2+i\epsilon}\bigg]
\end{align}

With the same simplifications that were used for the previous pole, we have
\begin{align}
\text{Res }[q^{z(2)}=-\omega_0+\omega_1]	&\approx\frac{(\cancel{q^{z(2)}+\omega_0-\omega_1})}{xP^+\cancel{(q^z +\omega_0-\omega_1)}}
		\frac{v(0,\textbf{q})e^{-i(-\omega_0+\omega_1)(z_1-z_0)}}{P^+(-\omega_0+\omega_1+\omega_0+\tilde{\omega}_m)}\nonumber\\
		&=\frac{1}{x(P^+)^2}\frac{v(0,\textbf{q})e^{-i(-\omega_0+\omega_1)(z_1-z_0)}}{\tilde{\omega}_m +\omega_1}
\end{align}

The third residue is given by
\begin{align}
\text{Res}[q^{z(3)}]	&= \lim_{q_1^z\rightarrow q_1^{z(3)}} 
			\Bigg[\frac{\cancel{(q_1^z+i\mu)}(4\pi\alpha_s)}{\cancel{(q_1^z+i\mu)}(q_1^z-i\mu)}
			\frac{1}{(p+k-q_1)^2-M^2+i\epsilon}\times\nonumber\\
		&\times\frac{e^{-iq)_1^z(z_1-z_0)}}{(k-q_1)^2-M^2+i\epsilon}\Bigg]\nonumber\\
		&\approx\frac{(4\pi\alpha_s)}{(-2i\mu_1)}\frac{1}{P^+(-i\mu_1+\omega_0+\tilde{\omega}_m)}
			\frac{e^{-\mu_1(z_1-z_0)}}{(P^+x)(-i\mu_1-\omega_1+\omega_0)}\nonumber\\
		&\approx\frac{4\pi\alpha_s}{(-2i\mu_1)}\frac{e^{-\mu_1(z_1-z_0)}}{\big(x(P^+)^2\big)}\frac{1}{(-i\mu_1)^2}e^{-\mu_1(z_1-z_0)}\nonumber\\
		&= \frac{1}{2}\frac{4\pi\alpha_s}{(-i\mu_1)^3}\frac{1}{x(P^+)^2}e^{-\mu_1(z_1-z_0)}
\end{align}

The integral $I_3$ is then (including a negative sign for the negative orientation of the contour)
\begin{alignat}{2}
I_3(p,k,\textbf{q}_1,&z_1&&-z_0)	\approx -\frac{2\pi i}{2\pi}
			\Bigg[\frac{4\pi\alpha_s\big(e^{i(\omega_0+\tilde{\omega}_m)(z_1-z_0)}-e^{i(\omega_0-\omega_1)(z_1-z_0)}\big)}
			{x(P^+)^2\mu_1^2(\omega_1+\tilde{\omega}_m)}\nonumber\\
		&	&&+\frac{1}{2}\frac{4\pi\alpha_s}{\mu_1^2}\frac{e^{-\mu_1(z_1-z_0)}}{(i\mu_1)}\frac{1}{x(P^+)^2}\Bigg]\nonumber\\
		& =	&&\frac{(-i)(4\pi\alpha_s)}{x(P^+)^2(\mu_1^2)}
			\Bigg[\frac{e^{i(\omega_0+\tilde{\omega}_m)(z_1-z_0)}-e^{i(\omega_0-\omega_1)(z_1-z_0)}}{(\omega_1+\tilde{\omega}_m)}
		+\frac{e^{-\mu_1(z_1-z_0)}}{(2i\mu_1)}	\Bigg]		
\end{alignat}
We see again here the effect described in \ref{MajorCancel}, resulting in the second term in the large brackets being suppressed in relation to the first, meaning that this integral remains unchanged under the small system assumption. The entire amplitude will also remain unchanged, we therefore have the Matrix element
\begin{alignat}{2}\label{ResM101}
\Amp_{1,0,1}	&\approx	&&J(p)e^{i(p+k)x_0}(-i)\int\frac{d2\textbf{q}_1}{(2\pi)^2}v(0,\textbf{q}_1)e^{-i\textbf{q}_1\cdot\textbf{b}_1}2ig_s
	\frac{\epsilon\cdot(\textbf{k}-\textbf{q}_1)}{(\textbf{k}-\textbf{q}_1)^2+M^2x^2+m_g^2} \times\nonumber\\
					&			&&\times \Big(e^{i(\omega_0+\tilde{\omega}_m)(z_1-z_0)}-e^{i(\omega_0-\omega_1)(z_1-z_0)}\Big)[c,a_1]T_{a_1}
\end{alignat}

\pagebreak

\section{$\Amp_{2,0,3}$}\label{CalcM203}

\begin{figure}
\centering
\includegraphics[scale=0.5]{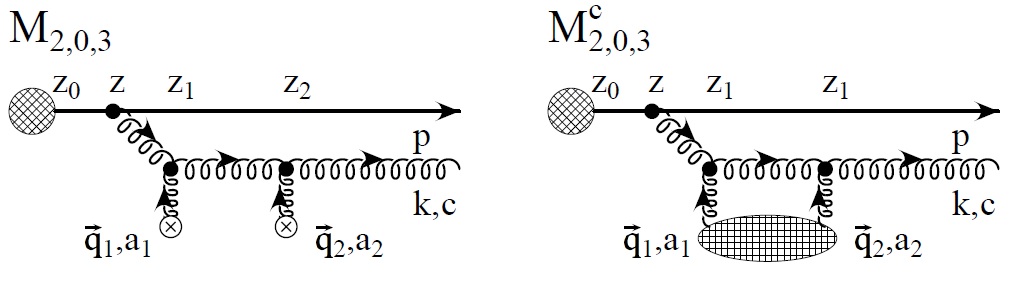}
\caption{$\Amp_{2,0,3}$ ``direct'' contributes to second order in opacity while $\Amp_{2,0,3}^c$ contributes to first order in opacity \cite{Djordjevic2004}. \label{M203diagram} }
\end{figure}

The results that these calculations are based on appear in Appendix C in \cite{Djordjevic2004}.  The relevant diagrams are shown in Figure \ref{M203diagram}.  The same procedure as described for the $\Amp_1$ diagrams is used to determine that the matrix element for $\Amp_{2,0,3}$ is given by
\begin{alignat}{2}
\Amp_{2,0,3}	&= &&\int\frac{d^4q_1}{(2\pi)^4}\int\frac{d^4q_2}{(2\pi)^4}iJ(p+k-q_1-q_2)e^{i(p+k-q_1-q_2)x_0}\times\nonumber\\
		& 	&&\times V(q_1)e^{iq_1x_1}V(q_2)e^{iq_2x_2}\Lambda_{12}(p,k,q_1,q_2)\times\nonumber\\
		&	&&\times i\vartriangle_M(p+k-q_1-q_2)(-i)\vartriangle_{m_g}(k-q_1-q_2)(-i)\vartriangle_{m_g}(k-q_2)\nonumber\\
		& \approx 	&&J(p+k)e^{i(p+k)x_0}\big[[c,a_2],a_1\big]\big(T_{a_2}(2)T_{a_1}(1)\big)\times\nonumber\\
		& 	&&\times (-i)\int\frac{d^2\textbf{q}_1}{(2\pi)^2}(-i)\int\frac{d^2\textbf{q}_2}{(2\pi)^2}2ig_s\epsilon\cdot(\textbf{k}-\textbf{q}_1-\textbf{q}_2)e^{-i\textbf{q}_1\cdot\textbf{b}_1}e^{-i\textbf{q}_2\cdot\textbf{b}_2}\times\nonumber\\
		&			&&\times \int\frac{dq^z_1}{2\pi}\int\frac{dq^z_2}{2\pi}\frac{(4E\omega)v(\vec{\textbf{q}_1})e^{-iq_1^z(z_1-z_0)}}{\big( (p+k-q_1-q_2)^2-M^2+i\epsilon\big)}\times\nonumber\\
		&	&&\times\frac{v(\vec{\textbf{q}_2})e^{-iq_2^z(z_2-z_0)}}{\big((k-q_1-q_2)^2-m_g^2+i\epsilon\big)\big((k-q_2)^2-m_g^2+i\epsilon\big)},
\end{alignat}
In order to evaluate the $q_1^z$ integral, it is convenient to rewrite the following
\begin{align*}
e^{-iq_1^z(z_1-z_0)}e^{-iq_2^z(z_2-z_0)}=e^{-i(q_1^z+q_2^z)(z_1-z_0)}e^{-iq_2^z(z_2-z_1)}
\end{align*}
The first longitudinal integral (keeping with the numbering in \cite{Djordjevic2004})
\begin{align}
I_2(p,k,\textbf{q}_1,\textbf{q}_2, z_1-z_0)	&= \int	\frac{dq_1^z}{2\pi}\frac{v(q_1^z,\textbf{q}_1)e^{-i(q_1^z+q_2^z)(z_1-z_0)}}{\big((p+k-q_1-q_2)^2-M^2+i\epsilon\big)}\times\nonumber\\
	&\times\frac{1}{\big((k-q_1-q_2)^2-m_g^2+i\epsilon\big)}
\end{align}
In the same way that $z_1-z_0>0$ determined that the contour be closed in the lower half of the $q_1^z$ plane, we again close the contour below the real axis here.  The poles are very similar to the poles for $\Amp_{1,0,1}$ and can be obtained quite simply by letting $q_1\rightarrow q_1+q_2$ or making a substitution into the results of $q_3 \equiv q_1+q_2$ which gives the poles as follows
\begin{align}
q_3^{z}	&=-\omega_0-\tilde{\omega}_m-i\epsilon\nonumber\\
q_1^{z(1)}+q_2^{z}	&=-\omega_0-\tilde{\omega}_m-i\epsilon\nonumber\\
q_1^{z(1)}	&=-q_2^{z}-\omega_0-\tilde{\omega}_m-i\epsilon,
\end{align}
and
\begin{align}\label{q3Example}
q_3^{z}	&=-\omega_0+\omega_3-i\epsilon\nonumber\\
q_1^{z(2)}+q_2^{z}	&=-\omega_0+\omega_{12}-i\epsilon\nonumber\\
q_1^{z(2)}	&=-q_2^{z}-\omega_0+\omega_{12}-i\epsilon
\end{align}
The third pole, from the potential, does not need any such a treatment and therefore has no $q_2^z$ dependence.  We therefore have that the poles are at:
\begin{align}
q_1^{z(1)}	&=-q_2^{z}-\omega_0-\tilde{\omega}_m-i\epsilon\nonumber\\
q_1^{z(2)}	&=-q_2^{z}-\omega_0+\omega_{12}-i\epsilon\nonumber\\
q_1^{z(3)}	&= -i\mu_1
\end{align}
The residues are therefore also similar.  From Section \ref{CalcM101} we have that 
\begin{align}
\text{Res }\big[I_2,q_1^{z(1)}] 	&=\frac{-v(-q_2^{z}-\omega_0-\tilde{\omega}_m,\textbf{q}_1)e^{i(\omega_0+\tilde{\omega}_m)(z_1-z_0)}}{x(P^+)^2(\tilde{\omega}_m+\omega_{12})}\\
\text{Res }\big[I_2,q_1^{z(2)}\big]]&=\frac{v(-q_2^{z}-\omega_0+\omega_{12})e^{i(\omega_0-\omega_{12})(z_1-z_0)}}{x(P^+)^2(\tilde{\omega}_m+\omega_{12})}
\end{align}

The residue due to the pole at $q_1^{z(3)}= -i\mu_1$ is given by
\begin{align}
\text{Res}[q_1^{z(3)}]&=\lim_{q_1^{z}\rightarrow q_1^{z(3)}}\Bigg[
				\frac{v(q_1^z,\textbf{q}_1)(q_1^z+i\mu_1)e^{-i(q_1^z+q_2^z)(z_1-z_0)}}
				{\big((p+k-q_1-q_2)^2+M^2+i\epsilon\big)\big((k-q_1-q_2)^2-m_g^2+i\epsilon\big)}\Bigg]
\end{align} 
The individual parts of which are calculated in sections \ref{factorpqk}, \ref{factorpqk} and \ref{factorkq} so that the residue of this pole is given by
\begin{align}
	\text{Res}[q_1^{z(3)}= -i\mu_1]&=\frac{4\pi\alpha_s}{-2i\mu_1}\frac{1}{P^+}\frac{1}{q_2^z-i\mu_1}\frac{1}{xP^+}\frac{1}{q_2^z-i\mu_1}
			e^{-i(q_2^z-i\mu_1)(z_1-z_0)}\nonumber\\
		&\approx -\frac{4\pi\alpha_s}{(2i\mu_1)}\frac{1}{x(P^+)^2}\frac{1}{(q_2^z-i\mu_1)^2}e^{-i(q_2^z-i\mu_1)(z_1-z_0)}
\end{align}

We can therefore calculate the integral $I_2$ (keeping in mind that the contour is once again negatively orientated)
\begin{alignat}{2}
I_2(p,k,&\textbf{q}_1,&&\textbf{q}_2, z_1-z_0)	= -\frac{2\pi i}{2\pi}\Bigg[\frac{1}{x(P^+)^2(\tilde{\omega}_m+\omega_{12})}\times\nonumber\\
		&	&&\times\Big( v(-q_2^{z}-\omega_0+\omega_{12},\textbf{q}_1)e^{i(\omega_0-\omega_{12})(z_1-z_0)}\nonumber\\
		&	&&-v(-q_2^{z}-\omega_0-\tilde{\omega}_m,\textbf{q}_1)e^{i(\omega_0+\tilde{\omega}_m)(z_1-z_0)}\Big)\nonumber\\
		&	&&-\frac{4\pi\alpha_s}{(2i\mu_1)}\frac{1}{x(P^+)^2}\frac{1}{(q_2^z-i\mu_1)^2}e^{-i(q_2^z-i\mu_1)(z_1-z_0)}	\Bigg]\nonumber\\
		&\approx &&\frac{i}{x(P^+)^2}\Bigg[\frac{1}{\tilde{\omega}_m+\omega_{12}}
				\Big(v(-q_2^{z}-\omega_0-\tilde{\omega}_m,\textbf{q}_1)e^{i(\omega_0+\tilde{\omega}_m)(z_1-z_0)}\nonumber\\
		&	&&-v(-q_2^{z}-\omega_0+\omega_{12},\textbf{q}_1)e^{i(\omega_0-\omega_{12})(z_1-z_0)}\Big)\nonumber\\
		&	&&+\frac{4\pi\alpha_s}{2i\mu_1}\frac{1}{(q_2^z-i\mu_1)^2}e^{-i(q_2^z-i\mu_1)(z_1-z_0)}\Bigg]\nonumber\\
		&\approx && \frac{i}{x(P^+)^2}\Bigg[\frac{v(q_2^z+\delta\omega,\textbf{q}_1)}{\tilde{\omega}_m+\omega_{12}}
				\Big(e^{i(\omega_0+\tilde{\omega}_m)(z_1-z_0)}-e^{i(\omega_0-\omega_{12})(z_1-z_0)}\Big)\nonumber\\
		&	&&+\frac{4\pi\alpha_s}{2i\mu_1}\frac{1}{(q_2^z-i\mu_1)^2}e^{-i(q_2^z-i\mu_1)(z_1-z_0)}\Bigg]\nonumber\\
\end{alignat}

Notice here that the potential is evaluated near $q_2^z$ which still needs to be integrated over. In light of the complexity of what follows, in order to be absolutely explicit, I will rewrite the Matrix Element here with the calculated $I_2$.  So, the matrix element becomes:

\begin{alignat}{2}
\Amp_{2,0,3}	&= && J(p+k)e^{i(p+k)x_0}\big[[c,a_2],a_1\big]\big(T_{a_2}(2)T_{a_1}(1)\big)\times\nonumber\\
		& 	&&\times (-i)\int\frac{d^2\textbf{q}_1}{(2\pi)^2}(-i)\int\frac{d^2\textbf{q}_2}{(2\pi)^2}2ig_s\epsilon\cdot(\textbf{k}-\textbf{q}_1-\textbf{q}_2)e^{-i\textbf{q}_1\cdot\textbf{b}_1}e^{-i\textbf{q}_2\cdot\textbf{b}_2}\times\nonumber\\
		&	&&\times 4E\omega\int\frac{dq_2^z}{(2\pi)}\frac{v(q_2^z,\textbf{q}_2)e^{-iq_2^z(z_2-z_0)}}{(k-q_2)^2-m_g^2+i\epsilon}\times\nonumber\\
		&	&&\times \frac{i}{x(P^+)^2}\Bigg[\frac{v(q_2^z+\delta\omega,\textbf{q}_1)}{\tilde{\omega}_m+\omega_{12}}
				\Big(e^{i(\omega_0+\tilde{\omega}_m)(z_1-z_0)}-e^{i(\omega_0-\omega_{12})(z_1-z_0)}\Big)+\nonumber\\
		&	&&+\frac{4\pi\alpha_s}{2i\mu_1}\frac{1}{(q_2^z-i\mu_1)^2}e^{-i(q_2^z-i\mu_1)(z_1-z_0)}\Bigg]\nonumber\\
\end{alignat}

\textbf{A note about $\delta\omega$:} We can `factorize' the potentials out if we take the $\omega$ contributions to be the same.  This has to do with the fact that, within the potential, $\omega$'s get neglected w.r.t. the $\mu$ (which is why the $v(\omega_1,\textbf{q}_1)\approx v(0,\textbf{q}_1)$).  And so, for any pole that contains $\omega$, we have that
\begin{align*}
v(-q_2^z-\omega_0-\tilde{\omega},\textbf{q}_1)&e^{i(\omega_0-\omega_{12})(z_1-z_0)}
	-v(-q_2^z-\omega_0-\tilde{\omega},\textbf{q}_1)e^{i(\omega_0+\tilde{\omega})(z_1-z_0)}	\\
	&\approx v(0,\textbf{q}_1)\big[e^{i(\omega_0-\omega_{12})(z_1-z_0)}-e^{i(\omega_0+\tilde{\omega})(z_1-z_0)}\big]
\end{align*}
Similarly, if the pole contains $\mu$, the $\mu$ dominates the $\omega$'s, so we have
\begin{align*}
	&\approx v(\mu,\textbf{q}_1)\big[e^{i(\omega_0-\omega_{12})(z_1-z_0)}-e^{i(\omega_0+\tilde{\omega})(z_1-z_0)}\big] 
\end{align*}
The question might arise as to whether or not this is still valid for other poles (being the ones that will be kept at a later stage) and some reflection will reveal that it does indeed since those poles were suppressed exponentially due to $(z_1-z_0)$ and not $\omega$.
Now we can more clearly see that the factorizing won't happen as easily as before.  So let us start by defining:
\begin{equation}
\{\star\}\equiv e^{i(\omega_0+\tilde{\omega}_m)(z_1-z_0)}-e^{i(\omega_0-\omega_{12})(z_1-z_0)},
\end{equation}
which allows one to redefine $I_3$.
\begin{alignat}{2}
I_3(k,\textbf{q}_1,\textbf{q}_2,z_2-z_1)&\equiv	&&\int\frac{dq_2^z}{(2\pi)}
				\frac{v(q_2^z,\textbf{q}_2)e^{-iq_2^z(z_2-z_1)}}{\big((k-q_2)^2-m_g^2+i\epsilon\big)}\times\nonumber\\
		&	&&\times\bigg[\frac{v(q_2^z+\delta\omega,\textbf{q}_1)\{\star\}}{\tilde{\omega}_m+\omega_{12}}
				+\frac{4\pi\alpha_s}{2i\mu_1}\frac{e^{-i(q_2^z-i\mu_1)(z_1-z_0)}}{(q_2^z-i\mu_1)^2}\bigg]
\end{alignat}

In the most general case (that includes the contact case), there are three contributing poles (two from the potentials and one from the propagator).  The poles from the potential are calculated in section \ref{polepotential}.  These poles contribute regardless of the system size, because the suppression in the large system approximation occurs due to the fact that $(z_1-z_0)$ is very large.  Here, this does not hold sway, since we will in fact take  $(z_2-z_1)$ to become very \textit{small}. The terms in $\delta\omega$ are neglected in the calculation.  The pole from the propagator is calculated in section \ref{polekq}.  \textbf{The only pole that is new in the small system limit is $+i\mu_1$ which is not considered because it lies in the upper half of the complex plane}. We therefore have the poles
\begin{align}
q_2^{z(1)}	&=-i\mu_1+\delta\omega\\
q_2^{z(2)}	&=-i\mu_2\\
q_2^{z(3)}	&=\omega_2-\omega_0-i\epsilon\\
q_2^{z(4)}	&=+i\mu_1, \qquad \text{(ignored)}
\end{align}
Note however that, although there are no extra poles, there is an extra term, which will change the other poles.

We calculate the residues individually:  For $q_2^{z(1)}=-i\mu_1$ there are three contributions to $I_3$:
\begin{alignat}{2}
\text{Res }\big[I_3, q_2^{z(1)}] &=&&\lim_{q_2^{z}\to q_2^{z(1)}}	\Bigg\{\frac{v(q_2^z,\textbf{q}_2)e^{-iq_2^z(z_2-z_1)}(q_2^z+i\mu_1-\delta\omega)}{\big((k-q_2)^2+m_g^2+i\epsilon\big)}\times\nonumber\\
		&	&&\times\bigg[\frac{v(q_2^z+\delta\omega,\textbf{q}_1)\{\star\}}{\tilde{\omega}_m+\omega_{12}}
				+\frac{4\pi\alpha_s}{2i\mu_1}\frac{e^{-i(q_2^z-i\mu_1)(z_1-z_0)}}{(q_2^z-i\mu_1)^2}\bigg]\Bigg\}
\end{alignat}

Since this pole comes from the potential inside the square brackets, the second term will evaluate to zero here because it does not contain anything to cancel.  We therefore have almost exactly the same pole, we just include here $\{\star\}$.  Using sections \ref{factorkq}, \ref{factorvMixed} and \ref{factorvSame} we have that \textbf{The first residue is not affected by the small separation distance approximation:}
\begin{align}
\text{Res }\big[I_3, q_2^{z(1)}\big]	&=\frac{1}{xP^+}\frac{e^{-\mu_1(z_2-z_1)}}{(-i\mu_1)}\frac{4\pi\alpha_s}{-2i\mu_1}\frac{4\pi\alpha_s}{(\mu_2^2-\mu_1^2)}\frac{\{\star\}}{\tilde{\omega}_m+\omega_{12}}\nonumber\\
\end{align}

For $q_2^{z(2)}	=-i\mu_2$ we have the following contributions

\begin{alignat}{2}
&\text{Res }\big[&&I_3, q_2^{z(2)}\big] =\lim_{q_2^{z}\to q_2^{z(2)}}	\Bigg\{\frac{v(q_2^z,\textbf{q}_2)e^{-iq_2^z(z_2-z_1)}(q_2^z+i\mu_2)}{\big((k-q)^2-m_g^2+i\epsilon\big)}\times\nonumber\\
		&	&&\times\bigg[\frac{v(q_2^z+\delta\omega,\textbf{q}_1)\{\star\}}{\tilde{\omega}_m+\omega_{12}}
				+\frac{4\pi\alpha_s}{2i\mu_1}\frac{e^{-i(q_2^z-i\mu_1)(z_1-z_0)}}{(q_2^z-i\mu_1)^2}\bigg]\Bigg\}
\end{alignat}

The results from sections \ref{factorkq}, \ref{factorvMixed} and \ref{factorvSame} allow for the calculation of the second residue:

\begin{alignat}{2}
\text{Res }\big[I_3, q_2^{z(2)=-i\mu_2}\big]&= &&\frac{4\pi\alpha_s}{-2i\mu_2}\frac{1}{xP^+}\frac{e^{-\mu_2(z_2-z_1)}}{(-i\mu_2)}\times\nonumber\\
			&	&&\times\bigg[\frac{4\pi\alpha_s}{(\mu_1^2-\mu_2^2)(\tilde{\omega}_m+\omega_{12})}\{\star\}
					-\frac{4\pi\alpha_s}{2i\mu_1}\frac{e^{-(\mu_2+i\mu_1)(z_1-z_0)}}{(\mu_2-\mu_1)^2}\bigg]\nonumber\\
			&= 	&&\frac{(4\pi\alpha_s)^2e^{-\mu_2(z_2-z_1)}}{(-2\mu_2^2)xP^+}\times\nonumber\\
			&	&&\bigg[\frac{\{\star\}}{(\mu_1^2-\mu_2^2)(\tilde{\omega}_m+\omega_{12})}
					-\frac{e^{-(\mu_2+i\mu_1)(z_1-z_0)}}{(2i\mu_1)(\mu_2-\mu_1)^2}	\bigg]		
\end{alignat}

Here we see again the effect that cancels terms, causing the second term in square brackets to be suppressed as compared to the first (Section \ref{MajorCancel}). Therefore, the second residue remains unchanged in the small system approximation.

\begin{align}
\text{Res }\big[I_3, q_2^{z(2)}\big]= -\frac{(4\pi\alpha_s)^2}{2\mu_2^2}\frac{\{\star\}}{xP^+}\frac{e^{-\mu_2(z_2-z_1)}}{(\mu_1^2-\mu_2^2)(\tilde{\omega}_m+\omega_{12})}
\end{align}

The third residue is probably the simplest, so for $q_2^{z(3)}=\omega_2-\omega_0-i\epsilon$
\begin{alignat}{2}
&\text{Res }&&\big[I_3, q_2^{z(3)}\big] =\lim_{q_2^{z}\to q_2^{z(3)}}	\Bigg\{\frac{v(q_2^z,\textbf{q}_2)e^{-iq_2^z(z_2-z_1)}(q_2^z-\omega_2+\omega_0)}{\big((k-q)^2-m_g^2+i\epsilon\big)}\times\nonumber\\
		&	&&\times\bigg[\frac{v(q_2^z+\delta\omega,\textbf{q}_1)\{\star\}}{\tilde{\omega}_m+\omega_{12}}
				+\frac{4\pi\alpha_s}{2i\mu_1}\frac{e^{-i(q_2^z-i\mu_1)(z_1-z_0)}}{(q_2^z-i\mu_1)^2}\bigg]\Bigg\}
\end{alignat}
where we have, using the same sections as for the previous poles that the residue due to the third pole (from the propagator) is then

\begin{alignat}{2}
\text{Res }\big[I_3, &q_2^{z(3)}&&=\omega_2-\omega_0\big]=\frac{4\pi\alpha_s}{xP^+}\frac{e^{-i(\omega_2-\omega_0)(z_2-z_1)}}{\mu_2^2}\times\nonumber\\
		&	&&\times\bigg[\frac{4\pi\alpha_s\{\star\}}{(\mu_1)^2(\tilde{\omega}_m+\omega_{12})}-\frac{4\pi\alpha_s}{2i\mu_1}
			\frac{e^{-\mu_1(z_1-z_0)}}{(\mu_1^2)}\bigg]\nonumber\\
		&=&&\frac{(4\pi\alpha_s)^2}{xP^+}\frac{e^{-i(\omega_2-\omega_0)(z_2-z_1)}}{\mu_1^2\mu_2^2}\times\nonumber\\
		&	&&\times\bigg[\frac{\{\star\}}{\tilde{\omega}_m+\omega_{12}}-\frac{e^{-\mu_1(z_1-z_0)}}{2i\mu_1}\bigg]
\end{alignat}

We see again the suppression effect from Section \ref{MajorCancel}, suppressing the second term relative to the first.  Therefore, the third residue remains unchanged under the small separation distance limit.
\begin{align}
\text{Res }\big[I_3, q_2^{z(3)=\omega_2-\omega_0}\big]&=\frac{(4\pi\alpha_s)^2}{xP^+}
			\frac{e^{-i(\omega_2-\omega_0)(z_2-z_1)}}{\mu_1^2\mu_2^2}\frac{\{\star\}}{\tilde{\omega}_m+\omega_{12}}
\end{align}

We now need to sum these terms in order to calculate the integral $I_3$.  
\begin{alignat}{2}\label{M203I3}
I_3(k,\textbf{q}_1,\textbf{q}_2,z_2-z_1)&=	&&\frac{1}{xP^+}\frac{e^{-\mu_1(z_2-z_1)}}{i\mu_1}
				\frac{4\pi\alpha_s}{-2i\mu_1}\frac{4\pi\alpha_s}{(\mu_2^2-\mu_1^2)}\{\star\}\nonumber\\
		&	&&-\frac{(4\pi\alpha_s)^2}{2\mu_2^2}\frac{\{\star\}}{xP^+}
				\frac{e^{-\mu_2(z_2-z_1)}}{(\mu_1^2-\mu_2^2)(\tilde{\omega}_m+\omega_{12})}\nonumber\\
		&	&&+\frac{(4\pi\alpha_s)^2}{xP^+}\frac{e^{-i(\omega_2-\omega_0)(z_2-z_1)}}{\mu_1^2\mu_2^2}
				\frac{\{\star\}}{\tilde{\omega}_m+\omega_{12}}\nonumber\\
		&=	&&\frac{(4\pi\alpha_2)^2(-i)\{\star\}}{xP^+(\tilde{\omega}_m+\omega_{12})}\Bigg[\frac{e^{-i(\omega_2-\omega_0)(z_2-z_1)}}{\mu_1^2\mu_2^2}\nonumber\\
		&	&&+\frac{1}{2(\mu_2^2-\mu_1^2)}\bigg\{\frac{e^{-\mu_1(z_2-z_1)}}{\mu_1}-\frac{e^{-\mu_2(z_2-z_1)}}{\mu_2^2}\bigg\}\Bigg]			
\end{alignat}

This seems difficult, but we need only consider two extreme limits (ignoring for the moment a common factor of $\frac{\{\star\}}{\tilde{\omega}_m+\omega_{12}}$):
\begin{enumerate}
\item The limit of well-separated scattering centres $z_2-z_1\gg 1/\mu$;
\item The special ``contact'' limit $z_2=z_1$ limit to compute unitary contributions
\end{enumerate}
For the first case in which the scattering centres are separated on the scale of the mean free path ($z_2-z_1\sim\lambda\gg 1/\mu$), equation (\ref{M203I3}) reduces to
\begin{align}
I_3(k,\textbf{q}_1,\textbf{q}_2,z_2-z_1\gg 1/\mu)\approx-\frac{i}{xP^+}v(0,\textbf{q}_1)v(0,\textbf{q}_2)e^{-i(\omega_2-\omega_0)(z_2-z_1)},
\end{align}
where the other terms have fallen away because the exponential terms have massively negative arguments.
For the second case in which $z_2-z_1=0$, consider the following fact that arises from $\mu_i^2\equiv \mu_{i\perp}=\textbf{q}_i^2+\mu^2$
\begin{align}
-\frac{(4\pi\alpha_s)^2}{2(\mu_2^2-\mu_1^2)}\bigg (\frac{1}{\mu_2^2}-\frac{1}{\mu_1^2}\bigg)
		&= -\frac{(4\pi\alpha_s)^2}{2(\mu^2-\textbf{q}_1^2-\mu^2+\textbf{q}_2^2)}\Bigg[\frac{1}{\mu^2-\textbf{q}_2}-\frac{1}{\mu^2-\textbf{q}_1}\Bigg]\nonumber\\
		&=\frac{(4\pi\alpha_s)^2}{2(\textbf{q}_2^2-\textbf{q}_1^2)}\Bigg[\frac{(\textbf{q}_2^2-\textbf{q}_1^2)}{(\mu^2-\textbf{q}_2)(\mu^2-\textbf{q}_1)}\Bigg]\nonumber\\
		&=\frac{1}{2}{v(0,\textbf{q}_1)v(0,\textbf{q}_2)}.
\end{align}
We therefore have that equation (\ref{M203I3}) reduces to 
\begin{align}
I_3(k,\textbf{q}_1,\textbf{q}_2,z_2=z_1)\approx-\frac{i}{xP^+}\frac{1}{2} v(0,\textbf{q}_1)v(0,\textbf{q}_2)
\end{align}
Which is exactly half (in strength) of the first case.

We can now return to the amplitude for $\Amp_{2,0,3}$ in the contact limit.  Note first that
\begin{align*}
\frac{4E\omega}{x(P^+)^2(\omega_{(12)}+\tilde{\omega_m})}\frac{1}{2k^+}&=\frac{4E\omega}{P^+\big[(\textbf{k}-\textbf{q}_1-\textbf{q}_2)^2+M^2x^2+m_g^2\big]}\frac{1}{2k^+}\nonumber\\
		&\approx \frac{1}{\big[(\textbf{k}-\textbf{q}_1-\textbf{q}_2)^2+M^2x^2+m_g^2\big]}\frac{4\big(\frac{1}{2}\cancel{E^+}\big)\big(\frac{1}{2}\cancel{xE^+}\big)}{2\cancel{E^+}\cancel{xE^+}}\\
		&=\frac{1}{2}\frac{1}{\big[(\textbf{k}-\textbf{q}_1-\textbf{q}_2)^2+M^2x^2+m_g^2\big]}
\end{align*}
We therefore have the amplitude:
\begin{alignat}{2}\label{ResM203}
\Amp_{2,0,3}	&\approx &&J(p)e^{i(p+k)x_0}(-i)\int\frac{d^2\textbf{q}_1}{(2\pi)^2}v(0,\textbf{q}_1)e^{-i\textbf{q}_1\cdot\textbf{b}_1}								(-i)\int\frac{d^2\textbf{q}_2}{(2\pi)^2}v(0,\textbf{q}_2)e^{-i\textbf{q}_2\cdot\textbf{b}_2}\times\nonumber\\
		&	&&\times\frac{1}{2}(2ig_s)\frac{\epsilon\cdot(\textbf{k}-\textbf{q}_1-\textbf{q}_2)}{\big[(\textbf{k}-\textbf{q}_1-\textbf{q}_2)^2+M^2x^2+m_g^2\big]}\big[[c,a_2],a_1\big]\big(T_{a_2}T_{a_1}\big)\times\nonumber\\
		&	&&\times\big\lbrace e^{i(\omega_0+\tilde{\omega}_m)(z_1-z_0)}-e^{i(\omega_0-\omega_{(12)})(z_1-z_0)}\big\rbrace
\end{alignat}

\pagebreak

\section{$\Amp_{2,0,0}$ and $\Amp_{2,2,0}$}

\subsection{$\Amp_{2,0,0}$}\label{CalcM200}
\begin{figure}
\centering
\includegraphics[scale=0.5]{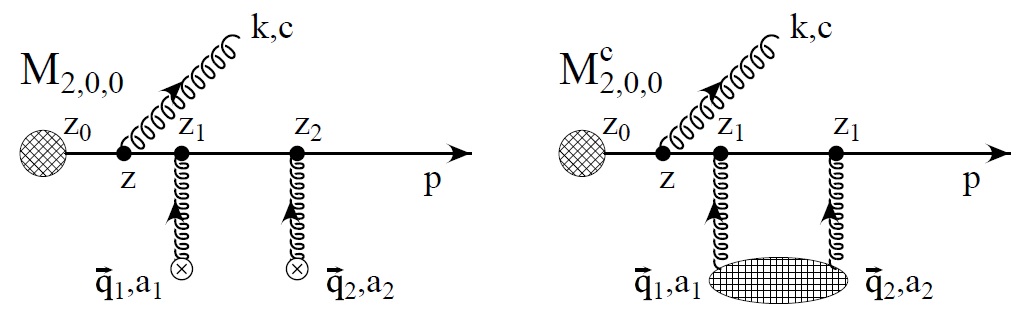}
\caption{$\Amp_{2,0,0}$ graphs for the well separated case as well as the contact limit \cite{Djordjevic2004}.\label{M200Diagram}}
\end{figure}
The Matrix element for $\Amp_{2,0,0}$ is calculated as before from the diagram in Figure \ref{M200Diagram}
\begin{alignat}{2}
\Amp_{2,0,0}	&= &&\int	\frac{d^4q_1}{(2\pi)^4}\frac{d^4q_2}{(2\pi)^4}iJ(p+k-q_1-q_2)e^{i(p+k-q_1-q_2)x_0}V(q_1)e^{iq_1x_1}\times\nonumber\\
		&	&&V(q_2)e^{iq_2x_2}(-i2E)^2ig_s(2p+k)_\mu\epsilon^\mu \times\nonumber\\
		&	&&\times i\vartriangle _M(p+k-q_1-q_2)i\vartriangle _M(p-q_1-q_2)i\vartriangle _M(p-q_2)a_1c(T_{a_2}T_{a_1})\nonumber\\
		&\approx	&& J(p)e^{i(p+k)x_0}(-i)\int\frac{d^2\textbf{q}_1}{(2\pi)^2}
				e^{-i\textbf{q}_1\cdot\textbf{b}_1}	(-i)\int\frac{d^2\textbf{q}_2}{(2\pi)^2}e^{-i\textbf{q}_2\cdot\textbf{b}_2}\times\nonumber\\
		&	&&\times\frac{2ig_s(\epsilon\cdot\textbf{k})}{x}a_2a_1c(T_{a_2}T_{a_1})(2E)^2\times\nonumber\\
		& 	&&\times\int\frac{dq_1^z}{2\pi}\frac{dq_2^z}{2\pi}
			\frac{v(\vec{\textbf{q}_1})e^{-iq_1^z(z_1-z_0)}}{\big((p+k-q_1-q_2)^2-M^2+i\epsilon\big)}\times\nonumber\\
		&	&&\times\frac{v(\vec{\textbf{q}_2})e^{-iq_2^z(z_2-z_0)}}{\big((p-q_1-q_2)^2-M^2+i\epsilon\big)\big((p-q_2)^2-M^2+i\epsilon\big)}
\end{alignat}
Here we define (again keeping with the numbering in \cite{Djordjevic2004})
\begin{align}
I_2(p,k,&\textbf{q}_1,\textbf{q}_2,z_1-z_0)\nonumber\\
	&=\int \frac{dq_1^z}{2\pi}\frac{v(q_1^z,\textbf{q}_1)e^{i(q_1^z-q_2^z)(z_1-z_0)}}{\big((p+k-q_1-q_2)^2-M^2+i\epsilon\big)\big((p-q_1-q_2)^2-M^2+i\epsilon\big)}
\end{align}
There are three poles, one from each propagator and one from the potential.  These are found by (as for $\Amp_{2,0,3}$) making the substitution $q_3^z=q_1^z+q_2^z$, using the results from sections \ref{polepqk} and \ref{polepq} and then `solving' for $q_1^z$.  This results in the following three poles:
\begin{align}
q_1^{z(1)}	&=-q_2^z-\omega_0-\tilde{\omega}_m -i\epsilon\\
q_1^{z(2)}	&=-q_2^z+x(\omega_{(12)}-\omega_0)-i\epsilon\approx-q_2^z-i\epsilon\\
q_1^{z(3)}	&=-i\mu_1
\end{align}
We calculate the residues individually.  The residue due to the pole at $q_1^z=q_1^{z(1)}$ is given by

\begin{align}
\text{Res }&\big[I_2,q_1^{z(1)}\big]\nonumber\\
	&=\lim_{q_1^z\to q_1^{z(1)}}\Bigg[\frac{v(q_1^z,\textbf{q}_1)e^{i(q_1^z-q_2^z)(z_1-z_0)}(q_1^z+q_2^z+\omega_0+\tilde{\omega}_m)}{\big((p+k-q_1-q_2)^2-M^2+i\epsilon\big)\big((p-q_1-q_2)^2-M^2+i\epsilon\big)}\Bigg]
\end{align}

Sections \ref{polepotential}, \ref{factorpqk} and \ref{factorpq} give that the first residue is 

\begin{align}
\text{Res }\big[I_2,q_1^{z(1)}\big]	&\approx\frac{v(-q_2^z,\textbf{q}_1)}{(P^+)^2(-\omega_0-\tilde{\omega}_m)}e^{i(\omega_0+\tilde{\omega}_m)(z_1-z_0)}
\end{align}

The second residue due, to the pole at $q_1^z=-q_2^z+x(\omega_{(12)}-\omega_0)$, is calculated as follows:
\begin{align}
\text{Res }&\big[I_2,q_1^{z(2)}\big]\nonumber\\
	&=\lim_{q_1^z\to q_1^{z(1)}}\Bigg[\frac{v(q_1^z,\textbf{q}_1)e^{-i(q_1^z+q_2^z)(z_1-z_0)}(q_1^z+q_2^z-x(\omega_{(12)}-\omega_0)\big)}{\big((p+k-q_1-q_2)^2-M^2+i\epsilon\big)\big((p-q_1-q_2)^2-M^2+i\epsilon\big)}\Bigg]
\end{align}

The individual contributions are calculated in sections \ref{polepotential}, \ref{factorpqk} and \ref{factorpq} to give
\begin{align}
\text{Res }\big[I_2,q_1^{z(2)}]	&\approx\frac{1}{(P^+)^2}\frac{v(-q_2^z,\textbf{q}_1)}{(\omega_0+\tilde{\omega}_m)}e^{i(0)(z_1-z_2)}\nonumber\\
	&\approx\frac{1}{(P^+)^2}\frac{v(-q_2^z,\textbf{q}_1)}{(\omega_0+\tilde{\omega}_m)}
\end{align}

The residue due to the pole at $q_1^{z(3)}=-i\mu_1$ is given by
\begin{align*}
\text{Res }&\big[I_2,q_1^{z(3)}\big]	\nonumber\\
&=\lim_{q_1^z\to q_1^{z(3)}}\Bigg[\frac{v(q_1^z,\textbf{q}_1)e^{-i(q_1^z+q_2^z)(z_1-z_0)}(q_1^z+i\mu_1)}{\big((p+k-q_1-q_2)^2-M^2+i\epsilon\big)\big((p-q_1-q_2)^2-M^2+i\epsilon\big)}\Bigg]
\end{align*}

with simplifications from the same sections giving
 
\begin{align}
\text{Res }\big[I_2,q_1^{z(3)}=-i\mu_1\big]	&\approx \frac{4\pi\alpha_s}{-2i\mu_1}\frac{1}{(P^+)^2}\frac{e^{-i(q_2^z-i\mu_1)(z_1-z_0)}}{(q_2^z-i\mu_1)^2}
\end{align}

We can now calculate the integral $I_2$:
\begin{alignat}{2}
I_2(p,k,\textbf{q}_1,\textbf{q}_2,z_1-z_0)&=&&\frac{-2\pi i}{2\pi}
			\Bigg[\frac{4\pi\alpha_s}{\big((q_2^z)^2+\mu_1^2\big)}\frac{e^{i(\omega_0+\tilde{\omega}_m)(z_1-z_0)}}{(P^+)^2(-\omega_0-\tilde{\omega}_m)}\nonumber\\
		&	&&+\frac{4\pi\alpha_s}{\big((q_2^z)^2+\mu_1^2\big)}\frac{1}{(P^+)^2}\frac{1}{(\omega_0+\tilde{\omega}_m)}\nonumber\\
		&	&&+ \frac{4\pi\alpha_s}{-2i\mu_1}\frac{1}{(P^+)^2}\frac{e^{-i(q_2^z-i\mu_1)(z_1-z_0)}}{(q_2^z-i\mu_1)^2}\Bigg]\nonumber\\
\end{alignat}

We also see here the suppression effect, so the second term is again suppressed relative to the first, and the integral stays the same:  
\begin{equation}
I_2(p,k,\textbf{q}_1,\textbf{q}_2,z_1-z_0)	=\frac{(4\pi\alpha_s)(i)}{(P^+)^2}\frac{e^{i(\omega_0+\tilde{\omega}_m)(z_1-z_0)}-1}{(\omega_0+\tilde{\omega}_m)(q_2^z-i\mu_1)^2}
\end{equation}

The matrix element with the calculated $I_2$:
\begin{alignat}{2}
\Amp_{2,0,0}		&\approx	&& J(p)e^{i(p+k)x_0}(-i)\int\frac{d^2\textbf{q}_1}{(2\pi)^2}
			e^{-i\textbf{q}_1\cdot\textbf{b}_1}		(-i)\int\frac{d^2\textbf{q}_2}{(2\pi)^2}e^{-i\textbf{q}_2\cdot\textbf{b}_2}\times\nonumber\\
	&	&&\frac{2ig_s(\epsilon\cdot\textbf{k})}{x}a_2a_1c(T_{a_2}T_{a_1})(2E)^2\times\nonumber\\
	& 	&&\times\frac{dq_2^z}{2\pi}	\frac{v(\vec{\textbf{q}_2})e^{-iq_2^z(z_2-z_1)}}{\big((p-q_2)^2-M^2+i\epsilon\big)}\frac{(4\pi\alpha_s)(i)}{(P^+)^2}\frac{e^{i(\omega_0+\tilde{\omega}_m)(z_1-z_0)}-1}{(\omega_0+\tilde{\omega}_m)(q_2^z-i\mu_1)^2}			
\end{alignat}

The remaining integral, $I_3$, is then defined as
\begin{align}
I_3(p,k,\textbf{q}_1,\textbf{q}_2,z_1-z_0)&=\int\frac{dq_2^z}{2\pi}\frac{v(-q_2^z,\textbf{q}_1)v(-q_2^z,\textbf{q}_2)e^{-iq_2^z(z_2-z_1)}}{\big((p-q_2)^2-M^2+i\epsilon\big)}\times\nonumber\\
	&\times\frac{e^{i(\omega_0+\tilde{\omega}_m)(z_1-z_0)}-1}{(\omega_0+\tilde{\omega}_m)(q_2^z-i\mu_1)^2}
\end{align}

Since we will specifically be looking at the contact limit, the poles from the potentials can not be neglected.  We therefore have three poles:  two from the two potentials and one from the propagator.  Also, there are no new poles from the short path length limit, and therefore the rest of the calculation of this diagram is identical to the large path length calculation.  By sections \ref{polepotential} and \ref{polepq}, these poles are
\begin{align}
q_2^{z(1)}	&=-i\mu_1\\
q_2^{z(2)}	&=-i\mu_2\\
q_2^{z(3)}	&=\frac{(\textbf{k}+\textbf{q})^2-\textbf{k}^2}{P^+}-i\epsilon
\end{align}

For the last pole, it will be sufficient to use $q_2^{z(3)}\approx-i\epsilon$, but the full results here for completeness until the residue for that pole is calculated. Again, we calculate the residues individually, starting with the residue due to the pole at $q_2^{z(1)}=-i\mu_1$, given by
\begin{align}
\text{Res }\big[I_3,q_2^{z(1)}]	&=\lim_{q_2^z\to q_2^{z(1)}}\bigg[\frac{v(-q_2^z,\textbf{q}_1)v(-q_2^z,\textbf{q}_2)(q_2^z+i\mu_1)}{\big((p-q_2)^2-M^2+i\epsilon\big)}e^{-iq_2^z(z_2-z_1)}\bigg],
\end{align}
and computed using sections \ref{factorvMixed}, \ref{factorvSame} and \ref{factorpq}:
 
\begin{align}
\text{Res }\big[I_3,q_2^{z(1)}=-i\mu_1]	&=\frac{(4\pi\alpha_s )}{-2i\mu_1}\frac{(4\pi\alpha_s )}{\mu_2^2-\mu_1^2}\frac{i}{P^+\mu_1}e^{-\mu_1(z_2-z_1)}\nonumber\\
	&=\frac{-(4\pi\alpha_s )^2}{2\mu_1^2(\mu_2^2-\mu_1^2)P^+}e^{-\mu_1(z_2-z_1)}
\end{align}

The residue for the pole at $q_2^{z(2)}	=-i\mu_2$ is calculated in much the same way with minor changes.  It is given by
\begin{align}
\text{Res }\big[I_3,q_2^{z(2)}]	&=\lim_{q_2^z\to q_2^{z(2)}}\bigg[\frac{v(-q_2^z,\textbf{q}_1)v(-q_2^z,\textbf{q}_2)(q_2^z+i\mu_2)}{\big((p-q_2)^2-M^2+i\epsilon\big)}e^{-iq_2^z(z_2-z_1)}\bigg],
\end{align}
and its parts are simplified with results from the same sections to give

\begin{align}
\text{Res }\big[I_3,q_2^{z(2)}=-i\mu_2]	&=\frac{-(4\pi\alpha_s )^2}{2\mu_2^2P^+(\mu_1^2-\mu_2^2)}e^{-\mu_2(z_2-z_1)}
\end{align}

The last pole is at $q_2^{z(3)}	=\frac{(\textbf{k}+\textbf{q})^2-\textbf{k}}{P^+}-i\epsilon$ and gives the residue
\begin{align}
\text{Res }\bigg[I_3,q_2^{z(3)}\bigg]	&=\lim_{q_2^z\to q_2^{z(3)}}\bigg[\frac{v(-q_2^z,\textbf{q}_1)v(-q_2^z,\textbf{q}_2)\big(q_2^z-\frac{(\textbf{k}+\textbf{q})^2-\textbf{k}}{P^+}\big)}{\big((p-q_2)^2-M^2+i\epsilon\big)}e^{-iq_2^z(z_2-z_1)}\bigg],
\end{align}
the individual contributions to which are also from the above section and give

\begin{align}
\text{Res }\bigg[I_3,q_2^{z(3)}=\frac{(\textbf{k}+\textbf{q})^2-\textbf{k}}{P^+}\bigg]	&=\frac{(4\pi\alpha_s)^2}{P^+\mu_1^2\mu_2^2}e^{-ix(\omega_2-\omega_0)(z_2-z_1)}\nonumber\\
	&=\frac{-1}{P^+}v(0,\textbf{q}_1)v(0,\textbf{q}_2)e^{-ix(\omega_2-\omega_0)(z_2-z_1)}
\end{align}

Which means that the integral $I_3$ becomes
\begin{alignat}{2}
I_3(p,k,\textbf{q}_1,\textbf{q}_2,z_1-z_0)&=-\frac{2\pi i}{2\pi}	&&\Bigg[\frac{-(4\pi\alpha_s )^2}{2\mu_1^2(\mu_2^2-\mu_1^2)P^+}e^{-\mu_1(z_2-z_1)}\nonumber\\
		&	&&\frac{-(4\pi\alpha_s )^2}{2\mu_2^2P^+(\mu_1^2-\mu_2^2)}e^{-\mu_2(z_2-z_1)}\nonumber\\
		&	&&\frac{-1}{P^+} v(0,\textbf{q}_1)v(0,\textbf{q}_2)e^{-ix(\omega_2-\omega_0)(z_2-z_1)}\Bigg]\nonumber\\
		&=	&&\frac{-i}{P^+}\bigg[v(0,\textbf{q}_1)v(0,\textbf{q}_2)e^{-ix(\omega_2-\omega_0)(z_2-z_1)}\nonumber\\
		&	&&+\frac{(4\pi\alpha_s )^2}{2(\mu_1^2-\mu_2^2)}\bigg(\frac{e^{-\mu_1(z_2-z_1)}}{\mu_1^2}-\frac{e^{-\mu_2(z_2-z_1)}}{\mu_2^2}\bigg)\bigg]
\end{alignat}

Again we consider two cases, the well separated and the contact limit. 
\begin{enumerate}
\item $z_2-z_1\gg 1/\mu$:  The two terms in the round brackets go to zero:
\begin{equation}
I_3\approx\frac{-i}{P^+}v(0,\textbf{q}_1)v(0,\textbf{q}_2)e^{-ix(\omega_2-\omega_0)(z_2-z_1)}
\end{equation}
\item $z_2=z_1$:  This case is slightly more complicated
\begin{align}
I_3=\frac{-i}{P^+}\bigg[v(0,\textbf{q}_1)v(0,\textbf{q}_2)+\frac{(4\pi\alpha_s )^2}{2(\mu_1^2-\mu_2^2)}\bigg(\frac{1}{\mu_1^2}-\frac{1}{\mu_2^2}\bigg)\bigg]
\end{align}

We can see that we get the factor of $1/2$ again in the following way.  Consider

\begin{align}\label{ReduceM200contact}
\frac{(4\pi\alpha_s )^2}{2(\mu_1^2-\mu_2^2)}\bigg(\frac{1}{\mu_1^2}-\frac{1}{\mu_2^2}\bigg)	&=\frac{(4\pi\alpha_s )^2}{2(\textbf{q}_1^2-\textbf{q}_2^2)}\bigg(\frac{1}{\mu^2-\textbf{q}_1^2}-\frac{1}{\mu^2-\textbf{q}_2^2}\bigg)\nonumber\\
		&=\frac{(4\pi\alpha_s )^2}{2\cancel{(\textbf{q}_1^2-\textbf{q}_2^2)}}\bigg(\frac{-\cancel{(\textbf{q}_1^2-\textbf{q}_2^2)}}{(\mu^2-\textbf{q}_1^2)(\mu^2-\textbf{q}_2^2)}\bigg)\nonumber\\
		&=-\frac{1}{2}v(0,\textbf{q}_1)v(0,\textbf{q}_2)
\end{align}

We therefore have that 
\begin{align}
I_3\approx\frac{1}{2} \frac{(-i)}{P^+}v(0,\textbf{q}_1)v(0,\textbf{q}_2)
\end{align}
\end{enumerate}
Finally, the amplitude can now be written by combining the above results (remembering that part of the result for $I_2$ is integrated over in $I_3$ and so should not be included twice):

\begin{alignat}{2}\label{ResM200}
\Amp_{2,0,0}		&\approx	&& J(p)e^{i(p+k)x_0}(-i)\int\frac{d^2\textbf{q}_1}{(2\pi)^2}e^{-i\textbf{q}_1\cdot\textbf{b}_1}		(-i)\int\frac{d^2\textbf{q}_2}{(2\pi)^2}e^{-i\textbf{q}_2\cdot\textbf{b}_2}\times\nonumber\\
		&	&&\times\frac{2ig_s(\epsilon\cdot\textbf{k}}{x}a_2a_1c(T_{a_2}T_{a_1})(2E)^2\times\nonumber\\
		& 	&&\times\int\frac{dq_1^z}{2\pi}\frac{dq_2^z}{2\pi}
				\frac{v(\vec{\textbf{q}_1})e^{-iq_1^z(z_1-z_0)}}{\big((p+k-q_1-q_2)^2-M^2+i\epsilon\big)}\times\nonumber\\
		&	&&\times\frac{v(\vec{\textbf{q}_2})e^{-iq_2^z(z_2-z_0)}}{\big((p-q_1-q_2)^2-M^2+i\epsilon\big)\big((p-q_2)^2-M^2+i\epsilon\big)}\\
		&\approx	&&J(p)e^{i(p+k)x_0}(-i)^2\int\frac{d^2\textbf{q}_1}{(2\pi)^2}\int\frac{d^2\textbf{q}_2}{(2\pi)^2}
				e^{-i(\textbf{q}_1+\textbf{q}_1)\cdot\textbf{b}_1}\times,\;(\textbf{b}_1\approx\textbf{b}_2)\nonumber\\
		&	&&\times\frac{2ig_s}{x}(\epsilon\cdot\textbf{k})a_2a_1c(T_{a_1}T_{a_2})(2E)^2\times\nonumber\\
		&	&&\times\frac{1}{(P^+)^2}\frac{1}{(\omega_0+\tilde{\omega}_m)}\Big(e^{i(\omega_0+\tilde{\omega}_m)(z_1-z_0)}-1\Big)\times\nonumber\\
		&	&&\times\frac{(-i)}{(P^+)}v(0,\textbf{q}_1)v(0,\textbf{q}_2)\times\begin{cases}
			1,& \text{separated}\\
			\frac{1}{2}, &\text{contact}\end{cases}\\
		&=	&&J(p)e^{i(p+k)x_0}\int\frac{d^2\textbf{q}_1}{(2\pi)^2}\int\frac{d^2\textbf{q}_2}{(2\pi)^2}v(0,\textbf{q}_1)
				v(0,\textbf{q}_2)e^{-i(\textbf{q}_1+\textbf{q}_1)\cdot\textbf{b}_1}\times\\
		&	&&\times\frac{-2ig_s(\epsilon\cdot\textbf{k})}{\textbf{k}^2+m_g^2+M^2x^2}\Big(e^{i(\omega_0+\tilde{\omega}_m)(z_1-z_0)}-1\Big)\nonumber\\
		&	&&\times a_2a_1cT_{a_2}T_{a_1}\begin{cases}
			1,& \text{separated}\\
			\frac{1}{2}, &\text{contact}\end{cases}\\
\end{alignat}

Where I have used that 
\begin{align}
\frac{x(2E)^2}{x(P^+)^2}&=1 \qquad \text{and}\nonumber\\
\frac{1}{xP^+}\frac{1}{\omega_0+\tilde{\omega}_m}&=\frac{1}{\textbf{k}^2+m_g^2+M^2x^2}
\end{align}
Therefore, this diagram, $\Amp_{2,0,0}$ remains unchanged under the small separation distance limit.

\subsection{$\Amp_{2,2,0}$}\label{CalcM220}

\begin{figure}
\centering
\includegraphics[scale=0.5]{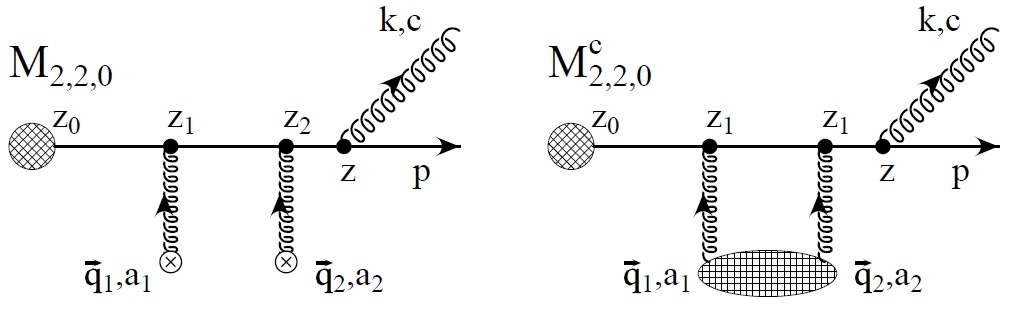}
\caption{$\Amp_{2,2,0}$ diagrams in both the well separated and contact limit case \cite{Djordjevic2004}.\label{M220Diagram}}
\end{figure}

Figure \ref{M220Diagram} shows the diagram for $\Amp_{2,2,0}$, the amplitude for which is very similar to that for $\Amp_{2,0,0}$ and differs only in the propagators (and the order of the color generator matrices in the result):

\begin{alignat}{2}
\Amp_{2,2,0}	&= &&\int	\frac{d^4q_1}{(2\pi)^4}\frac{d^4q_2}{(2\pi)^4}iJ(p+k-q_1-q_2)e^{i(p+k-q_1-q_2)x_0}\times\nonumber\\
		&	&&\times V(q_1)e^{iq_1x_1}V(q_2)e^{iq_2x_2}\times\nonumber(-i2E)^2ig_s(2p+k)_\mu\epsilon^\mu \times\nonumber\\
		&	&&\times i\vartriangle _M(p+k-q_1-q_2)i\vartriangle _M(p+k-q_2)i\vartriangle _M(p+k)a_1c(T_{a_2}T_{a_1})\nonumber\\
		&\approx	&& J(p)e^{ipx_0}(-i)\int\frac{d^2\textbf{q}_1}{(2\pi)^2}e^{-i\textbf{q}_1\cdot\textbf{b}_1}		
				(-i)\int\frac{d^2\textbf{q}_2}{(2\pi)^2}e^{-i\textbf{q}_2\cdot\textbf{b}_2}\frac{2ig_s(\epsilon\cdot\textbf{k})}{x}\times\nonumber\\
		&	&&\times ca_2a_1(T_{a_2}T_{a_1})(2E)^2\times\nonumber\\
		& 	&&\times\int\frac{dq_1^z}{2\pi}\frac{dq_2^z}{2\pi}
			\frac{v(\vec{\textbf{q}_1})e^{-iq_1^z(z_1-z_0)}}{\big((p+k-q_1-q_2)^2-M^2+i\epsilon\big)}\times\nonumber\\
		&	&&\times \frac{v(\vec{\textbf{q}_2})e^{-iq_2^z(z_2-z_0)}}{\big((p+k-q_2)^2-M^2+i\epsilon\big)\big((p+k)^2-M^2+i\epsilon\big)}
\end{alignat}

Notice that the last propagator does not depend on  $q_1$ or $q_2$ which greatly simplifies the calculations below but must be kept in mind in the final calculation of the amplitude.  We define the first integral 

\begin{align}
I_2(p,k,\textbf{q}_1,\textbf{q}_2,z_1-z_0)=\int \frac{dq_1^z}{2\pi}\frac{v(q_1^z,\textbf{q}_1)e^{-i(q_1^z+q_2^z)(z_1-z_0)}}{\big((p+k-q_1-q_2)^2-M^2+i\epsilon\big)}
\end{align}
There are two poles, one from the propagator and one from the potential (that had previously been neglected).  Using the results in section \ref{polepqk} along with an appropriate substitution in much the same way as was done to obtain equation (\ref{q3Example}), one can calculate the pole in the propagator, while section \ref{polepotential} describes the calculation of the pole at the potential.  The poles in question are then at
\begin{align}
q_1^{z(1)}	&=-q_2^z-\omega_0-\tilde{\omega}_m -i\epsilon\\
q_1^{z(2)}	&=-i\mu_1
\end{align}
The residue here is very similar to the corresponding residue in the amplitude of diagram $\mathcal{M_{2,0,0}}$ but does differ slightly:
\begin{align}
\text{Res }&\big[I_2,q_1^{z(1)}=-q_2^z-\omega_0-\tilde{\omega}_m\big]\nonumber\\
	&=\lim_{q_1^z\to q_1^{z(1)}}\Bigg[\frac{v(q_1^z,\textbf{q}_1)e^{-i(q_1^z+q_2^z)(z_1-z_0)}(q_1^z+q_2^z+\omega_0+\tilde{\omega}_m)}{\big((p+k-q_1-q_2)^2-M^2+i\epsilon\big)}\Bigg]
\end{align}

The individual contributions are calculated in sections \ref{polepotential} and \ref{factorpqk}, giving

\begin{align}
\text{Res }\big[I_2,q_1^{z(1)}=-q_2^z-\omega_0-\tilde{\omega}_m\big]	&=\frac{v(-q_2^z,\textbf{q}_1)}{P^+}e^{i(\omega_0+\tilde{\omega}_m)(z_1-z_0)}
\end{align}

The residue due to the second pole at $q_1^{z(2)}=-i\mu_1$ is given by

\begin{align}
\text{Res }\big[I_2,q_1^{z(2)}=-i\mu_1\big]	&=\lim_{q_1^z\to q_1^{z(2)}}\Bigg[\frac{v(q_1^z,\textbf{q}_1)e^{-i(q_1^z+q_2^z)(z_1-z_0)}(q_1^z+i\mu_1)}{\big((p+k-q_1-q_2)^2-M^2+i\epsilon\big)}\Bigg]
\end{align}
We can again calculate each part individually using results from the same sections, resulting in 
\begin{align}
\text{Res }\big[I_2,q_1^{z(2)}\big]	&=\frac{4\pi\alpha_s}{-2i\mu_1}\frac{e^{-i(q_2^z-i\mu_1)(z_1-z_0)}}{P^+(q_2^z-i\mu_1)}
\end{align}

Adding these two poles, the integral $I_2$ is given by (with the negative sign again because the contour is negatively orientated)
\begin{align}
I_2(p,k,&\textbf{q}_1,\textbf{q}_2,z_1-z_0)\nonumber\\
&=-\frac{2\pi i}{2\pi}\bigg[\frac{4\pi\alpha_s}{(q_2^z)^2+\mu_1^2}
			\frac{e^{i(\omega_0+\tilde{\omega}_m)(z_1-z_0)}}{P^+}
			-\frac{4\pi\alpha_s}{2i\mu_1}\frac{e^{-i(q_2^z-i\mu_1)(z_1-z_0)}}{P^+(q_2^z-i\mu_1)}\bigg]\nonumber\\
		&= \frac{-i}{P^+}\frac{4\pi\alpha_s}{(q_2^z-i\mu_1)}\bigg[\frac{e^{i(\omega_0+\tilde{\omega}_m)(z_1-z_0)}}{q_2^z+i\mu_1}
			-\frac{e^{-i(q_2^z-i\mu_1)(z_1-z_0)}}{2i\mu_1}\bigg]
\end{align}

The matrix element with the calculated $I_2$ is given by
\begin{alignat}{2}
\Amp_{2,2,0}		&\approx	&& J(p)e^{ipx_0}(-i)^2\int\frac{d^2\textbf{q}_1}{(2\pi)^2}e^{-i\textbf{q}_1\cdot\textbf{b}_1}
			\int\frac{d^2\textbf{q}_2}{(2\pi)^2}e^{-i\textbf{q}_2\cdot\textbf{b}_2}\frac{2ig_s(\epsilon\cdot\textbf{k})}{x}\times\nonumber\\
	&	&&\times ca_2a_1(T_{a_2}T_{a_1})(2E)^2\times\nonumber\\
	& 	&&\times\int\frac{dq_2^z}{2\pi}
			\frac{v(\vec{\textbf{q}_2})e^{-iq_2^z(z_2-z_1)}}{\big((p+k-q_2)^2-M^2+i\epsilon\big)\big((p+k)^2-M^2+i\epsilon\big)}\times\nonumber\\
	&	&&\times \frac{-i}{P^+}\frac{4\pi\alpha_s}{(q_2^z-i\mu_1)}\bigg[\frac{e^{i(\omega_0+\tilde{\omega}_m)(z_1-z_0)}}{q_2^z+i\mu_1}
			-\frac{e^{-i(q_2^z-i\mu_1)(z_1-z_0)}}{2i\mu_1}\bigg]
\end{alignat}

So we define the following $I_3$:
\begin{align}
I_3(p,k,&\textbf{q}_1,\textbf{q}_2,z_1-z_0)\nonumber\\
	&=\int\frac{dq_2^z}{2\pi}\frac{1}{(q_2^z-i\mu_1)}
			\bigg[\frac{e^{i(\omega_0+\tilde{\omega}_m)(z_1-z_0)}}{q_2^z+i\mu_1}-\frac{e^{-i(q_2^z-i\mu_1)(z_1-z_0)}}{2i\mu_1}\bigg]\times\nonumber\\
		&\times\frac{4\pi\alpha_s}{(q_2^z)^2+\mu_2^2}\frac{e^{-iq_2^z(z_2-z_1)}}{\big((p+k-q_2)^2-M^2+i\epsilon\big)}
\end{align}

Here the poles from the potentials can never be neglected because their contributions are not even exponentially suppressed in the large path length limit.  We therefore have three poles, two from the potentials (see section \ref{polepotential}) and one from the propagator (section \ref{polepqk}).  (The new pole has a positive sign and is therefore ignored because it is in the upper half of the complex plane.  However, there is a new term, and so all the poles have to be recalculated.  The poles are therefore

\begin{align}
q_2^{z(1)}	&=-i\mu_1\\
q_2^{z(2)}	&=-i\mu_2\\
q_2^{z(3)}	&=-\omega_0-\tilde{\omega}_m - i\epsilon
\end{align}

We calculate the residues in the same way as before - individually and by considering each contribution.  The residue due to the pole at $q_2^{z(1)}-i\mu_1$ is calculated as
\begin{align}
\text{Res }\big[I_3,&q_2^{z(1)}=-i\mu_1]	\nonumber\\
	&=\lim_{q_2^z\to q_2^{z(1)}}\bigg\{\frac{(q_2^z+i\mu_1)}{(q_2^z-i\mu_1)}
			\bigg[\frac{e^{i(\omega_0+\tilde{\omega}_m)(z_1-z_0)}}{q_2^z+i\mu_1}-\frac{e^{-i(q_2^z-i\mu_1)(z_1-z_0)}}{2i\mu_1}\bigg]\times\nonumber\\
		&\times\frac{4\pi\alpha_s}{(q_2^z)^2+\mu_2^2}\frac{e^{-iq_2^z(z_2-z_1)}}{\big((p+k-q_2)^2-M^2+i\epsilon\big)}\bigg\}
\end{align}

Each part can be evaluated individually using equation \eqref{factorvSame} and section \ref{factorpqk}, resulting in the residue
	\begin{align}
		\text{Res }\big[I_3,q_2^{z(1)}=-i\mu_1]	&\approx \frac{(4\pi\alpha_s )}{\mu_2^2-\mu_1^2}
			\frac{e^{-\mu_1(z_2-z_1)}}{P^+(-i\mu_1)}\frac{e^{i(\omega_0+\tilde{\omega}_m)(z_1-z_0)}}{-2i\mu_1}\nonumber\\
		&=-\frac{(4\pi\alpha_s )e^{-\mu_1(z_2-z_1)}e^{i(\omega_0+\tilde{\omega}_m)(z_1-z_0)}}{2P^+(\mu_2^2-\mu_1^2)\mu_1^2}
	\end{align}
This is exactly the result for the large path length calculation, the residue due to the pole at $q_2^{z(2)}=-i\mu_2$ is given by

	\begin{align}
		\text{Res }\big[I_3,&q_2^{z(2)}=-i\mu_2]	\nonumber\\
		&=\lim_{q_2^z\to q_2^{z(2)}}\bigg\{\frac{(q_2^z+i\mu_2)}{(q_2^z-i\mu_1)}
			\bigg[\frac{e^{i(\omega_0+\tilde{\omega}_m)(z_1-z_0)}}{q_2^z+i\mu_1}-\frac{e^{-i(q_2^z-i\mu_1)(z_1-z_0)}}{2i\mu_1}\bigg]\times\nonumber\\
		&\times\frac{4\pi\alpha_s}{(q_2^z)^2+\mu_2^2}\frac{e^{-iq_2^z(z_2-z_1)}}{\big((p+k-q_2)^2-M^2+i\epsilon\big)}\bigg\}
	\end{align}

Which can be calculated using the same results as above: 

	\begin{alignat}{2}
		\text{Res }&\big[I_3&&,q_2^{z(2)}=-i\mu_2]	=\frac{4\pi\alpha_s}{(-2i\mu_2)}\frac{e^{-\mu_2(z_2-z_1)}}{-i(\mu_1+\mu_2)}
			\frac{1}{P^+(-i\mu_2)}\times\nonumber\\
		&	&&\times \bigg[\frac{e^{i(\omega_0+\tilde{\omega}_m)(z_1-z_0)}}{i(\mu_1-\mu_2)}+\frac{e^{-(\mu_1+\mu_2)(z_1-z_0)}}{-2i\mu_1}\bigg]\nonumber\\
		&=	&&\frac{4\pi\alpha_s e^{-\mu_2(z_2-z_1)}}{2P^+\mu_2^2(\mu_1+\mu_2)}\bigg[\frac{e^{-(\mu_1+\mu_2)(z_1-z_0)}}{2\mu_1}-\frac{e^{i(\omega_0+\tilde{\omega}_m)(z_1-z_0)}}{(\mu_1-\mu_2)}\bigg]
	\end{alignat}

The residue due to the third pole at $q_2^{z(3)}=-\omega_0-\tilde{\omega}_m - i\epsilon$ is given by

\begin{align}
\text{Res }\big[I_3,&q_2^{z(3)}=-\omega_0-\tilde{\omega}_m]	\nonumber\\
	&=\lim_{q_2^z\to q_2^{z(3)}}\bigg\{\frac{(q_2^z+\omega_0+\tilde{\omega}_m)}{(q_2^z-i\mu_1)}
			\bigg[\frac{e^{i(\omega_0+\tilde{\omega}_m)(z_1-z_0)}}{q_2^z+i\mu_1}-\frac{e^{-i(q_2^z-i\mu_1)(z_1-z_0)}}{2i\mu_1}\bigg]\times\nonumber\\
		&\times\frac{4\pi\alpha_s}{(q_2^z)^2+\mu_2^2}\frac{e^{-iq_2^z(z_2-z_1)}}{\big((p+k-q_2)^2-M^2+i\epsilon\big)}\bigg\},
\end{align}

with the following contributions from sections \ref{factorvMixed} and \ref{factorpqk} to give
	\begin{alignat}{2}
		\text{Res }\big[I_3,q_2^{z(3)}\big]&=&&\frac{4\pi\alpha_s}{\mu_2^2}\frac{1}{P^+}\frac{e^{i(\omega_0+\tilde{\omega}_m)(z_2-z_1)}}{-i\mu_1}
			\bigg[\frac{e^{i(\omega_0+\tilde{\omega}_m)(z_1-z_0)}}{i\mu_1}-\frac{e^{-\mu_1(z_1-z_0)}}{2i\mu_1}\bigg]\nonumber\\
		&=	&&\frac{4\pi\alpha_se^{i(\omega_0+\tilde{\omega}_m)(z_2-z_1)}}{P^+\mu_1^2\mu_2^2}
			\bigg(e^{i(\omega_0+\tilde{\omega}_m)(z_1-z_0)}-\frac{1}{2}e^{-\mu_1(z_1-z_0)}\bigg)
	\end{alignat}

We combine the residues (keeping in mind that the contour is negatively orientated) to obtain the result for the second integral $I_3$.

\begin{alignat}{2}
I_3(p,k,&\textbf{q}_1,&&\textbf{q}_2,z_1-z_0)=-\frac{2\pi i}{2\pi}
			\Bigg\{-\frac{(4\pi\alpha_s )e^{-\mu_1(z_2-z_1)}e^{i(\omega_0+\tilde{\omega}_m)(z_1-z_0)}}{2P^+(\mu_2^2-\mu_1^2)\mu_1^2}\nonumber\\	
	&	&&+ \frac{4\pi\alpha_s e^{-\mu_2(z_2-z_1)}}{2P^+\mu_2^2(\mu_1+\mu_2)}\bigg[\frac{e^{-(\mu_1+\mu_2)(z_1-z_0)}}{2\mu_1}-\frac{e^{i(\omega_0+\tilde{\omega}_m)(z_1-z_0)}}{(\mu_1-\mu_2)}\bigg]\nonumber\\
	&	&&+	\frac{4\pi\alpha_se^{i(\omega_0+\tilde{\omega}_m)(z_2-z_1)}}{P^+\mu_1^2\mu_2^2}
			\bigg(e^{i(\omega_0+\tilde{\omega}_m)(z_1-z_0)}-\frac{1}{2}e^{-\mu_1(z_1-z_0)}\bigg)\Bigg\}\nonumber\\
	&=	&& \frac{(-i)4\pi\alpha_s}{P^+}\Bigg\{-\frac{e^{-\mu_1(z_2-z_1)}e^{i(\omega_0+\tilde{\omega}_m)(z_1-z_0)}}{2(\mu_2^2-\mu_1^2)\mu_1^2}\nonumber\\
	&	&&+ \frac{e^{-\mu_2(z_2-z_1)}}{2\mu_2^2(\mu_1+\mu_2)}
			\bigg(\frac{e^{-(\mu_1+\mu_2)(z_1-z_0)}}{2\mu_1}-\frac{e^{i(\omega_0+\tilde{\omega}_m)(z_1-z_0)}}{(\mu_2-\mu_1)}\bigg)\nonumber\\
	&	&&+ \frac{e^{i(\omega_0+\tilde{\omega}_m)(z_2-z_1)}}{\mu_1^2\mu_2^2}
			\bigg(e^{i(\omega_0+\tilde{\omega}_m)(z_1-z_0)}-\frac{1}{2}e^{-\mu_1(z_1-z_0)}\bigg)\Bigg\}
\end{alignat}

Before we can really look at the contact and large separation cases, it might be pertinent to simplify the curly brackets.  So, let us define:
\begin{alignat}{2}
\{\otimes\}&\equiv&&\Bigg\{-\frac{e^{-\mu_1(z_2-z_1)}e^{i(\omega_0+\tilde{\omega}_m)(z_1-z_0)}}{2(\mu_2^2-\mu_1^2)\mu_1^2}\nonumber\\
	&	&&+ \frac{e^{-\mu_2(z_2-z_1)}}{2\mu_2^2(\mu_1+\mu_2)}
			\bigg(\frac{e^{-(\mu_1+\mu_2)(z_1-z_0)}}{2\mu_1}-\frac{e^{i(\omega_0+\tilde{\omega}_m)(z_1-z_0)}}{(\mu_1-\mu_2)}\bigg)\nonumber\\
	&	&&+ \frac{e^{i(\omega_0+\tilde{\omega}_m)(z_2-z_1)}}{\mu_1^2\mu_2^2}
			\bigg(e^{i(\omega_0+\tilde{\omega}_m)(z_1-z_0)}-\frac{1}{2}e^{-\mu_1(z_1-z_0)}\bigg)\Bigg\}
\end{alignat}

Which would give the matrix element
\begin{alignat}{2}
\Amp_{2,2,0}		&\approx	&& J(p)e^{ipx_0}(-i)\int\frac{d^2\textbf{q}_1}{(2\pi)^2}e^{-i\textbf{q}_1\cdot\textbf{b}_1}
			(-i)\int\frac{d^2\textbf{q}_2}{(2\pi)^2}e^{-i\textbf{q}_2\cdot\textbf{b}_2}
			\frac{2ig_s(\epsilon\cdot\textbf{k})}{x}\times\nonumber\\
	&	&&\times ca_2a_1(T_{a_2}T_{a_1})(2E)^2 \frac{-i}{P^+}\frac{4\pi\alpha_s}{\big((p+k-q_2)^2-M^2+i\epsilon\big)}
			\frac{(-i)4\pi\alpha_s}{P^+}\Bigg\{\otimes\Bigg\}\nonumber\\
	&=	&& J(p)e^{ipx_0}(-i)\int\frac{d^2\textbf{q}_1}{(2\pi)^2}e^{-i\textbf{q}_1\cdot\textbf{b}_1}
			(-i)\int\frac{d^2\textbf{q}_2}{(2\pi)^2}e^{-i\textbf{q}_2\cdot\textbf{b}_2}
			\frac{2ig_s(\epsilon\cdot\textbf{k})}{x}\times\nonumber\\
	&	&& \times ca_2a_1(T_{a_2}T_{a_1})(2E)^2\frac{(4\pi\alpha_s)^2}{(P^+)^2}\frac{x}{m_g^2+\textbf{k}^2+x^2M^2}	(-1)\Bigg\{\otimes\Bigg\}
\end{alignat}

To simplify this, let's look at $\{\otimes\}$:
\begin{alignat}{2}
(-1)\{\otimes\}	&=&& \Bigg\{\frac{e^{-\mu_1(z_2-z_1)}e^{i(\omega_0+\tilde{\omega}_m)(z_1-z_0)}}{2(\mu_2^2-\mu_1^2)\mu_1^2}\nonumber\\
	&	&&- \frac{-e^{-\mu_2(z_2-z_1)}}{2\mu_2^2(\mu_1+\mu_2)}
			\bigg[\frac{e^{i(\omega_0+\tilde{\omega}_m)(z_1-z_0)}}{(\mu_1-\mu_2)}+\frac{e^{-(\mu_1+\mu_2)(z_1-z_0)}}{2\mu_1}\bigg]\nonumber\\
	&	&&- \frac{e^{i(\omega_0+\tilde{\omega}_m)(z_2-z_1)}}{\mu_1^2\mu_2^2}
			\bigg(e^{i(\omega_0+\tilde{\omega}_m)(z_1-z_0)}-\frac{1}{2}e^{-\mu_1(z_1-z_0)}\bigg)\Bigg\}\nonumber\\
	&=	&& \frac{e^{-\mu_1(z_2-z_1)}e^{i(\omega_0+\tilde{\omega}_m)(z_1-z_0)}}{2(\mu_2^2-\mu_1^2)\mu_1^2}\nonumber\\
	&	&&+\frac{e^{-\mu_2(z_2-z_1)}e^{i(\omega_0+\tilde{\omega}_m)(z_1-z_0)}}{2\mu_2^2(\mu_1+\mu_2)(\mu_1-\mu_2)}
			-\frac{e^{-\mu_2(z_2-z_1)}e^{-(\mu_1+\mu_2)(z_1-z_0)}}{2\mu_2^2(\mu_1+\mu_2)2\mu_1}\nonumber\\
	&	&&-\frac{e^{i(\omega_0+\tilde{\omega}_m)(z_2-z_1)}e^{i(\omega_0+\tilde{\omega}_m)(z_1-z_0)}}{\mu_1^2\mu_2^2}
			+\frac{1}{2}\frac{e^{i(\omega_0+\tilde{\omega}_m)(z_2-z_1)}e^{-\mu_1(z_1-z_0)}}{\mu_1^2\mu_2^2}
\end{alignat}

Here there are two terms that appear in the small separation limit that did not contribute to the large separation distance calculation.  The last term on the second line and the last term on the third line are both additional contributions that are exponentially suppressed in the large path length limit.  From this we also see that this expression still reduces to the large path length result.  One interesting thing that arose in the large path length limit was a factor of $1/2$ that appears in the contact limit.  We can check that this factor still appears if we take this expression for the short path length and look at the contact limit as well as the large path length limit.  The two `new' terms will be zero (large path length) while the exponents containing $(z_2-z_1)\rightarrow(z_1-z_1)=0$ become $1$ and the exponents containing $(\omega_0+\tilde{\omega}_m)(z_1-z_0)$ are factorized out, so that we have that
	\begin{alignat}{2}
		(-1)\{\otimes\}	^{\text{contact}}&\approx && e^{i(\omega_0+\tilde{\omega}_m)(z_1-z_0)}
					\bigg[\frac{1}{2\mu_1^2(\mu_2^2-\mu_1^2)}+\frac{1}{2\mu_2^2(\mu_1^2-\mu_2^2)}-\frac{1}{\mu_1^2\mu_2^2}\bigg]\nonumber\\
			&	&&+\frac{e^{-(\mu_1+\mu_2)(z_1-z_0)}}{4\mu_1\mu_2^2(\mu_1+\mu_2)}+\frac{e^{-\mu_1(z_1-z_0)}}{\mu_1^2\mu_1^2}
	\end{alignat}
Now, the square brackets in the first line are exactly the terms that appear in the large path length limit and they still produce the factor of $1/2$:
	\begin{align}
		&\frac{1}{2\mu_1^2(\mu_2^2-\mu_1^2)}+\frac{1}{2\mu_2^2(\mu_1^2-\mu_2^2)}-\frac{1}{\mu_1^2\mu_2^2}\nonumber\\
		&=\frac{1}{2(\mu_2^2-\mu_1^2)}\bigg(\frac{1}{\mu_1^2}-\frac{1}{\mu_2^2}\bigg)-\frac{1}{\mu_1^2\mu_2^2}\nonumber\\
		&=\frac{\cancel{(\mu_2^2-\mu_1^2)}}{2\cancel{(\mu_2^2-\mu_1^2)}\mu_1^2\mu_2^2}-\frac{1}{\mu_1^2\mu_2^2}\nonumber\\
		&=-\frac{1}{2}\frac{1}{\mu_1^2\mu_2^2}
	\end{align}
And so, although we are not surprised, we are relieved.  Suppose we now only consider the expression for the short path length in the contact limit, then we have that
	\begin{alignat}{2}
		(-1)\{\otimes\}^{\text{contact}}	&=&&-\frac{1}{2}\frac{1}{\mu_1^2\mu_2^2}e^{i(\omega_0+\tilde{\omega}_m)(z_1-z_0)}\nonumber\\
		&	&&+\frac{1}{2}\frac{e^{-\mu_1(z_1-z_0)}}{\mu_1^2\mu_2^2}-\frac{e^{-(\mu_1+\mu_2)(z_1-z_0)}}{2\mu_2^2(\mu_1+\mu_2)2\mu_1}
	\end{alignat}

Now, we would like for these new terms to mean something, so notice that we have
	\begin{align*}
		\frac{1}{4\mu_2^2\mu_1(\mu_1+\mu_2)}=\frac{1}{\mu_1^2\mu_2^2}\frac{\mu_1}{4(\mu_1+\mu_2)}
	\end{align*}
We also have that
	\begin{align*}
		e^{-(\mu_1+\mu_2)(z_1-z_0)}=e^{-\mu_1(z_1-z_0)}e^{-\mu_2(z_1-z_0)}
	\end{align*}
And so, the two small separation distance terms become:
	\begin{align*}
		\frac{1}{2}\frac{e^{-\mu_1(z_1-z_0)}}{\mu_1^2\mu_2^2}-\frac{e^{-(\mu_1+\mu_2)(z_1-z_0)}}{2\mu_2^2(\mu_1+\mu_2)2\mu_1}
		=\frac{1}{2}\frac{e^{-\mu_1(z_1-z_0)}}{\mu_1^2\mu_2^2}\bigg(1-\frac{\mu_1e^{-\mu_2(z_1-z_0)}}{2(\mu_1+\mu_2)}\bigg)
	\end{align*}

Which finally gives the matrix amplitude for $\Amp_{2,2,0}$:
\begin{alignat}{2}\label{ResM220}
\Amp_{2,2,0}		&\approx	&& J(p)e^{i(p+k)x_0}(-i)^2\int\frac{d^2\textbf{q}_1}{(2\pi)^2}\int\frac{d^2\textbf{q}_2}{(2\pi)^2}
					e^{-i(\textbf{q}_1\textbf{q}_2)\cdot\textbf{b}_1}\frac{2ig_s(\epsilon\cdot\textbf{k})}{x}\times\nonumber\\
				&	&&\times ca_2a_1(T_{a_2}T_{a_1})(2E)^2\times\nonumber\\
				&	&&\times\frac{(-1)(4\pi\alpha_s)^2}{(P^+)^2}\frac{x}{m_g^2+\textbf{k}^2+x^2M^2}	\times\nonumber\\
				&	&&\times\bigg[\frac{e^{i(\omega_0+\tilde{\omega}_m)(z_1-z_0)}}{\mu_1^2\mu_2^2}-\frac{e^{-\mu_1(z_1-z_0)}}{\mu_1^2\mu_2^2}\bigg(1-\frac{\mu_1e^{-\mu_2(z_1-z_0)}}{2(\mu_1+\mu_2)}\bigg)\bigg]\times\nonumber\\
				&	&&\times
						\begin{cases}
						1,& \text{separated}\\
						\frac{1}{2}, &\text{contact}
						\end{cases}\nonumber\\		
\end{alignat}

It is worth the effort to rewrite this expression to more closely match the form of the other two scattering centre diagrams since it will greatly minimize the work necessary when squaring and summing in the final energy loss calculation:
\begin{alignat}{2}
\Amp_{2,2,0}		&\approx	&& J(p)e^{i(p+k)x_0}\int\frac{d^2\textbf{q}_1}{(2\pi)^2}\int\frac{d^2\textbf{q}_2}{(2\pi)^2}
				e^{-i(\textbf{q}_1\textbf{q}_2)\cdot\textbf{b}_1}(-2ig_s)
				v(0,\textbf{q}_1)v(0,\textbf{q}_1)\times\nonumber\\
		&	&&\times ca_2a_1(T_{a_2}T_{a_1})\frac{(-1)(\epsilon\cdot\textbf{k})}{m_g^2+\textbf{k}^2+x^2M^2}\times\nonumber\\
		&	&&\times \bigg[e^{i(\omega_0+\tilde{\omega}_m)(z_1-z_0)}-e^{-\mu_1(z_1-z_0)}
				\bigg(1-\frac{\mu_1e^{-\mu_2(z_1-z_0)}}{2(\mu_1+\mu_2)}\bigg)\bigg]\times\nonumber\\
		&	&&\times\begin{cases}
						1,& \text{separated}\\
						\frac{1}{2}, &\text{contact}
						\end{cases}\nonumber\\	
\end{alignat}

\section{$\Amp_{2,0,1}$ and $\Amp_{2,0,1}$}

\subsection{$\Amp_{2,0,1}$}\label{CalcM201M202}

\begin{figure}
\centering
\includegraphics[scale=0.3]{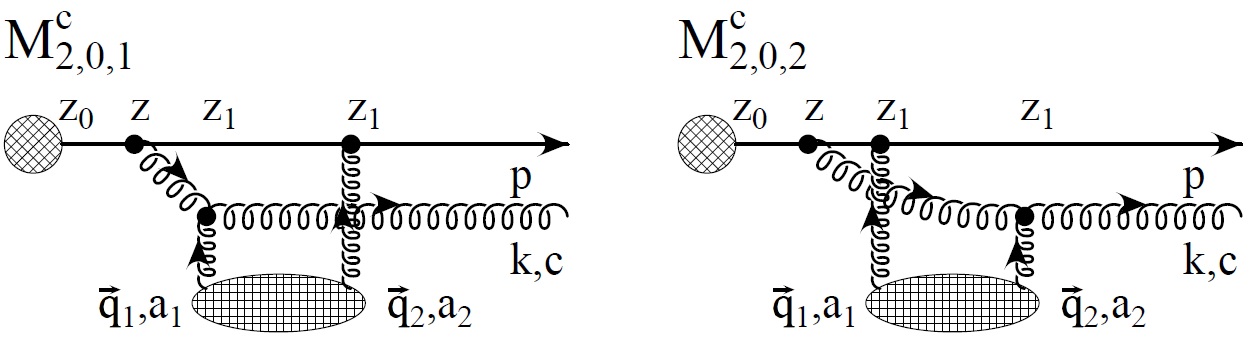}
\caption{Diagrams $\Amp_{2,0,1}$ and $\Amp_{2,0,2}$ showing only the special contact case \cite{Djordjevic2004}.\label{M201M202diagram}}
\end{figure}

Figure \ref{M201M202diagram} shows the diagram for the contact case for $\Amp_{2,0,1}$.  The matrix element is
\begin{alignat}{2}
\Amp_{2,0,1}	&=&&\int\frac{d^q_1}{(2\pi)^4}\frac{d^q_2}{(2\pi)^4}iJ(p+k-q_1-q_2)e^{i(p+k-q_1-q_2)x_0}V(q_1)e^{iq_1x_1}V(q_2)e^{iq_1x_2}\times\nonumber\\	
		&	&&\times(-iE^+)\Lambda_1i\vartriangle_M((p+k-q_1-q_2)(-i)\vartriangle_{m_g}(k-q_1)i\vartriangle_M(p-q_2)\nonumber\\
		&\approx &&J(p)e^{i(p+k)x_0}(-i)^2\int\frac{d^2\textbf{q}_1}{(2\pi)^2}e^{-i\textbf{q}_1\cdot\textbf{b}_1}\int\frac{d^2\textbf{q}_2}{(2\pi)^2}e^{-i\textbf{q}_2\cdot\textbf{b}_2}\times\nonumber\\
		&	&&\times 2ig_s(\epsilon\cdot(\textbf{k}-\textbf{q}_1)e^{i\omega_0 z_0}a_2[c,a_1](T_{a_2}T_{a_1})(2E)^2\times\nonumber\\
		&	&&\times\int	\frac{dq_1^z}{2\pi}\frac{dq_2^z}{2\pi}\frac{v(\vec{\textbf{q}_1})e^{-iq_1^z(z_1-z_0)}}
			{\big((p+k-q_1-q_2)^2-M^2+i\epsilon\big)}\times\nonumber\\
		&	&&\times\frac{v(\vec{\textbf{q}_2})e^{-iq_2^z(z_2-z_0)}}{\big((k-q_1)^2-m_g^2+i\epsilon\big)\big((p-q_2)^2-M^2+i\epsilon\big)}
\end{alignat}

We define the $q_1^z$ integral (keeping with the numbering in \cite{Djordjevic2004} as
\begin{align}
I_2(p,k,&\textbf{q}_1,\textbf{q}_2,z_1-z_0)\nonumber\\
	&=\int	\frac{dq_1^z}{2\pi}\frac{v(q_1^z,\textbf{q}_1)e^{-i(q_1^z+q_2^z)(z_1-z_0)}}{\big((p+k-q_1-q_2)^2-M^2+i\epsilon\big)\big((k-q_1)^2-m_g^2+i\epsilon\big)}
\end{align}

The three poles are calculated in sections \ref{polepotential}, \ref{polepqk} and \ref{polekq} (along with an appropriate substitution as was done to obtain equation (\ref{q3Example})). These three poles are then at
	\begin{align}
		q_1^{z(1)}	&=-q_2^z-\omega_0-\tilde{\omega}_m-i\epsilon\\
		q_1^{z(2)}	&=-\omega_0+\omega_1-i\epsilon\\
		q_1^{z(3)}	&=-i\mu_1
	\end{align}

The residue due to the first pole at $q_1^{z(1)}=-q_2^z-\omega_0-\tilde{\omega}_m-i\epsilon$ is calculated as follows
\begin{align}
\text{Res }\big[I_2, &q_1^{z(1)}=-q_2^z-\omega_0-\tilde{\omega}_m-i\epsilon]	\nonumber\\
	&=\lim_{q_1^z\to q_1^{z(1)}}\Bigg[\frac{(q_1^z+q_2^z+\omega_0+\tilde{\omega}_m)v(q_1^z,\textbf{q}_1)e^{-i(q_1^z+q_2^z)(z_1-z_0)}}{\big((p+k-q_1-q_2)^2-M^2+i\epsilon\big)\big((k-q_1)^2-m_g^2+i\epsilon\big)}\Bigg],
\end{align}
the contributions of which are calculated in sections \ref{factorvSame}, \ref{factorpqk} and \ref{factorkq} to give

\begin{align}
\text{Res }\big[I_2, q_1^{z(1)}]	&=\frac {v(-q_2^z-\omega_0-\tilde{\omega}_m,\textbf{q}_1)}{x(P^+)^2(q_2^z+\omega_1-\omega_0)}e^{-i(\omega_0+\tilde{\omega}_m)(z_1-z_0)}
\end{align}

The residue due to the pole at $q_1^{z(2)}=-\omega_0+\omega_1-i\epsilon$ is calculated with
\begin{align}
\text{Res }\big[I_2, &q_1^{z(2)}=-\omega_0+\omega_1]	\nonumber\\
	&=\lim_{q_1^z\to q_1^{z(1)}}\Bigg[\frac{(q_1^z+\omega_0-\omega_1)v(-\omega_0+\omega_1,\textbf{q}_1)e^{-i(q_1^z+q_2^z)(z_1-z_0)}}{\big((p+k-q_1-q_2)^2-M^2+i\epsilon\big)\big((k-q_1)^2-m_g^2+i\epsilon\big)}\Bigg]
\end{align}

Again we can most easily calculate this by looking at the individual contributions calculated in the above mentioned sections to give
\begin{align}
\text{Res }\big[I_2, q_1^{z(2)}=-\omega_0+\omega_1]	&\approx \frac{1}{xP^+}\frac{v(-\omega_0+\omega_1,\textbf{q}_1)}{P^+(\omega_1+\tilde{\omega}_m+q_2^z)}e^{i(q_2^z+\omega_o+\omega_1)(z_1-z_0)}
\end{align}

The residue due to the third pole 
	\begin{align}
		\text{Res }\big[I_2, &q_1^{z(3)}=-i\mu_1]\nonumber\\
			&=\lim_{q_1^z\to q_1^{z(3)}}
		\Bigg[\frac{(q_1^z+i\mu_1)v(q_1^z,\textbf{q}_1)e^{-i(q_1^z+q_2^z)(z_1-z_0)}}{\big((p+k-q_1-q_2)^2-M^2+i\epsilon\big)
		\big((k-q_1)^2-m_g^2+i\epsilon\big)}\Bigg]\nonumber\\
	&\approx\frac{4\pi\alpha_s}{-2i\mu_1}\frac{e^{-i(q_2^z-i\mu_1)(z_1-z_0)}}{x(P^+)^2(-i\mu_1)^2}=\frac{4\pi\alpha_s}{2i\mu_1^3}\frac{e^{-i(q_2^z-i\mu_1)(z_1-z_0)}}{x(P^+)^2}
\end{align} 

We therefore have that the integral $I_2$ become
\begin{alignat}{2}
I_2(p,k,\textbf{q}_1,\textbf{q}_2,z_1-z_0)	&\approx&&\frac{-2\pi i}{2\pi}\Bigg[\frac {v(-q_2^z-\omega_0-\tilde{\omega}_m,\textbf{q}_1)}
			{x(P^+)^2(q_2^z+\omega_1-\omega_0)}e^{-i(\omega_0+\tilde{\omega}_m)(z_1-z_0)}\nonumber\\
	&	&&-\frac{1}{xP^+}\frac{v(-\omega_0+\omega_1,\textbf{q}_1)}{P^+(\omega_1+\tilde{\omega}_m+q_2^z)}e^{i(q_2^z+\omega_o+\omega_1)(z_1-z_0)}\nonumber\\
	&	&&	+\frac{4\pi\alpha_s}{2i\mu_1^3}\frac{e^{-i(q_2^z-i\mu_1)(z_1-z_0)}}{x(P^+)^2}\Bigg]\nonumber\\
	&=	&& \frac{(-i)}{x(P^+)^2}\Bigg[\frac {v(-q_2^z-\omega_0-\tilde{\omega}_m,\textbf{q}_1)}{(q_2^z+\omega_1-\omega_0)}
			e^{-i(\omega_0+\tilde{\omega}_m)(z_1-z_0)}\nonumber\\
	&	&&-\frac{v(-\omega_0+\omega_1,\textbf{q}_1)}{(\omega_1+\tilde{\omega}_m+q_2^z)}e^{i(q_2^z+\omega_o+\omega_1)(z_1-z_0)}\nonumber\\
	&	&&+\frac{4\pi\alpha_s}{2i\mu_1^3}e^{-i(q_2^z-i\mu_1)(z_1-z_0)}\Bigg]
\end{alignat}

We also see here the major cancellation effect here, suppressing the small separation distance correction term in relation to the large path length terms.
	\begin{alignat}{2}
		I_2(p,k,\textbf{q}_1,\textbf{q}_2,z_1-z_0)	&\approx	&& \frac{(-i)}{x(P^+)^2}\Bigg[\frac {v(-q_2^z-\omega_0-\tilde{\omega}_m,\textbf{q}_1)}{(q_2^z+\omega_1-\omega_0)}
			e^{i(\omega_0+\tilde{\omega}_m)(z_1-z_0)}\nonumber\\
	&	&&-\frac{v(-\omega_0+\omega_1,\textbf{q}_1)}{(\omega_1+\tilde{\omega}_m+q_2^z)}e^{i(q_2^z+\omega_o+\omega_1)(z_1-z_0)}\Bigg]
	\end{alignat}

We now turn our attention to the $q_2^z$ integral which we define as
\begin{alignat}{2}
I_3(p,&k&&,\textbf{q}_1,\textbf{q}_2,z_1-z_0)	=\int\frac{dq_2^z}{2\pi}\frac{1}{\omega_1+\tilde{\omega}_m+q_2^z-i\epsilon}\frac{v(q_2^z,\textbf{q}_2)}{\big((p-q_2)^2-M^2+i\epsilon\big)}\nonumber\\
		&	&&\times\bigg[e^{-i(q_2^z(z_2-z_1)-(\omega_0+\tilde{\omega}_m)(z_1-z_0)}v(-q_2^z-\omega_0-\tilde{\omega}_m,\textbf{q}_1)\nonumber\\
		&	&&-e^{-i(q_2^z(z_2-z_0)-(\omega_1+\omega_0)(z_1-z_0)}v(-\omega_0+\omega_1,\textbf{q}_1)\bigg]\nonumber\\
		&=	&&\int\frac{dq_2^z}{2\pi}\frac{v(q_2^z,\textbf{q}_2)}{\omega_1+\tilde{\omega}_m+q_2^z-i\epsilon}\frac{e^{-iq_2^z(z_2-z_1)}}{\big((p-q_2)^2-M^2+i\epsilon\big)}\nonumber\\
		&	&&\times\bigg[e^{i(\omega_0+\tilde{\omega}_m)(z_1-z_0)}v(-q_2^z-\omega_0-\tilde{\omega}_m,\textbf{q}_1)\nonumber\\
		&	&&-e^{-i(q_2^z-\omega_1+\omega_0)(z_1-z_0)}v(-\omega_0+\omega_1,\textbf{q}_1)\bigg]
\end{alignat}

Notice here the difference between the first term in the exponential in  the second line and that of the first line.  It is important to note that one reads $(z_2-z_1)$ while the other is $(z_1-z_0)$.  Take care also to note that in the second exponential, the term $(z_2-z_0)$ appears which causes an exponential suppression in the large separation distance limit.  There are still only three poles here; two from potentials (note that the very last potential does not depend on $q_2^z$ and therefore does not contribute a pole), and one from the propagator. The singularity that arises in the first fraction is in the upper half of the complex plane and is therefore not considered. These poles are calculated in sections \ref{polepq} and \ref{polepotential} to be
\begin{align}
q_2^{z(1)}	&=-i\epsilon\\
q_2^{z(2)}	&=-i\mu_2\\
q_2^{z(3)}	&=-\omega_0-\tilde{\omega}_m-i\mu_1\approx-i\mu_1
\end{align}

The residue due to the first pole at $q_2^{z(1)}=-i\epsilon$
\begin{alignat}{2}
\text{Res }\big[I_3, q_2^{z(1)}=-i\epsilon]	&=\lim_{q_2^z\to q_2^{z(1)}}&&\Bigg[\frac{v(q_2^z,\textbf{q}_2)(q_2^z+i\epsilon)}{\omega_1+\tilde{\omega}_m+q_2^z-i\epsilon}\frac{e^{-iq_2^z(z_2-z_1)}}{\big((p-q_2)^2-M^2+i\epsilon\big)}\nonumber\\
		&	&&\times\bigg[e^{i(\omega_0+\tilde{\omega}_m)(z_1-z_0)}v(-q_2^z-\omega_0-\tilde{\omega}_m,\textbf{q}_1)\nonumber\\
		&	&&-e^{-i(q_2^z-\omega_1+\omega_0)(z_1-z_0)}v(-\omega_0+\omega_1,\textbf{q}_1)\bigg]\Bigg]
\end{alignat}

The individual contributions of which are detailed in sections \ref{factorvSame} and \ref{factorpq}, giving that the first residue for this integral evaluates to
\begin{align}
\text{Res }\big[I_3,& q_2^{z(1)}=-i\epsilon]	\nonumber\\
	&=\frac{1}{(\omega_1+\tilde{\omega}_m)}\frac{(4\pi\alpha_s)^2}{\mu_2^2\mu_1^2}\frac{1}{P^+}
			\big(e^{i(\omega_0+\tilde{\omega}_m)(z_1-z_0)}-e^{i(\omega_1+\omega_0)(z_1-z_0)}\big)
\end{align}

The residue due to the second pole (at $q_2^{z(2)}=-i\mu_2$) is given by 
\begin{alignat}{2}
\text{Res }\big[I_3, &q_2^{z(2)}&&=-i\mu_2]	=\lim_{q_2^z\to q_2^{z(1)}}
			\Bigg[\frac{v(q_2^z,\textbf{q}_2)(q_2^z+i\mu_2)}{\omega_1+\tilde{\omega}_m+q_2^z-i\epsilon}
			\frac{e^{-iq_2^z(z_2-z_1)}}{\big((p-q_2)^2-M^2+i\epsilon\big)}\nonumber\\
	&	&&\times\bigg[e^{i(\omega_0+\tilde{\omega}_m)(z_1-z_0)}v(-q_2^z-\omega_0-\tilde{\omega}_m,\textbf{q}_1)\nonumber\\
	&	&&-e^{-i(q_2^z-\omega_1+\omega_0)(z_1-z_0)}v(-\omega_0+\omega_1,\textbf{q}_1)\bigg]\Bigg]\nonumber\\
	&\approx &&-\frac{(4\pi\alpha_s)^2}{\mu_2^3P^+}\frac{e^{-\mu_2(z_2-z_1)}}{\mu_1^2-\mu_2^2}
				\big(e^{i(\omega_0+\tilde{\omega}_m)(z_1-z_0)}-e^{-\mu_2(z_1-z_0)}\big)\nonumber\\
\end{alignat}

Where we have used the fact that $\omega_0\ll\mu_2$.

The last residue is due to a pole at $ q_2^{z(3)}=-\omega_0-\tilde{\omega}_m-i\mu_1$:
\begin{alignat}{2}
\text{Res }\big[&I_3,&& q_2^{z(3)}=-\omega_0-\tilde{\omega}_m-i\mu_1]	\nonumber\\
		&=&&\lim_{q_2^z\to q_2^{z(3)}}\Bigg[\frac{v(q_2^z,\textbf{q}_2)(q_2^z+i\mu_1)}{\omega_1+\tilde{\omega}_m+q_2^z-i\epsilon}
			\frac{e^{-iq_2^z(z_2-z_1)}}{\big((p-q_2)^2-M^2+i\epsilon\big)}\nonumber\\
		&	&&\times\bigg[e^{i(\omega_0+\tilde{\omega}_m)(z_1-z_0)}v(-q_2^z-\omega_0-\tilde{\omega}_m,\textbf{q}_1)\nonumber\\
		&	&&-e^{-i(q_2^z-\omega_1+\omega_0)(z_1-z_0)}v(-\omega_0+\omega_1,\textbf{q}_1)\bigg]\Bigg]\nonumber\\
		&=&&\frac{4\pi\alpha_s}{\mu_2^2-\mu_1^2}\frac{e^{-\mu_1(z_2-z_1)}}{P^+(-i\mu_1)}
			\bigg[\frac{4\pi\alpha_s}{-2i\mu_1}e^{i(\omega_0+\tilde{\omega}_m)(z_1-z_0)}-\frac{4\pi\alpha_s}{-2i\mu_1}e^{-\mu_1(z_1-z_0)}\bigg]\nonumber\\
		&	&&\approx\frac{(4\pi\alpha_s)^2}{\mu_1^2-\mu_2^2}\frac{e^{-\mu_1(z_2-z_1)}}{P^+2\mu_1^3}
			\bigg[e^{i(\omega_0+\tilde{\omega}_m)(z_1-z_0)}-e^{-\mu_1(z_1-z_0)}\bigg]
\end{alignat}

We therefore have that the integral $I_3$ is approximately given by
\begin{alignat}{2}
I_3(p,&k,&&\textbf{q}_1,\textbf{q}_2,z_1-z_0)	\nonumber\\
	&\approx	&&\frac{-2\pi i}{2\pi}\Bigg[\frac{1}{(\omega_1+\tilde{\omega}_m)}\frac{(4\pi\alpha_s)^2}{\mu_2^2\mu_1^2}\frac{1}{P^+}
			\big(e^{i(\omega_0+\tilde{\omega}_m)(z_1-z_0)}-e^{i(\omega_1+\omega_0)(z_1-z_0)}\big)\nonumber\\
	&	&&-\frac{(4\pi\alpha_s)^2}{\mu_2^3P^+}\frac{e^{-\mu_2(z_2-z_1)}}{\mu_1^2-\mu_2^2}\big(e^{i(\omega_0+\tilde{\omega}_m)(z_1-z_0)}-e^{-\mu_2(z_1-z_0)}\big) 	\nonumber\\
	&	&&+\frac{(4\pi\alpha_s)^2}{\mu_1^2-\mu_2^2}\frac{e^{-\mu_1(z_2-z_1)}}{P^+2\mu_1^3}
			\big(e^{i(\omega_0+\tilde{\omega}_m)(z_1-z_0)}-e^{-\mu_1(z_1-z_0)}\big)\Bigg]	
\end{alignat}

This integral reduces to the short path length case and the dimensions are consistent.  Now, in the well separated case, the last two lines are no longer exponentially suppressed.  However, in the contact limit, they are equal but opposite in sign and therefore also cancel each other.  Therefore, for the contact case in which we are interested, they do not contribute and we may leave them out of the remainder of the calculation.  It is important to notice that there is no factor of $1/2$. The well separated case is not relevant for the short path length limit.

We therefore have the Matrix element, unchanged in the contact limit
\begin{alignat}{2}\label{ResM201}
\Amp_{2,0,1}^c	&=&&J(p)e^{i(p+k)x_0}(-i)^2\int	\frac{d^2\textbf{q}_1}{(2\pi)^2}e^{-i\textbf{q}_1\cdot\textbf{b}_1}
			v(0,\textbf{q}_1)v(0,\textbf{q}_2)\times\nonumber\\
	&	&&\times\int	\frac{d^2\textbf{q}_2}{(2\pi)^2}e^{-i\textbf{q}_2\cdot\textbf{b}_2}2ig_s
			\frac{\epsilon\cdot(\textbf{k}-\textbf{q}_1)}{(\textbf{k}-\textbf{q}_1)^2+M^2x^2+m_g^2}\times\nonumber\\
	&	&&\times \bigg[e^{i(\omega_0+\tilde{\omega}_m)(z_1-z_0)}-e^{i(\omega_0-\omega_1)(z_1-z_0)}\bigg]a_2[c,a_1](T_{a_1}T_{a_2})
\end{alignat}

\subsection{$\Amp_{2,0,2}$}
The result for this diagram is calculated in precisely the manner of $\Amp_{2,0,1}$ with a simple change of variable name that involves interchanging all $1$'s and $2's$.  Since scattering centres are identical then, we must symmetrise these two diagrams which is done by effectively multiplying each diagram by $1/2$.  This result still holds in the short path length limit since the change of variables applies across the board.

\begin{alignat}{2}\label{ResM202}
\Amp_{2,0,2}^c	&=&&J(p)e^{i(p+k)x_0}(-i)^2\int	\frac{d^2\textbf{q}_1}{(2\pi)^2}e^{-i\textbf{q}_1\cdot\textbf{b}_1}
			v(0,\textbf{q}_1)v(0,\textbf{q}_2)\times\nonumber\\
	&	&&\times\int	\frac{d^2\textbf{q}_2}{(2\pi)^2}e^{-i\textbf{q}_2\cdot\textbf{b}_2}
			 2ig_s\frac{\epsilon\cdot(\textbf{k}-\textbf{q}_1)}{(\textbf{k}-\textbf{q}_1)^2+M^2x^2+m_g^2}\times\nonumber\\
	&	&&\times\bigg[e^{i(\omega_0+\tilde{\omega}_m)(z_1-z_0)}-e^{i(\omega_0-\omega_1)(z_1-z_0)}\bigg]a_1[c,a_2](T_{a_2}T_{a_1})
\end{alignat}

It is also worth mentioning that the diagrams in in Appendix F of \cite{Djordjevic2004} will also be suppressed in the same manner in the short path length limit.

\chapter{Detailed calculation of Energy Loss Formula} 

\label{AppendixFormula} 

\lhead{\emph{Detailed calculation of Energy Loss Formula}} % Change X to a consecutive letter; this is for the header on each page - perhaps a shortened title

In this section I outline the process whereby one obtains an expression for the first order (in opacity) energy loss by using equation \eqref{ELossExp}.  

We will start with equation \ref{Magda10}

\begin{equation}
d^3N_g^{(1)}d^3N_J=\bigg(\frac{1}{d_T}\Tr\langle\vert\Amp_{1}\vert^2\rangle+\frac{2}{d_T}\Re\Tr\langle\Amp^*_0\Amp_2\rangle\bigg)\frac{d^3\vec{\textbf{p}}}{(2\pi)^32p^0}\frac{d^3\vec{\textbf{k}}}{(2\pi)^32\omega},
\end{equation}
where $\Amp_{1}$ is the sum of all diagrams with one interaction with a scattering centre and $\Amp_{2}$ the sum of all diagrams with two interactions with a single scattering centre.  Therefore, in order to calculate this quantity, we must sum the relevant diagrams first.   Once the summation is done, we square the amplitude (and take the real part of the trace), average over initial and sum over final states.  This means that there is an averaging over impact parameter (introducing a factor of $1/A_\perp$) and a sum over scattering centres (introducing a factor of $N$).

\section{Calculation of $\frac{1}{d_T}\Tr\langle\vert\Amp_{1}\vert^2\rangle$}

Using the results from sections \ref{CalcM100}, \ref{CalcM101} and \ref{CalcM110}, given in equations (\ref{ResM100}), (\ref{ResM101}) and (\ref{ResM110}) and performing a rearrangement of terms so as to group like phases together, we obtain the sum of the one scattering centre diagrams. 
\begin{alignat}{2}
\Amp_{1}	&= 	&&\Amp_{1,0,0}+\Amp_{1,1,0}+\Amp_{1,0,1}\nonumber\\
				&=	&&J(p)e^{i(p+k)x_0}(-i)(2ig_s)T_{a_1}\int
						\frac{d^2\textbf{q}_1}{(2\pi)^2}v(0,\textbf{q}_1)e^{-i\textbf{q}_1\cdot\textbf{b}_1}\times\nonumber\\
				&	&&\times\bigg[\frac{\boldsymbol\epsilon\cdot\textbf{k}}{\textbf{k}^2+m_g^2+M^2x^2}
						\big[e^{i(\omega_0+\tilde{\omega}_m)}-1\big]a_1c \nonumber\\
				&	&&-\frac{\boldsymbol\epsilon\cdot\textbf{k}}{m_g^2+\textbf{k}^2+x^2M^2}
						\bigg[e^{i(\omega_0+\tilde{\omega}_m)(z_1-z_0)}-\frac{1}{2}e^{-\mu_1(z_1-z_0)}\bigg] ca_1\nonumber\\
				&	&&+\frac{\boldsymbol\epsilon\cdot(\textbf{k}-\textbf{q}_1)}{(\textbf{k}-\textbf{q}_1)^2+M^2x^2+m_g^2}\times \nonumber\\
				&	&&\times\Big(e^{i(\omega_0+\tilde{\omega}_m)(z_1-z_0)}-e^{i(\omega_0-\omega_1)(z_1-z_0)}\Big)[c,a_1]T_{a_1}\bigg]
\end{alignat} 
The term that arises from the short path length generalization is the second term on the third to last line.  It is necessary to reorganise the big bracket - we do this so as to group terms with like exponential components.  One has to be careful with the color matrices $a$ and $c$ as they do not commute
\begin{alignat}{2}
\Amp_{1}	&= 	&&\Amp_{1,0,0}+\Amp_{1,1,0}+\Amp_{1,0,1}\nonumber\\
			&=	&&J(p)e^{i(p+k)x_0}(-i)(2ig_s)T_{a_1}\int\frac{d^2\textbf{q}_1}{(2\pi)^2}v(0,\textbf{q}_1)
						e^{-i\textbf{q}_1\cdot\textbf{b}_1}\times\nonumber\\
			&	&&\times\Bigg[\bigg(\frac{\boldsymbol\epsilon\cdot(\textbf{k}-\textbf{q}_1)}{(\textbf{k}-\textbf{q}_1)^2+M^2x^2+m_g^2}
					-\frac{\boldsymbol\epsilon\cdot\textbf{k}}{m_g^2+\textbf{k}^2+x^2M^2}\bigg)
					e^{i(\omega_0+\tilde{\omega}_m)(z_1-z_0)}[c,a_1]\nonumber\\
			&	&&-\frac{\boldsymbol\epsilon\cdot(\textbf{k}-\textbf{q}_1)}{(\textbf{k}-\textbf{q}_1)^2+M^2x^2+m_g^2}
					e^{i(\omega_0-\omega_1)(z_1-z_0)}[c,a_1]\nonumber\\
			&	&&-\frac{\boldsymbol\epsilon\cdot\textbf{k}}{m_g^2+\textbf{k}^2+x^2M^2}a_1c\bigg]\nonumber\\
			&	&&+\frac{1}{2}\frac{\boldsymbol\epsilon\cdot\textbf{k}}{m_g^2+\textbf{k}^2+x^2M^2}e^{-\mu_1(z_1-z_0)}ca_1\Bigg]\nonumber\\
\end{alignat}
To simplify the process above, consider the following short hand:
\begin{align}
f_k			&\equiv	\frac{\boldsymbol\epsilon\cdot\textbf{k}}{m_g^2+\textbf{k}^2+x^2M^2}\nonumber\\
f_q			&\equiv	\frac{\boldsymbol\epsilon\cdot(\textbf{k}-\textbf{q}_1)}{(\textbf{k}-\textbf{q}_1)^2+M^2x^2+m_g^2}\nonumber\\
\omega_{0m}	&\equiv (\omega_0-\tilde{\omega}_m)(z_1-z_0)\nonumber\\
\omega_{01}	&\equiv (\omega_0-\omega_1)(z_1-z_0)\nonumber\\
-\omega_{1m}	&\equiv \omega_{01}- \omega_{0m}=(\omega_0-\omega_1)(z_1-z_0)-(\omega_0-\tilde{\omega}_m)(z_1-z_0)\nonumber\\
			&=-(\omega_{1}--\tilde{\omega}_m)(z_1-z_0)\nonumber\\
\alpha_3	&\equiv \frac{1}{2} e^{-\mu\Delta z}.
\end{align}

Then we may write
\begin{alignat}{2}
\Amp_{1}	&= 	&&\Amp_{1,0,0}+\Amp_{1,1,0}+\Amp_{1,0,1}\nonumber\\
			&=	&&J(p)e^{i(p+k)x_0}(-i)(2ig_s)T_{a_1}\int\frac{d^2\textbf{q}_1}{(2\pi)^2}v(0,\textbf{q}_1)
					e^{-i\textbf{q}_1\cdot\textbf{b}_1}\times\nonumber\\
			&	&&\times \big(f_q-f_k(1-\alpha_3)\big)e^{i\omega_{0m}}[c,a]-f_qe^{i\omega_{01}}[c,a]-f_k(1-\alpha_3)ac,
\end{alignat}

which is a simpler matter to square.  The color matrices have to be handled with care; we will need a few results:
\begin{align}
[c,a][c,a]^*	&=(-1)[c,a][c,a]\nonumber\\
		&=-(ca-ac)(ca-ac)\nonumber\\
		&=-(caca-caac-acca+acac)\nonumber\\
		&=2ccaa-2caca, \qquad\text{(Section \ref{colorRes})}\\
		&\nonumber\\
[c,a]ca	&=(ca-ac)ca\nonumber\\
		&=caca-acca\nonumber\\
		&=acac-acca, \qquad \text{(change of variables in the first term)}\nonumber\\
		&=ac(ac-ca)=ac[a,c]\Rightarrow ca[a,c]=[c,a]ac\\
		&\nonumber\\
[c,a]ca	&=(ca-ac)ca=caca-acca\nonumber\\
		&=-(c^2a^2-caca)\\
		&\nonumber\\
acca	&=aacc \qquad, c^2\propto\textbf{1}
\end{align}

Lastly, we will use the following parameter but our result is slightly more general than the result used in \cite{Djordjevic2004};  On line \eqref{magdaAlpha}, the identity color matrix is assumed in \cite{Djordjevic2004} to be the identity matrix in the adjoint representation - effectively assuming that the hard parton is a gluon.  We will remain more general to allow for a later comparison of gluon vs.\ quark energy loss, at the expense of a slightly more complicated expression.
\begin{align}
\alpha	&\equiv	\Tr\big(c^2a^2-caca\big)\nonumber\\
		&=\Tr\big(c^2a^2-caca-if^{abc}cba\big)\nonumber\\
		&=\Tr(-if^{abc}cba\big)\nonumber\\
		&=\Tr(-i \quad 1/2 \quad iC_2(G)aa\big)\nonumber\\
		&=\frac{1}{2}C_2(G)\Tr\big(C_2(r) \hat{\mathbf{1}}\big)\label{magdaAlpha}\\
		&=\frac{1}{2}C_2(G)C_2(r)d(r)\label{alpha}
\end{align}

Armed with the above results and shorthand, we have compute the sum of the single interaction diagrams.  At this point we keep, for argument's sake, the phases.  At the end we will examine the result in the limit that $\mu_i\gg \omega_i$.  Noting that $\sum \epsilon_{\lambda_{i}}\cdot \epsilon_{\lambda_{j}}^*=\delta_{ij}$
%&	&&\times \big(f_q-f_k(1-\alpha_3)\big)e^{i\omega_{0m}}[c,a]-f_qe^{i\omega_{01}}[c,a]-f_k(1-\alpha_3)ac,
\begin{alignat}{3}
\frac{1}{d_T}\Tr&\langle\vert& \Amp_{1}&\vert^2\rangle =N\vert J(p)\vert^2(4g_s^2)\frac{1}{A_T}\int
				\frac{d^2\textbf{q}_1}{(2\pi)^2}\frac{C_2(T)}{d_A}\times\nonumber\\	
		&\times &\Bigg\{&2\alpha\bigg(f_q-f_k(1-\alpha_3)\bigg)\bigg(f_q-f_k(1-\alpha_3^*)\bigg)+2\alpha f_q^2\nonumber\\
		&	&-	&2\alpha\bigg[\bigg(f_q-f_k(1-\alpha_3)\bigg)f_qe^{i\omega_{1m}}+\bigg(f_q-f_k(1-\alpha_3^*)\bigg)f_qe^{-i\omega_{1m}}\bigg]\nonumber\\
		&	&-	&\alpha\bigg[\big(f_q-f_k(1-\alpha_3)\big)f_k(1-\alpha_3^*e^{-i\omega_{0m}}\big)e^{i\omega_{0m}}\nonumber\\
		&	&	&+\big(f_q-f_k(1-\alpha_3^*)\big)f_k(1-\alpha_3e^{i\omega_{0m}}\big)e^{-i\omega_{0m}}\bigg]\nonumber\\
		&	&+	&\alpha\bigg(f_qf_ke^{i\omega_{01}}(1-\alpha_3^*e^{-i\omega_{0m}})+f_qf_ke^{-i\omega_{01}}(1-\alpha_3e^{i\omega_{0m}})\bigg)\nonumber\\
		&	&+	&\Tr c^2a^2f_kf_q(1-\alpha_3e^{i\omega_{0m}})(1-\alpha_3^*e^{-i\omega_{0m}})\Bigg\}\\
		&=	&N&\vert J(p)\vert^2(4g_s^2)\frac{1}{A_T}\int
				\frac{d^2\textbf{q}_1}{(2\pi)^2}\frac{C_2(T)}{d_A}\times\nonumber\\	
		&	&\times	&\Bigg\{4\alpha f_q^2(1-\cos\omega_{0m})+2\alpha f_k^2(1-\cos\omega_{0m})\nonumber\\
		&	&	&-4\alpha f_kf_q(1-\cos\omega_{1m})+\alpha f_qf_k2\cos\omega_{01}+\Tr c^2a^2f_k^2+\nonumber\\
		&	&+	&e^{-\mu\Delta z}\bigg[f_k^2\Tr c^2a^2(\cos\omega_{0m}-1)-\Tr c^2a^2f_k^2\cos\omega_{0m}\nonumber\\
		&	&	&+f_kf_q\alpha(\cos\omega_{0m}-\cos\omega_{01})\bigg]\nonumber\\
		&	&+	&\frac{1}{4}f_k^2\Tr c^2a^2 e^{-2\mu\Delta z}\Bigg\} \label{M1Full,M1New}
	\end{alignat}

\section{Calculation of $\frac{2}{d_T}\Re\Tr\langle\Amp^*_0\Amp_2\rangle$}

The averaging over impact parameter plays a crucial simplifying role here, 
\begin{align}
\langle e^{-i(\textbf{q}-\textbf{q}')\cdot \textbf{b}}\rangle=\frac{(2\pi)^2}{A_\perp}\delta^2(\textbf{q}-\textbf{q}').
\end{align}

So, consider the results in equations (\ref{ResM203}), (\ref{ResM200}),  (\ref{ResM220}), (\ref{ResM201}) and \eqref{ResM201}. If they are added up and rearranged slightly, again with the purpose of grouping terms with similar exponential components, one finds that
\begin{alignat}{2}
\Amp_2	&=& \Amp&_{2,0,3}+\Amp_{2,0,0}+\Amp_{2,2,0}+\Amp_{2,0,1}+\Amp_{2,0,2}\nonumber\\
		&=	&\frac{1}{2}&J(p)e^{i(p+k)x_0}(-2ig_s)(T_{a_1}T_{a_2})\times\nonumber\\
		&	&\times&\Bigg[\frac{\boldsymbol\epsilon\cdot\textbf{k}}{\textbf{k}^2+M^2x^2+m_g^2}\big[[c,a_2],a_1\big]
				\big( e^{i(\omega_0+\tilde{\omega}_m)(z_1-z_0)}\nonumber\\
		&	&	&-e^{i(\omega_0-\omega_{(12)})(z_1-z_0)}\big)\nonumber\\
		&	&+&\frac{\boldsymbol\epsilon\cdot\textbf{k}}{\textbf{k}^2+M^2x^2+m_g^2}a_2a_1c\big(e^{i(\omega_0-\omega_{(12)})(z_1-z_0)}-1)\nonumber\\
		&	&-&\frac{\boldsymbol\epsilon\cdot\textbf{k}}{\textbf{k}^2+M^2x^2+m_g^2}ca_2a_1
				\Bigg(e^{i(\omega_0-\omega_{(12)})(z_1-z_0)}\nonumber\\
		&	&	&+e^{-\mu_1(z_1-z_0)}\bigg(1-\frac{1}{4}e^{-\mu_1(z_1-z_0)}\bigg)\Bigg)\nonumber\\
		&	&+	&\frac{\boldsymbol\epsilon\cdot(\textbf{k}-\textbf{q}_1)}{(\textbf{k}-\textbf{q}_1)^2+M^2x^2+m_g^2}a_2[c,a_1]
				\big(e^{i(\omega_0+\tilde{\omega}_m)(z_1-z_0)}\nonumber\\
		&	&	&-e^{i(\omega_0-\omega_{1})(z_1-z_0)}\big)\nonumber\\	
		&	&+	&\frac{\boldsymbol\epsilon\cdot(\textbf{k}-\textbf{q}_1)}{(\textbf{k}-\textbf{q}_1)^2+M^2x^2+m_g^2}a_1[c,a_2]
				\big(e^{i(\omega_0+\tilde{\omega}_m)(z_1-z_0)}\nonumber\\
		&	&	&-e^{i(\omega_0-\omega_{1})(z_1-z_0)}\big)\nonumber\\	
\end{alignat}

Here we have used the fact that
\begin{align*}
\mu_1^2	&=\textbf{q}_1^2+\mu^2=\textbf{q}_2^2+\mu^2=\mu_1^2.
\end{align*}

Now we need $\Amp_{0}^*$, which is given in \cite{Djordjevic2004} to be
\begin{align*}
\Amp_{0}	&\approx J(p)e^{ipx_0}(-2ig_s)\frac{\boldsymbol\epsilon\cdot\textbf{k}}{\textbf{k}^2+m_g^2+M^2x^2}e^{i\omega_0z_0}c
\end{align*}

Using the same shorthand as before, we may write

\begin{alignat}{3}
\langle\Amp_{0}^*\Amp_{2}&\rangle &=N&\vert J(p)\vert^2(4g_s^2)\frac{1}{A_T}\int	
			\frac{d^2\textbf{q}_1}{(2\pi)^2}\frac{C_2(T)}{d_A}\nonumber\\
		&	&\times &\bigg[f_k^2\big(2\alpha\cos\omega_{0m}-2\alpha-\Tr c^2a^2\big)\nonumber\\
		&	&	&+2\alpha f_kf_q(\cos\omega_{0m}-\cos\omega_{01})\nonumber\\
		&	&	&+e^{-\mu \Delta z}f_k^2\Tr c^2a^2\cos\omega_{0m}-\frac{1}{4} e^{-2\mu \Delta z}f_k^2\bigg]\label{M2Full}
\end{alignat}

\section{Addition of summed terms}
It now remains simply to add the contributions from equations \eqref{M1Full,M1New} and \eqref{M2Full} which, after a number of cancellations, gives
\begin{alignat}{3}
\frac{1}{d_T}\Tr\langle &\vert &\Amp_{1}&\vert^2\rangle+\frac{2}{d_T}\Re\Tr\langle\Amp^*_0\Amp_2\rangle	
		= N\vert J(p)\vert^2(4g_s^2)\frac{1}{A_T}\int	
			\frac{d^2\textbf{q}_1}{(2\pi)^2}\frac{C_2(T)}{d_A}\times\nonumber\\
		&\times	& \bigg[&-2f_q(f_k-f_q)(1-\cos\omega_{1m})\nonumber\\
		&		&+	&e^{-\mu \Delta z}\bigg(\alpha f_k^2(1-\cos\omega_{0m})+f_k^2(\cos\omega_{0m}-1)\Tr c^2 a ^2\nonumber\\
		&		&	&+f_kf_q\alpha(\cos\omega_{0m}-\cos\omega_{01})\bigg)\bigg]
\end{alignat}

\section{Final Calculation}

We now want to extract $dE_{ind}^{(1)}\equiv \omega d^3N_g$. From \cite{Djordjevic2004}, we know that 

\begin{align}\label{Magda8}
d^3N_J	&=d_R\vert J(p)\vert ^2\frac{d^3\vec{\textbf{p}}}{(2\pi)^22p^0},
\end{align}

which we can use, along with equation \eqref{Magda10}, (equation (10) from \cite{Djordjevic2004} to find
\begin{align}
dE_{ind}^{(1)}&=\omega d^3N_g =\cancel{\omega} \bigg(\frac{1}{d_T}\Tr\langle\vert\Amp_{1}\vert^2
			\rangle+\frac{2}{d_T}\Re\Tr\langle\Amp^*_0\Amp_2\rangle\bigg)
			\frac{\cancel{\frac{d^3\vec{\textbf{p}}}{(2\pi)^32p^0}}\frac{d^3\vec{\textbf{k}}}{(2\pi)^32\cancel{\omega}}}
			{d_R\vert J(p)\vert ^2\cancel{\frac{d^3\vec{\textbf{p}}}{(2\pi)^32p^0}}}			
\end{align}

We now do a kind of averaging of the position of the $z_1$ because the fraction of energy carried off by the radiated gluon seems to depend on the distance between the production ($z_0$) and the interaction with the medium ($z_1$):
\begin{align}
\bigg\langle\frac{dE}{dx}\bigg\rangle &=\int dz_1P(z_1)\frac{dE}{dx}(z_1)
&=\int	dz_1\frac{dE}{dx}(z_1)\exp{\bigg\{-2\frac{(z_1-z_0)}{L}\bigg\}\frac{2}{L}}
\end{align}

This is an assumed distribution.  It tries to take into account the rapidly exopanding medium so that the distance between scattering centres increases exponentially.  Noting that
\begin{align*}
d^3\vec{\textbf{k}}	&=dk^zd^2\textbf{k}=dk^0d^2\textbf{k}\nonumber\\
		&=d(xE^+)d^2\textbf{k}=E^+dxd^2\textbf{k}
		=2Edxd^2\textbf{k}
\end{align*}

For clarity, let
\begin{alignat}{3}
[\star]	&\equiv \bigg[&-&2f_q(f_k-f_q)(1-\cos\omega_{1m})\nonumber\\
		&		&+	&e^{-\mu \Delta z}\bigg(\alpha f_k^2(1-\cos\omega_{0m})+f_k^2(\cos\omega_{0m}-1)\Tr c^2 a ^2\nonumber\\
		&		&	&+f_kf_q\alpha(\cos\omega_{0m}-\cos\omega_{01})\bigg)\bigg]
\end{alignat}

We will also be using a `clever ``1'' ':  From \cite{Gyulassy2001}, we have a number of relations that will be useful.
\begin{itemize}
\item For a homogeneous rectangular target of thickness $L$, the density is $\rho=N/LA_\perp$.
\item Since the mean free path is $\lambda=1/\rho\sigma_{el}$, the opacity is simply $L/\lambda=N\sigma_{el}/A_\perp$.
\item In perturbation theory, $\mu/\lambda\approx 4\pi\alpha_s^2\rho$.
\end{itemize}
And so we can wrangle a convenient `one':
\begin{align}
\frac{\mu^2}{\lambda}	&=4\pi	\alpha_s^2\rho\nonumber\\
		&=\frac{4\pi\alpha_s^2N}{LA_\perp}\nonumber\\
\Rightarrow	1	&=\frac{\mu^2LA_\perp}{\lambda 4\pi\alpha_s^2 N}	
\end{align}
%------------------------------------------------
%\int dz_1			\frac{2}{L}\exp\bigg\{-2\frac{(z_1-z_0)}{L}
%---------------------------------------------------------------
We start with
\begin{align}
d&E_{ind}^{(1)}	=\omega d^3N_g = \frac{1}{d_R\cancel{\vert J(p)\vert^2}}\bigg(N\cancel{\vert J(p)\vert^2}(4g_s^2)
			\frac{C_2(T)c_R^2d_R}{A_\perp}\times\nonumber\\
		&\times\int\frac{d\textbf{q}_1}{(2\pi)^2}\vert v(\textbf{q}_1)\vert^2 [\star]
			\frac{d^2\vec{\textbf{k}}}{(2\pi)^3}\frac{1}{2}\int dz_1\bar{\rho}(z_1)\bigg)\nonumber\\
		&=\frac{4Ng_s^2(4\pi\alpha_s)}{A_\perp}\frac{C_2(T)}{d_A}C_R^2\int d^2\textbf{q}_1 \vert v(\textbf{q}_1)\vert^2\times\nonumber\\
		&\times E^+\int	d^2\textbf{k}dx\frac{1}{2}\frac{1}{(2\pi)^5}\int dz_1\bar{\rho}(z_1)[\star]\nonumber\\
		&=\frac{N\cancel{4}(4\pi\alpha_s)}{A_\perp}C_R\frac{1}{\cancel{2}}\int d^2\textbf{q}_1
			\frac{(4\pi\alpha_s)^2}{(\textbf{q}_1^2+\mu^2)^2}\times\\
		&\times E^+\int	d^2\textbf{k}dx\frac{1}{\cancel{2}}\frac{1}{(2\pi)^5}\int dz_1\bar{\rho}(z_1)\bigg\}[\star]\nonumber\\
\intertext{Here we multiply by the above strategic `1' and divide $dx$ out}
\frac{dE_{ind}^{(1)}}{dx}	&=\frac{\cancel{N}E^+C_R}{\cancel{A_\perp}}\int d^2\textbf{q}_1
			\frac{(4\pi\alpha_s)^3}{(\textbf{q}_1^2+\mu^2)^2}\times\\
		&\times\int	d^2\textbf{k}\frac{1}{(2\pi)^5}
			\frac{\mu^2L\cancel{A_\perp}}{\lambda 4\pi\alpha_s^2\cancel{N}}\frac{1}{(2\pi)^5}\int dz_1\bar{\rho}(z_1)\bigg\}[\star]\nonumber\\
		&=	\int d^2\textbf{q}_1\int	d^2\textbf{k}
			\frac{E^+C_R64\pi^3\alpha_s^3\mu^2L}{(\textbf{q}_1^2+\mu^2)^2 32\pi^6\lambda 4 \alpha_s^2}
			\times\int dz_1\frac{2}{L}\exp\bigg\{-2\frac{(z_1-z_0)}{L}\bigg\}[\star]\nonumber\\
\intertext{Using the fact that $E^+\approx 2E$}
		&=\frac{C_R\alpha_sLE}{\pi\lambda}	\int \frac{d^2\textbf{q}_1}{\pi}\frac{\mu^2}{(\mu^2+\textbf{q}_1^2)^2}
			\frac{d^2\textbf{k}}{4\pi}			dz_1\exp{\bigg\{-2\frac{(z_1-z_0)}{L}\bigg\}\frac{2}{L}}	[\star],\nonumber\\
\end{align}

which expands to
\begin{alignat}{3}
		=\frac{C_R\alpha_sLE}{\pi\lambda}&\times & \int& \frac{d^2\textbf{q}_1}{\pi}\frac{\mu^2}{(\mu^2+\textbf{q}_1^2)^2}
			\frac{d^2\textbf{k}}{\pi}			\int dz_1\exp{\bigg\{-2\frac{(z_1-z_0)}{L}\bigg\}\frac{2}{L}}	\times\nonumber\\
		&\times&\Bigg[&-2f_q(f_k-f_q)(1-\cos\omega_{1m})\nonumber\\
		&		&+	&\frac{1}{2}e^{-\mu \Delta z}\Bigg( f_k^2(1-\cos\omega_{0m})\bigg(1-\frac{\Tr c^2 a^2}{\alpha}\bigg)\nonumber\\
		&		&	&+f_kf_q\alpha\big(\cos\omega_{0m}-\cos\omega_{01}\big)\Bigg)\Bigg]							
\end{alignat}

At this point we pause to look at a few results.  Firstly, the $z_1$ integral for the first term in the square brackets (the leading order term derived by \cite{Djordjevic2004} is analytically known:
\begin{align}
\int^\infty_0 dz_1&\exp{\bigg\{-2\frac{(z_1-z_0)}{L}\bigg\}\frac{2}{L}}\big(1-\cos\big\{(\omega_1+\tilde{\omega}_m)(z_1-z_0)\big\}\big)\nonumber\\
		&=\frac{(\omega_1+\tilde{\omega}_m)^2L^2}{4+(\omega_1+\tilde{\omega}_m)^2L^2}\nonumber\\
		&=\frac{\cancel{L^2}\big[(\textbf{k}-\textbf{q}_1)^2+m_g^2+M^2x^2\big]^2}
			{\cancel{L^2}\big[(\textbf{k}-\textbf{q}_1)^2+m_g^2+M^2x^2\big]^2+\big(\frac{4p_zx}{L}\big)^2}
\end{align}

The $z_1$ integral can also be performed analytically for the next to leading order terms. They are cumbersome, but their form is insightful so I present them here. Term by term then, we have, for the next to leading order correction terms for the short separation distance generalization,
\begin{align}
\int^\infty_0 dz_1&\exp{\bigg\{-2\frac{(z_1-z_0)}{L}\bigg\}\frac{2}{L}}(\cos\{(\omega_0-\tilde{\omega}_m)(z_1-z_0)\}-1)\nonumber\\
	&=\frac{2L^2 (\omega_0-\tilde{\omega}_m)}{(2+L\mu_1)\big(4+4L\mu_1+L^2\big((\omega_0-\tilde{\omega}_m)^2+\mu_1^2\big)\big)}\\
\int^\infty_0 dz_1&\exp{\bigg\{-2\frac{(z_1-z_0)}{L}\bigg\}\frac{2}{L}}\times\nonumber\\
	&\times\big(\cos\{(\omega_0-\tilde{\omega}_m)(z_1-z_0)\}-\cos\{(\omega_0-\omega_1)(z_1-z_0)\}\big)\nonumber\\
	&=-\frac{2(2+L\mu_1)}{4+4L\mu_1+L^2\big((\omega_0-\omega_1)^2+\mu_1^2\big)}\nonumber\\
	&+\frac{2(2+L\mu_1)} {4+4L\mu_1+L^2\big((\omega_0-\tilde{\omega}_m)^2+\mu_1^2\big)}
\end{align}

Although further simplifications can be performed on the leading order terms and despite the fact that analytical expressions exist for the integrals over the separation distance, the expression becomes unnecessarily involved.  I therefore present here the full formula with corrections, but before the evaluation of the $z_1$ integral:

\begin{alignat}{2}
\Delta E_{ind}^{(1)}& =&&\frac{C_R\alpha_sLE}{\pi\lambda_g}	\int \frac{d^2\textbf{q}_1}{\pi}\frac{\mu^2}{(\mu^2+\textbf{q}_1^2)^2}
				\frac{d^2\textbf{k}}{4\pi} \int d\Delta z\bar{\rho}(\Delta z)\times\nonumber\\
		&\times && \Bigg[-\frac{2\big(1-\cos\big\{(\omega_1+\tilde{\omega}_m)\Delta z\big\}\big)}
				{(\textbf{k}-\textbf{q}_1)^2+M^2x^2+m_g^2}\times\nonumber\\
		&	&&\times\bigg(\frac{(\textbf{k}-\textbf{q}_1)\cdot\textbf{k}}{m_g^2+\textbf{k}^2+x^2M^2}
				-\frac{(\textbf{k}-\textbf{q}_1)^2}{(\textbf{k}-\textbf{q}_1)^2+M^2x^2+m_g^2}\bigg)\nonumber\\
		&	&&+\frac{1}{2}e^{-\mu\Delta z)}\Bigg\{\bigg(\frac{\textbf{k}}{m_g^2+\textbf{k}^2+x^2M^2}\bigg)^2\times\nonumber\\
		&	&&\times\bigg(1-\frac{2 C_R}{C_A}\bigg)
				\bigg(1-\cos\{(\omega_0-\tilde{\omega}_m)\Delta z\}\bigg)\nonumber\\
		&	&&+	\frac{\textbf{k}\cdot (\textbf{k}-\textbf{q}_1)}
				{\big(\textbf{k}^2+m_g^2+x^2M^2\big)\big((\textbf{k}-\textbf{q}_1)^2+M^2x^2+m_g^2\big)}\times\nonumber\\
		&	&&\times\big(\cos\{(\omega_0-\tilde{\omega}_m)\Delta z\}-\cos\{(\omega_0-\omega_1)\Delta z\}\big)\Bigg\}\Bigg]\end{alignat}

\addtocontents{toc}{\vspace{2em}} 

\backmatter

%----------------------------------------------------------------------------------------
%	BIBLIOGRAPHY
%----------------------------------------------------------------------------------------

\label{Bibliography}

\lhead{\emph{Bibliography}}
\bibliographystyle{unsrt}

\bibliography{Calculation,CalculationExtra}

\begin{thebibliography}{10}
\bibitem{Steinberg2014}
Peter Steinberg.
\newblock {What have we learned about the quark–gluon plasma with the ATLAS
  detector at the LHC?}
\newblock {\em Nucl. Phys. A}, 932:9--16, 2014.
\end{thebibliography}
\end{document}